\documentclass[%
 reprint,
superscriptaddress,
 amsmath,amssymb,
 aps,
 pre
]{revtex4-2}
\allowdisplaybreaks[1]
\usepackage{xr}
\UseRawInputEncoding
\usepackage{amsmath}
\usepackage[]{algorithm2e}
\usepackage{tikz}
\usetikzlibrary{shapes,arrows,positioning}
\usepackage{pgfplots}
\tikzset{decision/.style={diamond, draw, fill=yellow!20, 
    text width=10em, aspect=2, text badly centered, node distance=3cm, inner sep=0pt, minimum height=4em},
block/.style={rectangle, draw, fill=gray!20, 
    text width=15em, text badly centered, rounded corners, minimum height=4em},
small_block/.style={circle, draw, fill=white!20, 
    text width=0em, text centered, rounded corners, minimum height=0em},    
line/.style={draw, -latex'},
cloud/.style={draw, ellipse,fill=red!20, node distance=3cm,text width=15em,
    minimum height=6em, text centered,},}

\usepackage[caption=false]{subfig}

\usepackage{array}[=2016-10-06]

\usepackage{comment}
\usepackage{graphicx}
\usepackage{dcolumn}
\usepackage{bm}
\usepackage{xcolor}
\usepackage{soul}
\usepackage{ulem}
\usepackage[hidelinks,colorlinks=true,linkcolor=blue,citecolor=blue]{hyperref}

\makeatletter
\DeclareFontFamily{OMX}{MnSymbolE}{}
\DeclareSymbolFont{MnLargeSymbols}{OMX}{MnSymbolE}{m}{n}
\SetSymbolFont{MnLargeSymbols}{bold}{OMX}{MnSymbolE}{b}{n}
\DeclareFontShape{OMX}{MnSymbolE}{m}{n}{
    <-6>  MnSymbolE5
   <6-7>  MnSymbolE6
   <7-8>  MnSymbolE7
   <8-9>  MnSymbolE8
   <9-10> MnSymbolE9
  <10-12> MnSymbolE10
  <12->   MnSymbolE12
}{}
\DeclareFontShape{OMX}{MnSymbolE}{b}{n}{
    <-6>  MnSymbolE-Bold5
   <6-7>  MnSymbolE-Bold6
   <7-8>  MnSymbolE-Bold7
   <8-9>  MnSymbolE-Bold8
   <9-10> MnSymbolE-Bold9
  <10-12> MnSymbolE-Bold10
  <12->   MnSymbolE-Bold12
}{}

\let\llangle\@undefined
\let\rrangle\@undefined
\DeclareMathDelimiter{\llangle}{\mathopen}
                     {MnLargeSymbols}{'164}{MnLargeSymbols}{'164}
\DeclareMathDelimiter{\rrangle}{\mathclose}
                     {MnLargeSymbols}{'171}{MnLargeSymbols}{'171}
\makeatother

\newcommand{\plaind}{\mathrm{d}}
\newcommand{\dint}[1]{\mathchoice{\!\plaind#1\,}{\!\plaind#1\,}{\!\plaind#1\,}{\!\plaind#1\,}}

\newcommand{\ddintx}[2]{\mathchoice{\!\plaind^{#2}#1\,}{\!\plaind^{#2}#1\,}{\!\plaind^{#2}#1\,}{\!\plaind^{#2}#1\,}}

\newcommand{\dbar}{\text{\dj}}

\newcommand{\Dint}[1]{\!\!\mathcal{D}#1\,}
\DeclareFontFamily{U}{wncy}{}
\DeclareFontShape{U}{wncy}{m}{n}{<->wncyr10}{}
\DeclareSymbolFont{mcy}{U}{wncy}{m}{n}
\DeclareMathSymbol{\sha}{\mathord}{mcy}{"58}

\newcommand{\MCLineVsMathEnv}[2]{\mathchoice{#1}{#2}{#2}{#2}}
\newcommand{\dX}[1]{\plaind #1}

\newcommand{\ddX}[1]{\MCLineVsMathEnv{\frac{\plaind}{\plaind #1}}{\plaind/\plaind #1}}

\newcommand{\ddXX}[2]{\MCLineVsMathEnv{\frac{\plaind #1}{\plaind #2}}{\plaind #1/\plaind #2}}

\newcommand{\ppX}[1]{\MCLineVsMathEnv{\frac{\partial}{\partial #1}}{\partial/\partial #1}}

\newcommand{\ppXX}[2]{\MCLineVsMathEnv{\frac{\partial #1}{\partial #2}}{\partial #1/\partial #2}}

\usepackage{dsfont}

\newcommand{\gpset}[1]{\mathds{#1}}

\newcommand{\canetset}[1]{{\mathchoice {\hbox{$\sf\textstyle #1\kern-0.4em #1$}}
{\hbox{$\sf\textstyle #1\kern-0.4em #1$}}
{\hbox{$\sf\scriptstyle #1\kern-0.3em #1$}}
{\hbox{$\sf\scriptscriptstyle #1\kern-0.2em #1$}}}}

\def\nbZ{{\mathchoice {\hbox{$\sf\textstyle Z\kern-0.4em Z$}}
{\hbox{$\sf\textstyle Z\kern-0.4em Z$}}
{\hbox{$\sf\scriptstyle Z\kern-0.3em Z$}}
{\hbox{$\sf\scriptscriptstyle Z\kern-0.2em Z$}}}}

\newcommand{\gpvec}[1]{\mathbf{#1}}

\newcommand{\zerovec}{\gpvec{0}}

\newcommand{\ident}{\mathds{1}}

\newcommand{\BC}{\mathcal{B}}
\newcommand{\CC}{\mathcal{C}}

\newcommand{\GC}{\mathcal{G}}

\newcommand{\LC}{\mathcal{L}}
\newcommand{\MC}{\mathcal{M}}
\newcommand{\NC}{\mathcal{N}}
\newcommand{\OC}{\mathcal{O}}

\newcommand{\WC}{\mathcal{W}}

\newcommand{\mutilde}{\tilde{\mu}}

\newcommand{\fourth}{\mathchoice{\frac{1}{4}}{(1/4)}{\frac{1}{4}}{(1/4)}}

\newcommand{\Exp}[1]{\operatorname{exp}\left(#1\right)}
\renewcommand{\exp}[1]{\mathchoice
{\mathrm{e}^{#1}}
{\operatorname{exp}(#1)}
{\operatorname{exp}\left(#1\right)}
{\operatorname{exp}\left(#1\right)}}

\newcommand{\elabel}[1]{\label{eq:#1}}
\newcommand{\eref}[1]{(\ref{eq:#1})}
\newcommand{\Eref}[1]{\mbox{Eq.~(\ref{eq:#1})}}
\newcommand{\Erefs}[1]{\mbox{Eqs.~(\ref{eq:#1})}}

\newcommand{\seclabel}[1]{\label{sec:#1}}

\newcommand{\Sref}[1]{Sec.~\ref{sec:#1}}

\newcommand{\SMref}[1]{\mbox{Suppl.~\ref{sec:#1}} \cite{PRRsupplement}}
\newcommand{\SMrefLocal}[1]{Suppl.~\ref{sec:#1}}

\newcommand{\Fref}[2][]{Fig.~\ref{fig:#2}#1}

\newcommand{\tlabel}[1]{\label{tab:#1}}
\newcommand{\Tref}[1]{Tab.~\ref{tab:#1}}

\newcommand{\latin}[1]{{\it #1}}
\newcommand{\ie}{\latin{i.e.}\@\xspace}

\newcommand{\cf}{\latin{cf.}\@\xspace}

\newlength \standardfigwidth
\DeclareMathAlphabet{\matheub}{U}{eur}{m}{n}
\newcommand{\unitX}[1]{\matheub{#1}}

\newcommand{\ai}[4]{{\left[\begin{smallmatrix}\textcolor{cyan}{#1}&\textcolor{cyan}{#2}\\\textcolor{orange}{#3}&\textcolor{orange}{#4}\end{smallmatrix}\right]}}

\usepackage{tikz}
\usetikzlibrary{decorations.pathmorphing}
\usetikzlibrary{decorations.markings}
\usetikzlibrary{arrows,shapes,decorations,automata,backgrounds,petri}
\usetikzlibrary{calc}
\usetikzlibrary{patterns}
\usetikzlibrary{decorations.text}

\newcommand\Overline[2][1pt]{
    \begin{tikzpicture}[baseline=(a.base)]
      \node[inner xsep=0pt,inner ysep=1.5pt] (a) {$#2$};
      \draw[line width= #1] (a.north west) -- (a.north east);
    \end{tikzpicture}
    }

\makeatletter
\let\save@mathaccent\mathaccent
\newcommand*\if@single[3]{
  \setbox0\hbox{${\mathaccent"0362{#1}}^H$}
  \setbox2\hbox{${\mathaccent"0362{\kern0pt#1}}^H$}
  \ifdim\ht0=\ht2 #3\else #2\fi
  }
\newcommand*\rel@kern[1]{\kern#1\dimexpr\macc@kerna}
\newcommand*\wideaccent[2]{\@ifnextchar^{{\wide@accent{#1}{#2}{0}}}{\wide@accent{#1}{#2}{1}}}
\newcommand*\wide@accent[3]{\if@single{#2}{\wide@accent@{#1}{#2}{#3}{1}}{\wide@accent@{#1}{#2}{#3}{2}}}
\newcommand*\wide@accent@[4]{
  \begingroup
  \def\mathaccent##1##2{
    \let\mathaccent\save@mathaccent
    \if#42 \let\macc@nucleus\first@char \fi
    \setbox\z@\hbox{$\macc@style{\macc@nucleus}_{}$}
    \setbox\tw@\hbox{$\macc@style{\macc@nucleus}{}_{}$}
    \dimen@\wd\tw@
    \advance\dimen@-\wd\z@
    \divide\dimen@ 3
    \@tempdima\wd\tw@
    \advance\@tempdima-\scriptspace
    \divide\@tempdima 10
    \advance\dimen@-\@tempdima
    \ifdim\dimen@>\z@ \dimen@0pt\fi
    \rel@kern{0.6}\kern-\dimen@
    \if#41
      #1{\rel@kern{-0.6}\kern\dimen@\macc@nucleus\rel@kern{0.4}\kern\dimen@}
      \advance\dimen@0.4\dimexpr\macc@kerna
      \let\final@kern#3
      \ifdim\dimen@<\z@ \let\final@kern1\fi
      \if\final@kern1 \kern-\dimen@\fi
    \else
      #1{\rel@kern{-0.6}\kern\dimen@#2}
    \fi
  }
  \macc@depth\@ne
  \let\math@bgroup\@empty \let\math@egroup\macc@set@skewchar
  \mathsurround\z@ \frozen@everymath{\mathgroup\macc@group\relax}
  \macc@set@skewchar\relax
  \let\mathaccentV\macc@nested@a
  \if#41
    \macc@nested@a\relax111{#2}
  \else
    \def\gobble@till@marker##1\endmarker{}
    \futurelet\first@char\gobble@till@marker#2\endmarker
    \ifcat\noexpand\first@char A\else
      \def\first@char{}
    \fi
    \macc@nested@a\relax111{\first@char}
  \fi
  \endgroup
}
\makeatother

\newcommand\doubleoverline[1]{\overline{\overline{#1}}}

\makeatletter
\newcommand{\spave}[2][]{\@ifempty{#1}{}{\left.}\Overline[1pt]{#2}\@ifempty{#1}{}{\right|_{#1}}}
\newcommand{\thinspave}[2][]{\@ifempty{#1}{}{\left.}\overline{#2}\@ifempty{#1}{}{\right|_{#1}}}
\newcommand{\dave}[2][]{\mathchoice
{\left\llangle #2 \right\rrangle_{#1}}
{\left\llangle #2 \right\rrangle_{#1}}
{\llangle #2\rrangle_{#1}}
{\llangle #2\rrangle_{#1}}}
\newcommand{\ave}[2][]{\mathchoice
{\left\langle #2 \right\rangle_{#1}}
{\left\langle #2 \right\rangle_{#1}}
{\langle #2\rangle_{#1}}
{\langle #2\rangle_{#1}}}
\newcommand{\aveS}[3][]{\mathchoice
{\left\langle #2 \right\rangle_{#1}^{#3}}
{\left\langle #2 \right\rangle_{#1}^{#3}}
{\langle #2\rangle_{#1}^{#3}}
{\langle #2\rangle_{#1}^{#3}}}
\newcommand{\have}[2][]{\@ifempty{#1}{}{\left.}\widehat{#2}\@ifempty{#1}{}{\right|_{#1}}}

\newcommand\doverline[1]{
\tikz[baseline=(nodeAnchor.base)]{
    \node[inner sep=0] (nodeAnchor) {$#1$};
    \draw[line width=0.1ex,line cap=round]
        ($(nodeAnchor.north west)+(0.0em,0.2ex)$)
            --
        ($(nodeAnchor.north east)+(0.0em,0.2ex)$)
        ($(nodeAnchor.north west)+(0.0em,0.5ex)$)
            --
        ($(nodeAnchor.north east)+(0.0em,0.5ex)$)
    ;
}}

\newcommand{\Xbave}[2][]{\@ifempty{#1}{}{\left.}\overline{\overline{#2}}\@ifempty{#1}{}{\right|_{#1}}}
\newcommand{\bave}[2][]{\@ifempty{#1}{}{\left.}\doverline{#2}\@ifempty{#1}{}{\right|_{#1}}}
\makeatother

\newcommand{\Xave}[2][]{\thinspave[#1]{#2}}
\newcommand{\XIave}[2][]{\dave{#2}_{#1}}
\newcommand{\WSTARave}[2][\!*]{\ave[#1]{#2}}
\newcommand{\WSTARaveS}[3][\!*]{\aveS[#1]{#2}{#3}}
\newcommand{\Wave}[2][]{\ave[#1]{#2}}

\newcommand{\PprobX}{\mathds{P}}

\newcommand{\PprobXW}{\mathds{P}}
\newcommand{\PprobWSTAR}{\mathds{P}_{\!*}}
\newcommand{\PprobW}{\mathds{P}}

\newcommand{\NormWSTAR}{\NC_{\!*}}

\makeatletter
\newcommand{\creatX}[2][]{a^{\@ifempty{#1}{\dagger}{\dagger\,#1}}\@ifempty{#2}{}{(#2)}}
\makeatother
\newcommand{\creatDS}{\tilde{a}}
\makeatletter
\newcommand{\creatDSX}[2][]{\@ifempty{#1}{\creatDS}{\creatDS^{#1}}\@ifempty{#2}{}{(#2)}}
\makeatother

\makeatletter
\newcommand{\annihX}[2][]{a\@ifempty{#1}{}{^{#1}}\@ifempty{#2}{}{(#2)}}
\newcommand{\kernel}[1]{K\@ifempty{#1}{}{\,\!\!\left(#1\right)}}
\newcommand{\kernelDeri}[1]{K'\@ifempty{#1}{}{\left(#1\right)}}

\newcommand{\tableofappendixcontents}{\@starttoc{toa}}
\newcommand{\l@toact}[2]{\@dottedtocline{1}{1.5em}{5.5em}{#1}{#2}}
\newcommand{\l@toactspace}[2]{\ \hfil}
\makeatother

\newcommand{\tilden}{\widetilde{n}}

\newcommand{\tildeq}{\widetilde{q}}

\newcommand{\xdot}{\dot{x}}

\newcommand{\sdots}{\ifmmode\mathinner{\ldotp\kern-0.2em\ldotp\kern-0.2em\ldotp}\else.\kern-0.13em.\kern-0.13em.\fi}

\usepackage{xspace}

\newcommand{\transform}{\mathcal{T}}

\newcommand{\transMatLin}{\MC}
\newcommand{\transMatNonLin}{\BC}

\newcommand{\Transition}{\WC}

\newcommand{\EPR}{\dot{S}}
\newcommand{\EPRfull}{\EPR_{x,w}}

\newcommand{\EPRpartX}{\EPR_{x}}
\newcommand{\SymParity}{\mathcal{P}}

\newcommand{\SymTime}{\mathcal{T}}

\newcommand{\SymParityTime}{\mathcal{PT}}

\newcommand{\xpath}{\{x(t)\}}

\newcommand{\Rxpath}{\{x(T-t)\}}

\newcommand{\wpath}{\{w(t)\}}
\newcommand{\wdpath}{\{w'(t)\}}

\newcommand{\wddpath}{\{w''(t)\}}
\newcommand{\wdddpath}{\{w'''(t)\}}
\newcommand{\Mwpath}{\{-w(t)\}}
\newcommand{\Rwpath}{\{w(T-t)\}}

\makeatletter
\newcommand{\pushright}[1]{\ifmeasuring@#1\else\omit\hfill$\displaystyle#1$\fi\ignorespaces}
\newcommand{\pushleft}[1]{\ifmeasuring@#1\else\omit$\displaystyle#1$\hfill\fi\ignorespaces}
\makeatother

\makeatletter
\newcommand{\corrStar}[1][]{\@ifempty{#1}{}{}\CC\@ifempty{#1}{}{^{(#1)}}}
\makeatother
\newcommand{\gpmatrix}[1]{\bm{\mathsf{#1}}}

\newcommand{\cbb}{\text{(c)}}

\newcommand{\articleTitle}{Partial Entropy production of active particles with hidden states in potentials}

\newcommand{\potential}[1]{V\left(#1\right)}
\newcommand{\potentialPrime}[1]{V'\left(#1\right)}
\newcommand{\potentialDoublePrime}[1]{V''\left(#1\right)}

\newcommand{\potentialx}{\potential{x}}
\newcommand{\potentialxt}{\potential{x(t)}}
\newcommand{\potentialxtn}[1]{\potentialPrime{x(t_{#1})}}
\newcommand{\potentialxtnShort}[1]{\potentialPrime{x_{#1}}}

\newcommand{\potentialxtzero}{\potential{x(0)}}
\newcommand{\potentialxtT}{\potential{x(T)}}
\newcommand{\potentialprimex}{\potentialPrime{x}}
\newcommand{\potentialprimext}{\potentialPrime{x(t)}}
\newcommand{\potentialprimextR}{\potentialPrime{x(T-t)}}
\newcommand{\potentialdoubleprimex}{\potentialDoublePrime{x}}
\newcommand{\potentialdoubleprimext}{\potentialDoublePrime{x(t)}}
\newcommand{\potentialdoubleprimextR}{\potentialDoublePrime{x(T-t)}}

\newcommand{\xplusVn}[1]{(\xdot(t_{#1})+\potentialxtn{#1})}
\newcommand{\xplusVnShort}[1]{(\xdot(t_{#1})+\potentialxtnShort{#1})}
\newcommand{\xplusVnShortR}[1]{(-\xdot(t_{#1})+\potentialxtnShort{#1})}
\newcommand{\xdotn}[1]{\xdot(t_{#1})}
\newcommand{\Cancel}[2][black]{{\color{#1}\cancel{\color{black}#2}}}

\newcommand{\CancelMan}[2][]{%
  \tikz[baseline=(X.base),inner sep=0pt]{%
    \node (X) {$#2$};
    \tikzset{#1}
    \draw[draw=#1, overlay, shorten >=-2pt, shorten <=-2pt] 
      (X.south west) -- (X.north east);
  }%
}
\newcommand{\DoubleCancelMan}[3]{%
  \tikz[baseline=(X.base),inner sep=0pt]{%
    \node (X) {$#3$};
    \draw[draw=#1, line width=0.5pt, overlay, shorten >=-2pt, shorten <=-2pt]
      ([xshift=0pt]X.south west) -- ([xshift=-5pt]X.north east);
    \draw[draw=#2, line width=0.5pt, overlay, shorten >=-2pt, shorten <=-2pt]
      ([xshift=5pt]X.south west) -- ([xshift=0pt]X.north east);
  }%
}

\newcommand{\bCancelMan}[2][]{%
  \tikz[baseline=(X.base),inner sep=0pt]{%
    \node (X) {$#2$};
    \tikzset{#1}
    \draw[draw=#1, overlay, shorten >=-2pt, shorten <=-2pt] 
      (X.north west) -- (X.south east);
  }%
}
\newcommand{\bDoubleCancelMan}[3]{%
  \tikz[baseline=(X.base),inner sep=0pt]{%
    \node (X) {$#3$};
    \draw[draw=#1, line width=0.5pt, overlay, shorten >=-2pt, shorten <=-2pt]
      ([xshift=20pt]X.north west) -- ([xshift=0pt]X.south east);
    \draw[draw=#2, line width=0.5pt, overlay, shorten >=-2pt, shorten <=-2pt]
      ([xshift=0pt]X.north west) -- ([xshift=-20pt]X.south east);
  }%
}
\newcommand{\bTripleCancelMan}[4]{%
  \tikz[baseline=(X.base),inner sep=0pt]{%
    \node (X) {$#4$};
    \draw[draw=#1, line width=0.5pt, overlay, shorten >=-2pt, shorten <=-2pt]
      ([xshift=40pt]X.north west) -- ([xshift=0pt]X.south east);
    \draw[draw=#2, line width=0.5pt, overlay, shorten >=-2pt, shorten <=-2pt]
      ([xshift=20pt]X.north west) -- ([xshift=-20pt]X.south east);
    \draw[draw=#3, line width=0.5pt, overlay, shorten >=-2pt, shorten <=-2pt]
      ([xshift=0pt]X.north west) -- ([xshift=-40pt]X.south east);
  }%
}

\newcommand{\triplePartial}[3]{
\partial_{t_#1}\partial_{t_#2}\partial_{t_#3}
}

\usepackage{cancel}

\makeatletter
\newcommand{\customlabel}[2]{%
   \protected@write \@auxout {}{\string \newlabel {#1}{{#2}{\thepage}{#2}{#1}{}} }%
   \hypertarget{#1}{\hspace{0pt}}
}
\makeatother

\begin{document}

\customlabel{sec:OU_corr_star}{SI}
\customlabel{sec:correlators_one_dot}{SII}
\customlabel{sec:four_time_correlator}{SIII}
\customlabel{sec:rnt_epr_derivation}{SIV}
\customlabel{sec:sixth_order}{SV}
\customlabel{eq:forms_of_Ld}{S27}

\preprint{}

\title{\articleTitle}

\author{Jacob Knight}
\email{jwk21@ic.ac.uk}
\affiliation{Department of Mathematics, Imperial College London, South Kensington, London SW7 2BZ, UK}

\author{Gunnar Pruessner}
\email{g.pruessner@imperial.ac.uk}
\affiliation{Department of Mathematics, Imperial College London, South Kensington, London SW7 2BZ, UK}

\date{26 May 2026}

\begin{abstract}
Partially observed stochastic systems can appear (almost) time-reversal symmetric while in fact operating far from equilibrium. The present work extends the perturbative framework introduced in [Phys. Rev. Lett. \textbf{136}, 198302 (2026)] to calculate in a generic confining potential the \emph{partial entropy production}, which quantifies the time-reversal asymmetry of a generic active particle with hidden self-propulsion. Focusing on the harmonic case, we apply our framework to reproduce an exact result for the partial entropy production rate of an active Ornstein-Uhlenbeck particle and to derive the partial entropy production rate of a run-and-tumble particle.
\end{abstract}

\maketitle

Broken time-reversal symmetry (TRS) always implies that a system is out of equilibrium. However, the converse is not true: a system which is out-of-equilibrium may \emph{appear} fully or almost fully time-reversible if it cannot be observed with complete resolution, in particular when (some of) its internal degrees of freedom remain hidden. This presents a problem to experimentalists aiming to answer the key question of whether a system operates at equilibrium or is subject to some non-equilibrium driving. This is of particular relevance when this becomes a question of observation, \ie which observables are enquired to determine its ``non-equilibriumness''.
Since many experiments involve observing complex systems at a mesoscopic level, particularly in biology, this problem is ubiquitous across a range of fields \cite{Roldan2021-pj, harunariUncoveringNonequilibriumUnresolved2024a}. 

Stochastic processes, both physical and theoretical, can appear identical, similar or very different upon time-reversal. When a stochastic process is observed in full microscopic detail, its time-irreversibility as quantified by the Kullback-Leibler divergence between the probability densities
of the forward and time-reversed trajectories, \Eref{EP_full_potential} below, is proportional to its physical entropy production rate (EPR). This allows thermodynamic bounds to be placed on the speed and accuracy with which such processes can occur \cite{dechantThermodynamicBoundsCorrelation2023,leightonJensenBoundEntropy2024,itoUniversalRelationsBounds2025}. When such a process is instead observed at a mesoscopic level (\ie partially hidden), the Kullback-Leibler divergence of distributions of the \emph{visible} variables represents only a \emph{lower bound} to the entropy production rate \cite{dieballPerspectiveTimeIrreversibility2025, cocconi2022scaling}. Adopting terminology introduced by us in \cite{KnightKavehPruessner:2026}, this lower bound is subsequently referred to as the \emph{partial EPR} of the process.

Hiding a single degree of freedom can \emph{completely} restore TRS to the visible dynamics of an out-of-equilibrium system, such as a free run-and-tumble (RnT) particle \cite{KnightKavehPruessner:2026} with its internal self-propulsion state hidden. Despite the fact that the underlying system is away from equilibrium, the particle dynamics will therefore yield a partial EPR of zero and yet may still have work extracted from, even with only its position accessible to observation.

\begin{figure}[t]
    \centering
    \includegraphics[width=1.0\linewidth]{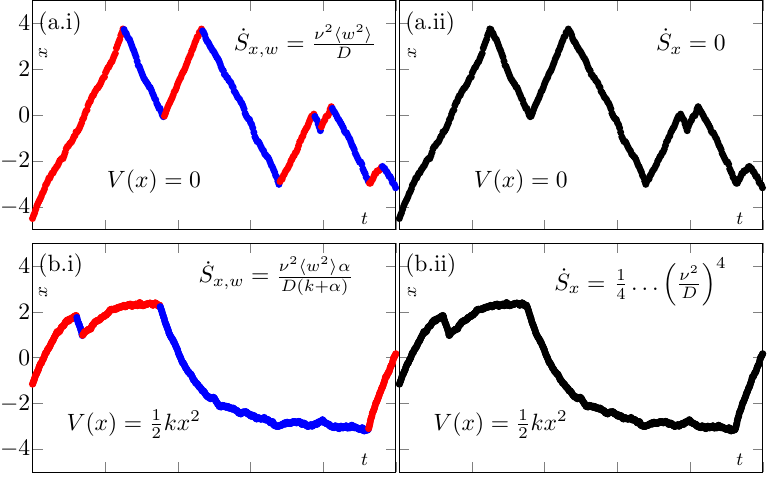}
    \caption{Run-and-tumble (RnT) process with (i) fully resolved (red corresponding to self-propulsion towards the right, blue to self-propulsion to the left) and (ii) hidden self-propulsion velocity, in the left and right columns respectively. Rows correspond to a: a) free, and b) harmonically confined, RnT particle. In a), trajectories break time-reversal symmetry (TRS) when fully resolved, as revealed by the effective swapping of colours under mirroring along the vertical axis. They obey TRS when the self-propulsion state is hidden. The process therefore has positive total EPR $\EPRfull>0$ and zero partial EPR $\EPRpartX=0$. In b), trajectories break TRS even with hidden self-propulsion, immediately visible by the kinks not being invariant under mirroring, so both the total and partial EPRs are positive. The relevant EPR is shown in each panel.}
    \label{fig:rnt2_schematic}
\end{figure}

In the present work we derive the partial EPR of a class of canonical stochastic processes, namely \emph{confined} active particles whose position, but not self-propulsion, can be observed. 
We thereby generalise the framework of \cite{KnightKavehPruessner:2026} to include the presence of a confining potential. We validate our approach by recovering the vanishing partial EPR of active Ornstein-Uhlenbeck particles in a harmonic potential, \Sref{AOUPs}, demonstrating that this is a unique feature of Gaussian self-propulsion and the harmonic potential. We further calculate, for the first time, the exact leading order of the partial EPR of harmonically confined RnT particles. These are subtly but visibly out of equilibrium, as they slow down when they move away from the potential minimum and speed up in the approach, \Fref{rnt2_schematic}.

In contrast to \cite{KnightKavehPruessner:2026}, we find that the lowest order allowed by physical considerations, which is already made up of $15$ terms, vanishes and thus is not the leading order of the partial EPR.
We consequently consider a further $14+84$ terms, which do not vanish and thus produce the desired key result. We expect that knowledge of the leading order terms and their precise structure will allow experimentalists to characterise non-equilibrium systems when only partial information is available \cite{Roldan2021-pj}.

Our approach is to start from a complete description of the fully observable system, represented by joint path probability distributions of the position and self-propulsion of the active particle. We then marginalise over the hidden self-propulsion, representing the partially observable system through the path probability distribution of the particle position \emph{only}. Although marginalising over the hidden variable is generally a hard problem, we arrive at a perturbative expression of the partial EPR of generic active particles in generic potentials. The term attributed to self-propulsion may also be considered as a correlated or non-Gaussian noise, as it can arise from an active bath \cite{mestresRealizationNonequilibriumThermodynamic2014,dibelloBrownianParticlePoissonShotNoise2024,tucciModelingActiveNonMarkovian2022a}. 

The scenario we are concerned with  is illustrated in \Fref{rnt2_schematic}, which shows trajectories of a free and harmonically confined RnT particle in rows (a) and (b) respectively. In the first column (i), both the position and the self-propulsion direction (represented by the line colour) of the particles are visible. Both processes are easily distinguishable from their time-reversed trajectories (not shown), which are obtained by mirroring the trajectory across the vertical axis. The full, physical EPRs, $\EPRfull$, are indicated in the figures. In the second column (ii), the self-propulsion directions of the run-and-tumble particles are hidden. This renders the visible trajectory of the free run-and-tumble particle completely time-reversal symmetric, with a partial EPR $\EPRpartX$ of zero \cite{KnightKavehPruessner:2026}. The calculation of the partial EPR for a run-and-tumble particle in a harmonic potential, displayed in (b.ii), is the subject of the present work. It is obvious that the partial EPR of an RnT in a harmonic potential cannot be zero: The kinks in the trajectory have a direction to them that reverses un-physically under mirroring, suggesting that particles rush up a potential wall and creep down when they reverse direction.

\section{Key derivation}
In the following, we study 
the Langevin equation of a particle in a 
static potential \cite{TailleurCates:2008,CatesTailleur:2013}: 
\begin{subequations}
\elabel{langevin_with_potential}
\begin{align}
        \xdot(t)&= \nu w(t) - \partial_x \potentialxt + \xi(t)\\
        \XIave{\xi(t) \xi(t')} &= 2 D \delta(t-t')\;.\elabel{noise_correlator_with_potential} 
\end{align}
\end{subequations}
This describes an overdamped active particle at position $x(t)$ in a potential $\potentialx$, which we assume to allow for the existence of a (unique) steady state. 
The self-propulsion is given by $\nu w(t)$, where $\nu$ has dimensions of velocity and $w(t)$ is a steady-state dimensionless stochastic process independent of $x(t)$. The particle is subject to Gaussian white noise $\xi(t)$ with diffusivity $D$ and vanishing mean, $\XIave{\xi}=0$, where $\XIave{\bullet}$ denotes the average over the noise.

The approach taken here is the same as in \cite{KnightKavehPruessner:2026}. Key definitions are repeated so that the present paper can be read as a self-contained piece of work. 

\begin{widetext}
The 
full steady-state EPR \cite{gaspard2004time,Sekimoto:2010,cocconi2020entropy} of \Eref{langevin_with_potential} depends on the ratio of the path probabilities of forward and reversed trajectories of $\{x(t)\}$ and $\{w(t)\}$ \cite[][Eq.~(2)]{KnightKavehPruessner:2026}
\begin{equation}\elabel{EP_full_potential}
    \EPRfull = \lim_{T \to \infty} \frac{1}{T} \int\Dint{x}\!\!\int\Dint{w}\,\PprobXW[\xpath,\wpath] \ln\left(\frac{\PprobXW[\xpath, \wpath]}{\PprobXW[\Rxpath, \Rwpath]}\right)\;,
\end{equation}
where $T$ is the trajectory duration. Trajectories are initialised according to steady-state distributions, negating transient contributions arising from initial conditions (which vanish anyway due to the limit $T \to \infty$). 
$\PprobXW[\xpath,\wpath]$ denotes a normalised measure for the steady-state probability density of a joint path $[\xpath,\wpath]$ for all $t\in[0,T]$ to occur in process \Eref{langevin_with_potential}, while $\PprobXW[\Rxpath,\Rwpath]$ is the probability density for the reverse path. These path probabilities are written using an Onsager-Machlup functional \cite{Taeuber:2014} based on \Eref{langevin_with_potential},
\begin{equation}\elabel{OM_with_potential}
    \PprobXW[\xpath|\wpath]\propto \Exp{-\frac{1}{4D}\int_0^T \!\!\!\dint{t}\,\left[(\xdot(t)+\potentialprimext -\nu w(t))^2 
    - 2 D \potentialdoubleprimext
    \right]}\;.
\end{equation}
Primes represent derivatives with respect to $x$, \ie $\potentialprimext = \partial_x|_{x=x(t)}\potentialx$
and $\potentialdoubleprimex = \partial^2_x|_{x=x(t)}\potentialx$. 
The final term in the exponent is the so-called ``Stratonovich term''.
Here and subsequently we use the Stratonovich convention. The corresponding time-reversed path probability is given by
\begin{align}\elabel{OM_reversed_with_potential}
    \PprobXW[\Rxpath|\wpath]&\propto \Exp{-\frac{1}{4D}\int_0^T \!\!\!\dint{t}\,\left[(\xdot(T-t)+\potentialprimextR -\nu w(t))^2 - 2 D\potentialdoubleprimextR\right]}\;,\notag\\
    &\propto \Exp{-\frac{1}{4D}\int_0^T \!\!\!\dint{t}\,\left[(-\xdot(t)+\potentialprimext -\nu w(T-t))^2 - 2 D\potentialdoubleprimext\right]}\;,
\end{align}
where the second line is obtained by substituting $T-t$ by the dummy variable $t$. For the
case that only the trajectory of $\{x(t)\}$ and \emph{not} $\{w(t)\}$
can be observed, we define the partial EPR as
\begin{equation}\elabel{partial_EP_defn_with_pot}
    \EPRpartX = \lim_{T \to \infty} \frac{1}{T} \int \mathcal{D}x 
    \PprobX[\xpath]
    \ln\left(
    \frac{\PprobX[\xpath]}{\PprobX[\Rxpath]}\right)\;. 
\end{equation}
following \cite[][Eq.~(5)]{KnightKavehPruessner:2026}. Expressing $\PprobX[\xpath]$ as the joint distribution $\PprobXW[\xpath,\wpath]$ marginalised over $\wpath$ yields 
\begin{equation}\elabel{Pprobx_from_conditional_with_potential}
    \PprobX[\xpath] =\int \Dint{w} \PprobXW[\xpath|\wpath]\PprobW[\wpath] \;,
\end{equation}
with the conditional path probability given by \Eref{OM_with_potential}. The partial EPR can now be expressed as
\cite[][Eq.~(S11)]{KnightKavehPruessnerSUPPL:2026}
\begin{equation}\label{eq:partial_EP_1_with_potential}
\begin{split}
    \EPRpartX &= \lim_{T \to \infty} \frac{1}{T} \int \mathcal{D}x \int\mathcal{D}w'\PprobXW[\xpath|\wdpath]\PprobW[\wdpath]\ln\left(\frac{\int\mathcal{D}w''\PprobXW[\xpath|\wddpath]\PprobW[\wddpath]}{\int\mathcal{D}w'''\PprobXW[\Rxpath|\wdddpath]\PprobW[\wdddpath]}\right)\;, 
    \end{split}
\end{equation}
where $w'$, $w''$ and $w'''$ are dummy variables. The ratio inside the logarithm can be rewritten using \Erefs{OM_with_potential} and \eref{OM_reversed_with_potential}, 
\begin{equation}\elabel{partial_EP_2_with_potential} 
    \begin{split}
    \frac{\PprobX[\xpath]}{\PprobX[\Rxpath]}
    &=
\frac
{
\int \Dint{w} 
\Exp{-\frac{1}{4D}\int_0^T \dint{t}\,\left[(\xdot(t)+\potentialprimext -\nu w(t))^2\right]}
\PprobW[\wpath]
}
{
\int \Dint{w} 
\Exp{-\frac{1}{4D}\int_0^T \dint{t}\,\left[(-\xdot(t)+\potentialprimext -\nu w(T-t))^2\right]}
\PprobW[\wpath]
}
\;, 
    \end{split}
\end{equation}
where the $w$-independent ``Stratonovich'' terms in the numerator and denominator have cancelled one another. The remaining $w$-independent terms in \Eref{partial_EP_2_with_potential} produce an essentially constant prefactor:
\begin{equation}\label{eq:potential_prefactor} 
    \begin{split}
        \frac{\Exp{-\frac{1}{4D} \int_0^T \dint{t} [\xdot(t)+\potentialprimext]^2 }}{\Exp{-\frac{1}{4D} \int_0^T \dint{t} [-\xdot(t)+\potentialprimext]^2 }} &= \Exp{-\frac{4}{4D} \int_0^T \dint{t} \xdot(t)\potentialprimext} \;, \\
        &= \Exp{-\frac{[\potentialxtT-\potentialxtzero]}{D}} \;,
    \end{split}
\end{equation}
which is reminiscent of a Boltzmann factor. We thus arrive at the unwieldy expression, 
\begin{equation}\elabel{partial_EP_2_with_potential_2} 
    \begin{split}
    \frac{\PprobX[\xpath]}{\PprobX[\Rxpath]}
    &=\Exp{-\frac{[\potentialxtT-\potentialxtzero]}{D}}\\
    &\hspace{0.4cm}\times
\frac
{
\int \Dint{w} 
\Exp{\frac{\nu}{4D} 
\int_0^T \dint{t}
\left[2w(t) (\xdot(t)+\potentialprimext) - (\nu w(t))^2 \right]
}
\PprobW[\wpath]
}
{
\int \Dint{w} 
\Exp{\frac{\nu}{4D}  
\int_0^T \dint{t}
\left[2w(T-t) (-\xdot(t)+\potentialprimext) - (\nu w(t))^2 \right]
}
\PprobW[\wpath]
}\;,
    \end{split}
\end{equation}
where the substitution $T-t \to t$ has been made in the second term in the integral in the denominator, $\int \dint{t} (\nu w(T-t))^2=\int \dint{t} (\nu w(t))^2$ . 
The Boltzmann-like prefactor \Eref{potential_prefactor} becomes an additive contribution to \Eref{partial_EP_1_with_potential} which vanishes after division by $T$ due to the limit $T \to \infty$. In analogy with 
Eq.~(7) of \cite{KnightKavehPruessner:2026},
the partial EPR can thus be written as
\begin{gather}\elabel{partial_EP_3_with_potential}
    \EPRpartX = \lim_{T \to \infty} \frac{1}{T} \int \Dint{x} \PprobX[\xpath]
    \ln\left(\frac{\WSTARave{\Exp{\frac{\nu}{2D} \int_0^T \dint{t} w(t) (\xdot(t)+\potentialprimext)}}}{\WSTARave{\Exp {\frac{\nu}{2D} \int_0^T \dint{t} w(T-t) (-\xdot(t)+\potentialprimext)}}}\right)\\
\text{with }\qquad 
\elabel{overline_def_main_with_pot}
    \WSTARave{\bullet}
    =\frac{1}{\NormWSTAR(\nu)}\int\Dint{w}\bullet 
    \Exp{-\frac{\nu^2}{4D} \int_0^T\dint{t} w(t)^2 }
    \;\PprobW[\wpath] =
    \int\Dint{w}\bullet \PprobWSTAR[\wpath] \;.
\end{gather}
The normalisation (ratio) of the modified path probability distribution  $\PprobWSTAR[\wpath]\propto\exp{-\nu^2\int_0^T\dint{t}w^2/(4D)}\PprobW[\wpath]$ of $\wpath$ is given by $\NormWSTAR(\nu)$, \SMref{OU_corr_star}. 
To leading order $\nu^0$ of the exponential, $\WSTARave{\bullet}$ is the expectation $\Wave{\bullet}$ under $\PprobW[\wpath]$,
\begin{equation}\elabel{def_bave}
\Wave{\bullet}=\int\Dint{w}\bullet\PprobW[\wpath] + \OC(\nu)\;.
\end{equation}
The logarithm of the expectation of an exponential in \Eref{partial_EP_3_with_potential} can be recognised as the cumulant-generating function of $\PprobWSTAR$, with conjugate variable $(\xdot(t)+\potentialprimext)$. In the following, we use an over-line to indicate expectations over the density $\PprobX[\xpath]$,
\begin{equation}\elabel{xdot_barred_with_potential}
    \Xave{\xplusVn{1}\sdots\xplusVn{n}} = 
    \int\Dint{x} \Big(\xplusVn{1}\sdots\xplusVn{n}\Big)\  \PprobX[\xpath]\;.
\end{equation}
With that, \Eref{partial_EP_3_with_potential} can be rewritten in terms of cumulants of $w(t)$ and correlators of $x(t)$ and functions thereof,
\begin{multline}\elabel{partial_EP_expansion_with_potential}
    \EPRpartX 
    = \lim_{T \to \infty} \frac{1}{T} \sum_{n=1}^\infty \frac{1}{n!}\left(\frac{\nu}{2D}\right)^n \int_0^T \dint{t_1}\!\!\sdots\dint{t_n}\Big[
    \ \Xave{\xplusVnShort{1}\sdots\xplusVnShort{n}}
    \WSTARaveS{w(t_1)\sdots w(t_n)}{\cbb}\\
    \ - \Xave{\xplusVnShortR{1}\sdots\xplusVnShortR{n}}
    \WSTARaveS{w(T-t_1)\sdots w(T-t_n)}{\cbb}\
    \Big] \;,
\end{multline}
where the superscripted $\cbb$ indicates the cumulants and we introduce the shorthand $\potentialxtnShort{i}=\potentialxtn{i}$. \Eref{partial_EP_expansion_with_potential} is the central result of the present work.
When $\potentialprimex=0$, the leading-order contribution to the partial EPR $\EPRpartX$ depends solely on symmetries of the self-propulsion $w(t)$ \cite[][Eq.~(13)]{KnightKavehPruessner:2026}. This is not the case when $\potentialprimex \neq 0$ due to the additional terms it produces in the integrands of \Erefs{partial_EP_3_with_potential} and \eref{partial_EP_expansion_with_potential}, which do not change sign under the parity reversal $\xdot(t)\to -\xdot(t)$. Self-propulsion symmetries still result in cancellations of terms in \Eref{partial_EP_expansion_with_potential} in a more \emph{ad-hoc} manner, discussed below and tabulated in \Tref{w_symmetries}.

\section{Results}
\subsection{Symmetries and cancellations}\seclabel{sym_and_cancellations}
\subsubsection{Cancellations due to self-propulsion dynamics}\seclabel{cancel_w}
In this subsection we highlight symmetries and other properties of $w(t)$ which result in cancellations of terms in \Eref{partial_EP_expansion_with_potential}. The relevant properties are summarised in \Tref{w_symmetries}.

\begin{table*}[t]
    \centering
\begin{tabular}{c|b{11cm}|c}
\hline
\textbf{Property of $w(t)$} & \textbf{Vanishing terms} & \textbf{Notation} \\
\hline
\parbox{45mm}{Zero-mean\\$\Wave{w(t)} = 0$} & All terms at order $n=1$.& \CancelMan[green]{\bullet} \\
\hline
\parbox{45mm}{Time-translation invariance\\$\Wave{w(t_1)\sdots w(t_n)}$\\$ = \Wave{w(t_1 + \Delta t)\sdots w(t_n + \Delta t)}$} & Terms containing \newline $\WSTARaveS{w(t)}{\cbb}-\WSTARaveS{w(T-t)}{\cbb}$ and 
$\WSTARaveS{w(t_1)w(t_2)}{\cbb}-\WSTARaveS{w(T-t_1)w(T-t_2)}{\cbb}$.& $[\bCancelMan[gray]{\bullet - \bullet}]$ \\
\hline
\parbox{45mm}{Parity symmetry\\$\PprobW[\wpath]=\PprobW[\Mwpath]$\\$=\PprobW[\SymParity\wpath]$} & All terms $\WSTARaveS{w(t_1)\sdots w(t_n)}{\cbb}$ with odd $n$.& \CancelMan[blue]{\bullet} \\
\hline
\parbox{45mm}{Time-reversal symmetry\\$\PprobW[\wpath]=\PprobW[\Rwpath]$\\$=\PprobW[\SymTime\wpath]$} & Terms containing $\WSTARaveS{w(t_1)\sdots w(t_n)}{\cbb} - \WSTARaveS{w(T-t_1)\sdots w(T-t_n)}{\cbb}$ at all orders.& $[\bCancelMan[red]{\bullet - \bullet}]$ \\
\hline
\parbox{45mm}{Parity-time symmetry\\$\PprobW[\wpath]=\PprobW[-\Rwpath]$\\$=\PprobW[\SymParityTime\wpath] $} & Terms containing $\WSTARaveS{w(t_1)\sdots w(t_n)}{\cbb} + \WSTARaveS{w(T-t_1)\sdots w(T-t_n)}{\cbb}$ at odd orders and $\WSTARaveS{w(t_1)\sdots w(t_n)}{\cbb} - \WSTARaveS{w(T-t_1)\sdots w(T-t_n)}{\cbb}$ at even orders.& $[\bCancelMan[violet]{\bullet \pm \bullet}]$ \\
\hline
\end{tabular}
    \caption{Various properties of the hidden self-propulsion $w(t)$ result in cancellations of terms in the expansion of the partial EPR $\EPRpartX$,  \Eref{partial_EP_expansion_with_potential}. The zero-mean and time-translation invariance properties have effects only at orders $n=1$ and $n=1,2$ respectively, while the various symmetries of $w(t)$ also affect higher order terms. Cancellations for $n\leq4$ are shown explicitly in \Eref{colourful_cancellations}; notation corresponding to each type of cancellation is shown in the third column of this table.
    }
    \tlabel{w_symmetries}
\end{table*}

We highlight their effects by writing the first four orders in the expansion \Eref{partial_EP_expansion_with_potential} explicitly: 
\begin{align}\elabel{colourful_cancellations}
    \EPRpartX &= \lim_{T \to \infty} \frac{1}{T} \Bigg[\left(\frac{\nu}{2D}\right)\int_0^T \dint{t}\left( \Xave{\xdotn{1}} \left[\bCancelMan[violet]{\DoubleCancelMan{blue}{green}{\WSTARaveS{w(t)}{\cbb}} + \DoubleCancelMan{blue}{green}{\WSTARaveS{w(T-t)}{\cbb}}}\right] + \Xave{\potentialxtnShort{1}} \left[\bDoubleCancelMan{red}{gray}{\DoubleCancelMan{blue}{green}{\WSTARaveS{w(t)}{\cbb}} - \DoubleCancelMan{blue}{green}{\WSTARaveS{w(T-t)}{\cbb}}}\right]\right)\notag\\
    &+ \frac{1}{2!}\left(\frac{\nu}{2D}\right)^2\int_0^T \dint{t_1}\dint{t_2}\bigg( \left[\Xave{\xdotn{1}\xdotn{2}}+\Xave{\potentialxtnShort{1}\potentialxtnShort{2}}\right] \left[\bTripleCancelMan{red}{gray}{violet}{\WSTARaveS{w(t_1)w(t_2)}{\cbb} - \WSTARaveS{w(T-t_1)w(T-t_2)}{\cbb}}\right] \notag\\
    &\hspace{4cm}+ \left[\Xave{\xdotn{1}\potentialxtnShort{2}}+\Xave{\potentialxtnShort{1}\xdotn{2}}\right] \left[\WSTARaveS{w(t_1)w(t_2)}{\cbb} + \WSTARaveS{w(T-t_1)w(T-t_2)}{\cbb}\right]\bigg)\notag\\
    &+ \frac{1}{3!}\left(\frac{\nu}{2D}\right)^3\int_0^T \dint{t_1}\!\!\sdots\dint{t_3}\bigg( \left[\Xave{\xdotn{1}\xdotn{2}\xdotn{3}}+3\Xave{\xdotn{1}\potentialxtnShort{2}\potentialxtnShort{3}}\right] \notag\\
    &\hspace{5cm}\times\left[\bCancelMan[violet]{\CancelMan[blue]{\WSTARaveS{w(t_1)w(t_2)w(t_3)}{\cbb}} + \CancelMan[blue]{\WSTARaveS{w(T-t_1)w(T-t_2)w(T-t_3)}{\cbb}}}\right] \notag\\
    &\hspace{4.48cm}+ \left[3\Xave{\xdotn{1}\xdotn{2}\potentialxtnShort{3}}+\Xave{\potentialxtnShort{1}\potentialxtnShort{2}\potentialxtnShort{3}}\right] \notag\\
    &\hspace{5cm}\times\left[\bCancelMan[red]{\CancelMan[blue]{\WSTARaveS{w(t_1)w(t_2)w(t_3)}{\cbb}} - \CancelMan[blue]{\WSTARaveS{w(T-t_1)w(T-t_2)w(T-t_3)}{\cbb}}}\right]\bigg) \notag\\
    &+ \frac{1}{4!}\left(\frac{\nu}{2D}\right)^4\int_0^T \dint{t_1}\!\!\sdots\dint{t_4}\bigg( \left[\Xave{\xdotn{1}\sdots\xdotn{4}}+6\Xave{\xdotn{1}\xdotn{2}\potentialxtnShort{3}\potentialxtnShort{4}}+\Xave{\xdotn{1}\sdots\potentialxtnShort{4}}\right] \notag\\
    &\hspace{5cm}\times\left[\bDoubleCancelMan{violet}{red}{\WSTARaveS{w(t_1)\sdots w(t_4)}{\cbb} - \WSTARaveS{w(T-t_1)\sdots w(T-t_4)}{\cbb}}\right] \notag\\
    &\hspace{4.48cm}+ \left[4\Xave{\xdotn{1}\xdotn{2}\xdotn{3}\potentialxtnShort{4}}+4\Xave{\xdotn{1}\potentialxtnShort{2}\potentialxtnShort{3}\potentialxtnShort{4}}\right] \notag\\
    &\hspace{5cm}\times\left[\WSTARaveS{w(t_1)\sdots w(t_4)}{\cbb} + \WSTARaveS{w(T-t_1)\sdots w(T-t_4)}{\cbb}\right]\bigg) + \ldots 
    \;.
\end{align}
Strikethroughs in different colours correspond to different reasons for
cancellation, as detailed in \Tref{w_symmetries}. 
Unlike
in the case of free particles $V(x)=0$, full $\SymParityTime$-symmetry of
$w(t)$ does \textit{not} result in zero partial EPR. This corresponds
with the intuition developed by considering an RnT particle in a
potential as shown in \Fref{rnt2_schematic}. Nonetheless, the self-propulsion processes which result in the most significant cancellations of terms in \Eref{colourful_cancellations} are indeed those which satisfy $\SymParity$-, $\SymTime$- and $\SymParityTime$- symmetries. 

\subsubsection{Cancellations due to spatial dynamics}\seclabel{cancel_x}
Further cancellations can arise due to properties of $\xdot(t)$ and for specific choices of $\potentialx$. In particular, our demands that $w(t)$ is in steady state and $V(x)$ constant in time result in a steady state distribution $\PprobX[x(t)]$, such that
\begin{equation}
    \Xave{\xdot(t)}= \frac{\mathrm{d}}{\mathrm{d}t} \Xave{x(t)}=0\;,
\end{equation} 
resulting in the cancellation of the first term at order $n=1$ in \Eref{colourful_cancellations}. In the special case of a \textit{harmonic} potential $V(x)\propto x^2$, a further cancellation occurs,
\begin{subequations}\elabel{harmonic_cancellation}
\begin{align}
    \Xave{\xdotn{1}\potentialxtnShort{2}}+\Xave{\potentialxtnShort{1}\xdotn{2}} &\propto \Xave{\xdotn{1} x(t_2)}+\Xave{x(t_1)\xdotn{2}}\;,\\
    &= \left(\frac{\mathrm{d}}{\mathrm{d}t_1} + \frac{\mathrm{d}}{\mathrm{d}t_2}\right)\Xave{x(t_1) x(t_2)}\;,\\
    &=0,
\end{align}
\end{subequations}
since the two-time correlation function at steady state depends only on $t_2-t_1$, \ie $\Xave{x(t_1) x(t_2)} = f(|t_2-t_1|)$, \cf \SMref{correlators_one_dot}. 
As a consequence of \Eref{harmonic_cancellation} and \Tref{w_symmetries}, self-propulsion processes with $\SymParity$- and $\SymTime
$- (and thus $\SymParityTime$-) symmetries that take place in a \emph{harmonic} potential have leading-order partial EPR of (at least) $n=4$ in the expansion \Eref{partial_EP_expansion_with_potential}.

In the subsequent sections, we derive the leading-order partial EPR $\EPRpartX$ for two canonical models: an active Ornstein-Uhlenbeck particle (AOUP) \cite{martin2021statistical} and a run-and-tumble (RnT) particle \cite{TailleurCates:2008}. Both processes have $\SymParity$- and $\SymTime
$- (and thus $\SymParityTime$-) symmetric self-propulsion and thus have $\EPRpartX=0$ without a potential, but in general $\EPRpartX>0$ in the presence of a potential.
\subsection{Active Ornstein-Uhlenbeck Particle}\seclabel{AOUPs}
Active Ornstein-Uhlenbeck particles (AOUPs) are defined by the Langevin equations \cite{szamel2014,FodorETAL:2016,BothePruessner:2021}
\begin{subequations}\elabel{AOUPs_langevin_space}
  \begin{align}
    \xdot(t)&= \nu w(t) - \partial_x \potentialxt + \xi(t)\;,\\
    \XIave{\xi(t) \xi(t')} &= 2 D \delta(t-t')\;,
  \end{align}
\end{subequations}
with the dynamics of the self-propulsion velocity $w(t)$ governed by an Ornstein-Uhlenbeck process,
\begin{subequations}\elabel{AOUPs_langevin_vel}
  \begin{align}
    \dot{w}(t)&= -\mu w(t) + \eta(t)\;,\\
    \langle\eta(t) \eta(t')\rangle &= 2 D_w \delta(t-t')\;.
  \end{align}
\end{subequations}
with Gaussian, uncorrelated noise $\eta(t)$, which results in a Gaussian distribution $\PprobW[\wpath]$ and therefore Gaussian $\PprobWSTAR[\wpath]$ \cite[][\SMrefLocal{OU_corr_star}]{vanKampen:1992,PRRsupplement}. As a consequence, cumulants of $w(t)$ vanish above second order. As  we choose $\Wave{\eta}=0$, the first moment (and thus cumulant) of $w(t)$ is also zero, and as $w(t)$ is time-translation invariant, $\WSTARaveS{w(t_1)w(t_2)}{\cbb}=\WSTARaveS{w(T-t_1)w(T-t_2)}{\cbb}$. The expansion \Eref{partial_EP_expansion_with_potential} therefore contains a single contribution:
\begin{align}\elabel{second_term_full}
    \EPRpartX &= \frac{1}{2!}\left(\frac{\nu}{2D}\right)^2  \lim_{T \to \infty} \frac{2}{T} \int_0^T \dint{t_1}\dint{t_2} \left[\Xave{\xdotn{1}\potentialxtnShort{2}}+\Xave{\potentialxtnShort{1}\xdotn{2}}\right] \WSTARaveS{w(t_1)w(t_2)}{\cbb}\;.
\end{align}
The term in the square brackets vanishes in the special case of a harmonic potential $V(x)\propto x^2$ by \Eref{harmonic_cancellation}, resulting in $\EPRpartX=0$ for a harmonically confined AOUP \cite{DabelowBoEichhorn:2021}. This is not the case for general potentials.  The two-time cumulant of $w(t)$ with respect to $\PprobWSTAR[w(t)]$ is derived in 
\SMref{OU_corr_star}: 
\begin{subequations}\elabel{starred_corr_aoup}
    \begin{align}
        \WSTARaveS{w(t_1)w(t_2)}{\cbb} 
        &= \frac{D_w}{\mutilde}\exp{-\mutilde |t_2 - t_1|}\;,
    \end{align}
\end{subequations}
where $\mutilde^2 = \mu^2 +  \nu^2 D_w/D$. As such, the partial EPR of an AOUP is given by:
\begin{subequations}\elabel{aoup_EPR}
\begin{align}
    \EPRpartX &= \frac{\nu^2 D_w}{2\mutilde D^2}\lim_{T \to \infty} \frac{1}{T} \int_0^T \dint{t_1}\dint{t_2} \Xave{\xdotn{1}\potentialxtnShort{2}}\;\exp{-\mutilde |t_2 - t_1|}\;,\\
    &= \frac{\nu^2 D_w}{\mutilde D^2} \int_0^\infty \dint{t} \Xave{\dot{x}(0)\potentialprimext}\;\exp{-\mutilde t}\;,
\end{align}
\end{subequations}
where in the second equality we make the substitution $t = t_2-t_1$.
This result reproduces the expressions presented in \cite{dabelowIrreversibilityActiveMatter2019, DabelowBoEichhorn:2021, capriniEntropyProductionOrnstein2019}, which were obtained by direct evaluation of the path integral \Eref{Pprobx_from_conditional_with_potential} for the special case of Gaussian dynamics of $w(t)$. 

\subsection{Run-and-Tumble particle}
In this section, we derive the leading-order partial EPR of an RnT particle in a harmonic potential.
RnT dynamics is defined by a self-propulsion velocity which switches between two discrete values $w(t)\in\{w_+,w_-\}$ with Poissonian rates $\{\alpha_+,\alpha_-\}$, as discussed and illustrated in Ref.~\cite{KnightKavehPruessner:2026}. In this work we confine our attention to the symmetric case 
$\alpha_+=\alpha_-:=\alpha/2$ 
and $w_+ = - w_- = 1$, so that $w(t)^2/D$ is constant and $\PprobWSTAR[\wpath]=\PprobW[\wpath]$, as $\exp{-\nu^2T/(4D)}$ cancels with the normalisation, \Eref{overline_def_main_with_pot}. This dynamics has $\Wave{w}=0$ and is in steady state as well as $\SymParity$- and $\SymTime
$-, and hence $\SymParityTime$-, symmetric, resulting in cancellation of all terms as highlighted in \Eref{colourful_cancellations}. As a result, the leading-order contribution to $\EPRpartX$ will in general be at order $n=2$ in \Eref{partial_EP_expansion_with_potential}:
\begin{equation}
    \EPRpartX = \lim_{T \to \infty} \frac{1}{T} \frac{1}{2!}\left(\frac{\nu}{2D}\right)^2\int_0^T \dint{t_1}\dint{t_2} 2\WSTARaveS{w(t_1)w(t_2)}{\cbb}\left[\Xave{\xdotn{1}\potentialxtnShort{2}}+\Xave{\potentialxtnShort{1}\xdotn{2}}\right] + \OC\left(\frac{\nu^4}{D^4}\right)\;.
\end{equation}
However, we focus on the case of confinement in a harmonic potential $V(x) = k x^2/2$, which is analytically tractable since the required correlators are accessible through the field-theoretic framework of \cite{garcia2021run}. By \Eref{harmonic_cancellation}, this results in a leading-order contribution at $n=4$:
\begin{multline}\elabel{rnt_epr_correlators_harmonic}
    \EPRpartX = \lim_{T \to \infty} \frac{1}{T} \frac{1}{4!}\left(\frac{\nu}{2D}\right)^4\int_0^T \dint{t_1}\!\!\sdots\dint{t_4} 2 \WSTARaveS{w(t_1)\sdots w(t_4)}{\cbb}  \\
    \hspace{5cm}\times\left[4k\Xave{\xdotn{1}\xdotn{2}\xdotn{3}x(t_4)}+4k^3\Xave{\xdotn{1}x(t_2)x(t_3)x(t_4)}\right]+ \OC\left(\frac{\nu^{6}}{D^{6}}\right)\;,
\end{multline}
where we have invoked the $\SymTime$- symmetry of $w(t)$ to rewrite $\WSTARaveS{w(T-t_1)\sdots w(T-t_4)}{\cbb} = \WSTARaveS{w(t_1)\sdots w(t_4)}{\cbb}$. We demonstrate in 
\SMref{correlators_one_dot} that sums of correlation functions of the form $\Xave{\xdotn{1}x(t_2)\sdots x(t_n)}$ across all permutations of the times $t_1,\sdots,t_n$ vanish due to the time-translation invariance of the $x$-dynamics, similar to \Eref{harmonic_cancellation}. Using this and symmetrising with respect to the times $t_1, \sdots, t_4$ yields a leading-order EPR of
\begin{equation}\elabel{rnt_epr_correlators}
    \EPRpartX = \lim_{T \to \infty} \frac{1}{T} \frac{1}{4!}\left(\frac{\nu}{2D}\right)^4\int_0^T \dint{t_1}\!\!\sdots\dint{t_4} 2 \WSTARaveS{w(t_1)\sdots w(t_4)}{\cbb}  4k \frac{1}{4}
    \LC_d \Xave{x(t_1)x(t_2)x(t_3)x(t_4)}+ \OC\left(\frac{\nu^{6}}{D^{6}}\right)\;,
\end{equation}
with the triple-derivative operator
\begin{equation}\elabel{def_Ld}
    \LC_d = \triplePartial{1}{2}{3}+\triplePartial{1}{2}{4}+\triplePartial{1}{3}{4}+\triplePartial{2}{3}{4}
\end{equation}
in the notation $\partial_{t_n} = \partial/\partial t_n$. This symmetrisation produces a factor of $1/4$ in \Eref{rnt_epr_correlators}. To evaluate \Eref{rnt_epr_correlators}, we are required to calculate the moment and cumulant functions of $x(t)$ and $w(t)$ in the integrand of \Eref{rnt_epr_correlators} explicitly. 

For the present choice of dynamics, $w(t)$ is given by a symmetric telegraph process with $w_+=-w_-=1$. Moments of this process $w(t)$ are calculated using the method of propagator matrices described in Suppl.~SIII\,A \cite{KnightKavehPruessnerSUPPL:2026} of \cite{KnightKavehPruessner:2026}. This yields 
\begin{subequations}\elabel{w_cumulant}
\begin{align}
    \WSTARaveS{w(t_1)\sdots w(t_4)}{\cbb} =& \Wave{w(t_1)\sdots w(t_4)}^{\cbb} \\
    =& \Wave{w(t_1)\sdots w(t_4)} - \Wave{w(t_1) w(t_2)}\Wave{w(t_3) w(t_4)}\notag \\
    &- \Wave{w(t_1) w(t_3)}\Wave{w(t_2) w(t_4)}- \Wave{w(t_1) w(t_4)}\Wave{w(t_2) w(t_3)}\;,\\
    =& - 2 e^{-\alpha(t_4 + t_3 - t_2-t_1)}\;,
\end{align}
\end{subequations}
where we use $\Wave{w(t_1)\sdots w(t_4)} = e^{-\alpha(t_4 - t_3 + t_2-t_1)}$ and $\Wave{w(t_1) w(t_2)}= e^{-\alpha(t_2 - t_1)}$ with time-ordering $t_4\ge t_3\ge t_2\ge t_1$.

Correlators of $x(t)$ are significantly more involved. These are calculated in 
\SMref{four_time_correlator}:
\begin{multline}
\elabel{triple_diff_correlator}
\mathcal{L}_d \Xave{x(t_1) x(t_2) x(t_3)x(t_4)}
= \frac{4 \alpha \nu^4 }{(k+\alpha)(k-\alpha)(\alpha+3k)(\alpha-3k)}
   \bigg[
      12k^2 e^{-k(t_4+t_3+t_2-3t_1)}
\\
      + (\alpha - 3k)(k+\alpha)
         \Big(
            e^{-\alpha(t_4-t_1)-k(t_3+t_2-2t_1)}
            +e^{-\alpha(t_3-t_1)-k(t_4+t_2-2t_1)}
            +e^{-\alpha(t_2-t_1)-k(t_4+t_3-2t_1)}
         \Big)
\\
      + (\alpha + 3k)(k-\alpha) \, e^{-\alpha(t_2-t_1)-k(t_4+t_3-2t_2)}
   \bigg] 
\,.
\end{multline}
In \SMref{sixth_order} we demonstrate that the relevant time-derivatives of $\Xave{x(t_1)\sdots x(t_6)}$ are at least of order $\nu^{4}D$, 
giving rise to terms $\propto(\nu/D)^6\nu^{4}D$ in the expansion of the partial EPR. 
The integral \Eref{rnt_epr_correlators} is evaluated in 
\SMref{rnt_epr_derivation} and finally yields 
\begin{equation}\elabel{rnt2_epr}
    \EPRpartX = 
    \left(\frac{\nu^2}{D}\right)^4 
    \frac{\alpha k }{4(\alpha+k)^3(\alpha+3k)^2} + \OC\left(\frac{\nu^{10}}{D^5}\right)\;,
\end{equation}
Accounting for the different contributions to the orders in $\nu$ and $D$ is difficult, because the expansion \Eref{partial_EP_expansion_with_potential} produces pre-factors $(\nu/D)^n$, but the $\Xave{x(t_1)\sdots x(t_n)}$ are of different orders in $\nu$, namely $D^{n/2}(\nu^2/D)^m$ with $0\le 2m\le n$ in the case of an RnT particle in a harmonic potential, and unless $w(t)^2$ is constant, as is the case here, the starred expectations of $w(t)$ produce terms of any order in $\nu$. Terms overall odd in $\nu$ cannot generally be ruled out if both potential and self-propulsion process are asymmetric, rendering the partial EPR dependent on the sign of $\nu$.

\end{widetext}

\section{Discussion and Outlook}

The key result of this paper is the expansion \Eref{partial_EP_expansion_with_potential}, a series representation of the partial EPR of an active particle. For active particles with no net self-propulsion, \ie $\Wave{w(t)}=0$, the first term in the expansion vanishes, resulting in a partial EPR of order $n=2$ 
or higher. These processes can be considered zero-mean non-Markovian fluctuations. However, even for active particles with $\Wave{w}\ne0$, the leading order is $n>1$ in a confining potential as in steady state $\Xave{\xdot}=0$ to all orders in $\nu$. As such, \Eref{partial_EP_expansion_with_potential} demonstrates that 
in such current-free processes, the leading order of the partial EPR is beyond that of the full EPR,
which for most active particles is of order $\nu^2/D$ \cite{cocconi2020entropy,PruessnerGarcia-Millan:2025}.

In general, compared to their free counterpart,
introducing a potential \textit{increases} the time-reversal asymmetry of purely spatial trajectories of particles, \ie their partial EPR, since it provides a preferred direction of travel at each point in space. This can be seen from \Eref{colourful_cancellations}, which suggests that the leading order contribution due to the potential is generally positive. Intuition is provided by the run-and-tumble (RnT) dynamics illustrated in \Fref{rnt2_schematic}. In the absence of a confining potential, forwards and reversed trajectories can be distinguished only when both their position and their self-propulsion is visible. When the self-propulsion is hidden, forwards and reversed trajectories become indistinguishable, \Fref{rnt2_schematic}a.ii. This is not true for a harmonically confined particle. Spatial trajectories alone break time-reversal symmetry in this case, since the particle moves quickly down the potential and slowly up the other side. In reversed trajectories, particles move slowly down the potential and climb quickly upwards. However, the \textit{extent} to which time-reversal symmetry is broken decreases when the self-propulsion is hidden, as quantified by the partial EPR, as \Eref{rnt2_epr} is $\propto \nu^8/D^4$ ($n=4$), while the full EPR is $\propto\nu^2/D$.

The presence of a potential in \Eref{langevin_with_potential} means that symmetries of $w(t)$ do not fully determine the leading-order partial EPR. Since the expansion with a potential, \Eref{partial_EP_expansion_with_potential}, contains all the terms of the expansion without a potential in addition to new terms, the partial EPR of a particle in a potential is of an order $n$ in the expansion that is generally lower or equal to that of a free particle. For instance, the asymmetric telegraph process with $\Wave{w}=0$ described 
in Ref.~\cite[][Suppl.~SIII\,A]{KnightKavehPruessnerSUPPL:2026}
has leading-order partial EPR $n=2$ in the expansion \Eref{partial_EP_expansion_with_potential}, unlike the corresponding free particle whose leading-order partial EPR appears at $n=3$. In the special case of a harmonic potential, the leading-order EPR \textit{does} enter at $n=3$ (\cf \Eref{harmonic_cancellation}) but contains an extra term compared to the free case.

An AOUP in a harmonic potential represents the \textit{unique} nontrivial choice of $w(t)$ and $V(x)$ that produces zero partial EPR \cite{dabelowIrreversibilityActiveMatter2019, capriniEntropyProductionOrnstein2019} in the steady state:
No distribution other than a Gaussian contains a finite number of non-zero cumulants \cite{marcinkiewiczProprieteLoiGauss1939}, making the Ornstein-Uhlenbeck process a unique choice of $w(t)$ that gives rise to a finite number of terms in \Eref{partial_EP_expansion_with_potential}. The single term that arises, given in \Eref{second_term_full}, comes from the second cumulant of $w(t)$ and is non-zero for any confining potential except for $V(x) \propto x^2$. Any other choice of self-propulsion process represents a non-Gaussian noise in the dynamics of $x(t)$ \cite{dibelloBrownianParticlePoissonShotNoise2024,tucciModelingActiveNonMarkovian2022a,mestresRealizationNonequilibriumThermodynamic2014}. The non-vanishing persistence of the noise combined with the presence of a potential results in $x(t)$ violating TRS. 

That the lowest order contribution to the partial EPR allowable by physical considerations, $\nu^6/D^4$, vanishes may run contrary to na{\"i}ve expectations and earlier findings \cite{KnightKavehPruessner:2026}. The next order, $\nu^8/D^4$, which turns out to be non-zero, comprises many more terms.
As detailed in \SMref{rnt_epr_derivation} and \SMref{sixth_order} this meant that the first $14$ terms had to be calculated and then a further $84$. As the resulting terms follow the pattern $(\nu^2/D)^n$, one may speculate whether \emph{only} these terms contribute, in line with \cite{KnightKavehPruessner:2026}.

It is striking that the leading-order partial EPR of an RnT particle in a harmonic potential \Eref{rnt2_epr} appears at order $(\nu^2/D)^4$, while the full EPR is proportional to $\nu^2/D$. However, such high-order partial EPRs have already been identified in \cite{KnightKavehPruessner:2026}, in particular for active particles with parity-symmetric, time-asymmetric driving, such as allowing $w(t)$ to undergo diffusion with stochastic resetting. 

Calculating the correlators needed to evaluate leading-order partial EPRs in the present framework is challenging. Doi-Peliti field theory \cite{Doi:1976,Peliti:1985,Cardy:2008,garcia2021run,garcia2020interactions,PruessnerGarcia-Millan:2025}, used for the case of a harmonically confined RnT particle, represents a powerful tool for such calculations. Conventional probabilistic methods based on Fokker-Planck equations can also be used if more convenient \cite{vanKampen:1992,Pavliotis2014}. Correlation functions can also be determined through Monte Carlo simulations and partial EPR calculated through numerical integration \cite[][SIV]{KnightKavehPruessnerSUPPL:2026}. Expressing time-irreversibility, which is challenging to measure numerically, in terms of correlation functions is an appealing feature of the present framework. 

As discussed in \cite{KnightKavehPruessner:2026}, a thermodynamic interpretation of the partial EPR would be an interesting subject of future research \cite{dieballPerspectiveTimeIrreversibility2025}. In particular, this framework could be extended to time-varying potentials to allow for the investigation of the entropy production associated with work extraction protocols \cite{schuettler, garciamillan2025optimalclosedloopcontrolactive} or to potentials which depend on both the observable and the hidden variable \cite{ghosalInferringEntropyProduction2022}. 

\begin{acknowledgments}
The authors thank L Cocconi, J Fry, R Garcia-Millan, H J Jensen, S Loos and C Roberts for useful discussions. JK acknowledges support from the Engineering and Physical Sciences Research Council (grant number 2620369).
\end{acknowledgments}

\bibliography{references}

\providecommand{\noopsort}[1]{}\providecommand{\singleletter}[1]{#1}%
\begin{thebibliography}{42}%
\makeatletter
\providecommand \@ifxundefined [1]{%
 \@ifx{#1\undefined}
}%
\providecommand \@ifnum [1]{%
 \ifnum #1\expandafter \@firstoftwo
 \else \expandafter \@secondoftwo
 \fi
}%
\providecommand \@ifx [1]{%
 \ifx #1\expandafter \@firstoftwo
 \else \expandafter \@secondoftwo
 \fi
}%
\providecommand \natexlab [1]{#1}%
\providecommand \enquote  [1]{``#1''}%
\providecommand \bibnamefont  [1]{#1}%
\providecommand \bibfnamefont [1]{#1}%
\providecommand \citenamefont [1]{#1}%
\providecommand \href@noop [0]{\@secondoftwo}%
\providecommand \href [0]{\begingroup \@sanitize@url \@href}%
\providecommand \@href[1]{\@@startlink{#1}\@@href}%
\providecommand \@@href[1]{\endgroup#1\@@endlink}%
\providecommand \@sanitize@url [0]{\catcode `\\12\catcode `\$12\catcode
  `\&12\catcode `\#12\catcode `\^12\catcode `\_12\catcode `\%12\relax}%
\providecommand \@@startlink[1]{}%
\providecommand \@@endlink[0]{}%
\providecommand \url  [0]{\begingroup\@sanitize@url \@url }%
\providecommand \@url [1]{\endgroup\@href {#1}{\urlprefix }}%
\providecommand \urlprefix  [0]{URL }%
\providecommand \Eprint [0]{\href }%
\providecommand \doibase [0]{https://doi.org/}%
\providecommand \selectlanguage [0]{\@gobble}%
\providecommand \bibinfo  [0]{\@secondoftwo}%
\providecommand \bibfield  [0]{\@secondoftwo}%
\providecommand \translation [1]{[#1]}%
\providecommand \BibitemOpen [0]{}%
\providecommand \bibitemStop [0]{}%
\providecommand \bibitemNoStop [0]{.\EOS\space}%
\providecommand \EOS [0]{\spacefactor3000\relax}%
\providecommand \BibitemShut  [1]{\csname bibitem#1\endcsname}%
\let\auto@bib@innerbib\@empty
\bibitem [{\citenamefont {Rold{\'a}n}\ \emph {et~al.}(2021)\citenamefont
  {Rold{\'a}n}, \citenamefont {Barral}, \citenamefont {Martin}, \citenamefont
  {Parrondo},\ and\ \citenamefont {J{\"u}licher}}]{Roldan2021-pj}%
  \BibitemOpen
  \bibfield  {author} {\bibinfo {author} {\bibfnamefont {{\'E}.}~\bibnamefont
  {Rold{\'a}n}}, \bibinfo {author} {\bibfnamefont {J.}~\bibnamefont {Barral}},
  \bibinfo {author} {\bibfnamefont {P.}~\bibnamefont {Martin}}, \bibinfo
  {author} {\bibfnamefont {J.~M.~R.}\ \bibnamefont {Parrondo}},\ and\ \bibinfo
  {author} {\bibfnamefont {F.}~\bibnamefont {J{\"u}licher}},\ }\bibfield
  {title} {\bibinfo {title} {Quantifying entropy production in active
  fluctuations of the hair-cell bundle from time irreversibility and
  uncertainty relations},\ }\href@noop {} {\bibfield  {journal} {\bibinfo
  {journal} {New J. Phys.}\ }\textbf {\bibinfo {volume} {23}},\ \bibinfo
  {pages} {083013} (\bibinfo {year} {2021})}\BibitemShut {NoStop}%
\bibitem [{\citenamefont
  {Harunari}(2024)}]{harunariUncoveringNonequilibriumUnresolved2024a}%
  \BibitemOpen
  \bibfield  {author} {\bibinfo {author} {\bibfnamefont {P.~E.}\ \bibnamefont
  {Harunari}},\ }\bibfield  {title} {\bibinfo {title} {Uncovering
  nonequilibrium from unresolved events},\ }\href
  {https://doi.org/10.1103/PhysRevE.110.024122} {\bibfield  {journal} {\bibinfo
   {journal} {Phys. Rev. E}\ }\textbf {\bibinfo {volume} {110}},\ \bibinfo
  {pages} {024122} (\bibinfo {year} {2024})}\BibitemShut {NoStop}%
\bibitem [{\citenamefont {Dechant}\ \emph {et~al.}(2023)\citenamefont
  {Dechant}, \citenamefont {{Garnier-Brun}},\ and\ \citenamefont
  {Sasa}}]{dechantThermodynamicBoundsCorrelation2023}%
  \BibitemOpen
  \bibfield  {author} {\bibinfo {author} {\bibfnamefont {A.}~\bibnamefont
  {Dechant}}, \bibinfo {author} {\bibfnamefont {J.}~\bibnamefont
  {{Garnier-Brun}}},\ and\ \bibinfo {author} {\bibfnamefont {S.-i.}\
  \bibnamefont {Sasa}},\ }\bibfield  {title} {\bibinfo {title} {Thermodynamic
  {{Bounds}} on {{Correlation Times}}},\ }\href
  {https://doi.org/10.1103/PhysRevLett.131.167101} {\bibfield  {journal}
  {\bibinfo  {journal} {Phys. Rev. Lett.}\ }\textbf {\bibinfo {volume} {131}},\
  \bibinfo {pages} {167101} (\bibinfo {year} {2023})}\BibitemShut {NoStop}%
\bibitem [{\citenamefont {Leighton}\ and\ \citenamefont
  {Sivak}(2024)}]{leightonJensenBoundEntropy2024}%
  \BibitemOpen
  \bibfield  {author} {\bibinfo {author} {\bibfnamefont {M.~P.}\ \bibnamefont
  {Leighton}}\ and\ \bibinfo {author} {\bibfnamefont {D.~A.}\ \bibnamefont
  {Sivak}},\ }\bibfield  {title} {\bibinfo {title} {Jensen bound for the
  entropy production rate in stochastic thermodynamics},\ }\href
  {https://doi.org/10.1103/PhysRevE.109.L012101} {\bibfield  {journal}
  {\bibinfo  {journal} {Phys. Rev. E}\ }\textbf {\bibinfo {volume} {109}},\
  \bibinfo {pages} {L012101} (\bibinfo {year} {2024})}\BibitemShut {NoStop}%
\bibitem [{\citenamefont {Ito}\ \emph {et~al.}(2025)\citenamefont {Ito},
  \citenamefont {Xu}, \citenamefont {Jiang}, \citenamefont {Rold{\'a}n},
  \citenamefont {{Rica-Alarc{\'o}n}}, \citenamefont {Mart{\'i}nez},\ and\
  \citenamefont {Watanabe}}]{itoUniversalRelationsBounds2025}%
  \BibitemOpen
  \bibfield  {author} {\bibinfo {author} {\bibfnamefont {K.}~\bibnamefont
  {Ito}}, \bibinfo {author} {\bibfnamefont {G.-H.}\ \bibnamefont {Xu}},
  \bibinfo {author} {\bibfnamefont {C.}~\bibnamefont {Jiang}}, \bibinfo
  {author} {\bibfnamefont {{\'E}.}~\bibnamefont {Rold{\'a}n}}, \bibinfo
  {author} {\bibfnamefont {R.~A.}\ \bibnamefont {{Rica-Alarc{\'o}n}}}, \bibinfo
  {author} {\bibfnamefont {I.~A.}\ \bibnamefont {Mart{\'i}nez}},\ and\ \bibinfo
  {author} {\bibfnamefont {G.}~\bibnamefont {Watanabe}},\ }\bibfield  {title}
  {\bibinfo {title} {Universal relations and bounds for fluctuations in
  quasistatic small heat engines},\ }\href
  {https://doi.org/10.1038/s42005-025-01961-1} {\bibfield  {journal} {\bibinfo
  {journal} {Commun. Phys.}\ }\textbf {\bibinfo {volume} {8}},\ \bibinfo
  {pages} {60} (\bibinfo {year} {2025})}\BibitemShut {NoStop}%
\bibitem [{\citenamefont {Dieball}\ and\ \citenamefont
  {Godec}(2025)}]{dieballPerspectiveTimeIrreversibility2025}%
  \BibitemOpen
  \bibfield  {author} {\bibinfo {author} {\bibfnamefont {C.}~\bibnamefont
  {Dieball}}\ and\ \bibinfo {author} {\bibfnamefont {A.}~\bibnamefont
  {Godec}},\ }\bibfield  {title} {\bibinfo {title} {Perspective: {{Time}}
  irreversibility in systems observed at coarse resolution},\ }\href
  {https://doi.org/10.1063/5.0251089} {\bibfield  {journal} {\bibinfo
  {journal} {J. Chem. Phys.}\ }\textbf {\bibinfo {volume} {162}},\ \bibinfo
  {pages} {090901} (\bibinfo {year} {2025})}\BibitemShut {NoStop}%
\bibitem [{\citenamefont {Cocconi}\ \emph {et~al.}(2022)\citenamefont
  {Cocconi}, \citenamefont {Salbreux},\ and\ \citenamefont
  {Pruessner}}]{cocconi2022scaling}%
  \BibitemOpen
  \bibfield  {author} {\bibinfo {author} {\bibfnamefont {L.}~\bibnamefont
  {Cocconi}}, \bibinfo {author} {\bibfnamefont {G.}~\bibnamefont {Salbreux}},\
  and\ \bibinfo {author} {\bibfnamefont {G.}~\bibnamefont {Pruessner}},\
  }\bibfield  {title} {\bibinfo {title} {Scaling of entropy production under
  coarse graining in active disordered media},\ }\href@noop {} {\bibfield
  {journal} {\bibinfo  {journal} {Phys. Rev. E}\ }\textbf {\bibinfo {volume}
  {105}},\ \bibinfo {pages} {L042601} (\bibinfo {year} {2022})}\BibitemShut
  {NoStop}%
\bibitem [{\citenamefont {Knight}\ \emph {et~al.}(2026)\citenamefont {Knight},
  \citenamefont {Kaveh},\ and\ \citenamefont
  {Pruessner}}]{KnightKavehPruessner:2026}%
  \BibitemOpen
  \bibfield  {author} {\bibinfo {author} {\bibfnamefont {J.}~\bibnamefont
  {Knight}}, \bibinfo {author} {\bibfnamefont {F.}~\bibnamefont {Kaveh}},\ and\
  \bibinfo {author} {\bibfnamefont {G.}~\bibnamefont {Pruessner}},\ }\bibfield
  {title} {\bibinfo {title} {Self-propulsion symmetries determine entropy
  production of active particles with hidden states},\ }\href
  {https://doi.org/10.1103/xbk2-ggcf} {\bibfield  {journal} {\bibinfo
  {journal} {Phys. Rev. Lett.}\ }\textbf {\bibinfo {volume} {136}},\ \bibinfo
  {pages} {198302} (\bibinfo {year} {2026})}\BibitemShut {NoStop}%
\bibitem [{\citenamefont {Mestres}\ \emph {et~al.}(2014)\citenamefont
  {Mestres}, \citenamefont {Martinez}, \citenamefont {{Ortiz-Ambriz}},
  \citenamefont {Rica},\ and\ \citenamefont
  {Roldan}}]{mestresRealizationNonequilibriumThermodynamic2014}%
  \BibitemOpen
  \bibfield  {author} {\bibinfo {author} {\bibfnamefont {P.}~\bibnamefont
  {Mestres}}, \bibinfo {author} {\bibfnamefont {I.~A.}\ \bibnamefont
  {Martinez}}, \bibinfo {author} {\bibfnamefont {A.}~\bibnamefont
  {{Ortiz-Ambriz}}}, \bibinfo {author} {\bibfnamefont {R.~A.}\ \bibnamefont
  {Rica}},\ and\ \bibinfo {author} {\bibfnamefont {E.}~\bibnamefont {Roldan}},\
  }\bibfield  {title} {\bibinfo {title} {Realization of nonequilibrium
  thermodynamic processes using external colored noise},\ }\href
  {https://doi.org/10.1103/PhysRevE.90.032116} {\bibfield  {journal} {\bibinfo
  {journal} {Phys. Rev. E}\ }\textbf {\bibinfo {volume} {90}},\ \bibinfo
  {pages} {032116} (\bibinfo {year} {2014})}\BibitemShut {NoStop}%
\bibitem [{\citenamefont {Di~Bello}\ \emph {et~al.}(2024)\citenamefont
  {Di~Bello}, \citenamefont {Majumdar}, \citenamefont {Marathe}, \citenamefont
  {Metzler},\ and\ \citenamefont
  {Rold{\'a}n}}]{dibelloBrownianParticlePoissonShotNoise2024}%
  \BibitemOpen
  \bibfield  {author} {\bibinfo {author} {\bibfnamefont {C.}~\bibnamefont
  {Di~Bello}}, \bibinfo {author} {\bibfnamefont {R.}~\bibnamefont {Majumdar}},
  \bibinfo {author} {\bibfnamefont {R.}~\bibnamefont {Marathe}}, \bibinfo
  {author} {\bibfnamefont {R.}~\bibnamefont {Metzler}},\ and\ \bibinfo {author}
  {\bibfnamefont {{\'E}.}~\bibnamefont {Rold{\'a}n}},\ }\bibfield  {title}
  {\bibinfo {title} {Brownian particle in a poisson-shot-noise active bath:
  Exact statistics, effective temperature, and inference},\ }\href
  {https://doi.org/https://doi.org/10.1002/andp.202300427} {\bibfield
  {journal} {\bibinfo  {journal} {Ann. Phys.}\ }\textbf {\bibinfo {volume}
  {536}},\ \bibinfo {pages} {2300427} (\bibinfo {year} {2024})}\BibitemShut
  {NoStop}%
\bibitem [{\citenamefont {Tucci}\ \emph {et~al.}(2022)\citenamefont {Tucci},
  \citenamefont {Rold{\'a}n}, \citenamefont {Gambassi}, \citenamefont
  {Belousov}, \citenamefont {Berger}, \citenamefont {Alonso},\ and\
  \citenamefont {Hudspeth}}]{tucciModelingActiveNonMarkovian2022a}%
  \BibitemOpen
  \bibfield  {author} {\bibinfo {author} {\bibfnamefont {G.}~\bibnamefont
  {Tucci}}, \bibinfo {author} {\bibfnamefont {{\'E}.}~\bibnamefont
  {Rold{\'a}n}}, \bibinfo {author} {\bibfnamefont {A.}~\bibnamefont
  {Gambassi}}, \bibinfo {author} {\bibfnamefont {R.}~\bibnamefont {Belousov}},
  \bibinfo {author} {\bibfnamefont {F.}~\bibnamefont {Berger}}, \bibinfo
  {author} {\bibfnamefont {R.~G.}\ \bibnamefont {Alonso}},\ and\ \bibinfo
  {author} {\bibfnamefont {A.~J.}\ \bibnamefont {Hudspeth}},\ }\bibfield
  {title} {\bibinfo {title} {Modeling {{Active Non-Markovian Oscillations}}},\
  }\href {https://doi.org/10.1103/PhysRevLett.129.030603} {\bibfield  {journal}
  {\bibinfo  {journal} {Phys. Rev. Lett.}\ }\textbf {\bibinfo {volume} {129}},\
  \bibinfo {pages} {030603} (\bibinfo {year} {2022})}\BibitemShut {NoStop}%
\bibitem [{\citenamefont {Tailleur}\ and\ \citenamefont
  {Cates}(2008)}]{TailleurCates:2008}%
  \BibitemOpen
  \bibfield  {author} {\bibinfo {author} {\bibfnamefont {J.}~\bibnamefont
  {Tailleur}}\ and\ \bibinfo {author} {\bibfnamefont {M.~E.}\ \bibnamefont
  {Cates}},\ }\bibfield  {title} {\bibinfo {title} {Statistical mechanics of
  interacting run-and-tumble bacteria},\ }\href
  {https://doi.org/10.1103/PhysRevLett.100.218103} {\bibfield  {journal}
  {\bibinfo  {journal} {Phys. Rev. Lett.}\ }\textbf {\bibinfo {volume} {100}},\
  \bibinfo {pages} {218103} (\bibinfo {year} {2008})}\BibitemShut {NoStop}%
\bibitem [{\citenamefont {Cates}\ and\ \citenamefont
  {Tailleur}(2013)}]{CatesTailleur:2013}%
  \BibitemOpen
  \bibfield  {author} {\bibinfo {author} {\bibfnamefont {M.~E.}\ \bibnamefont
  {Cates}}\ and\ \bibinfo {author} {\bibfnamefont {J.}~\bibnamefont
  {Tailleur}},\ }\bibfield  {title} {\bibinfo {title} {When are active brownian
  particles and run-and-tumble particles equivalent? consequences for
  motility-induced phase separation},\ }\href
  {https://doi.org/10.1209/0295-5075/101/20010} {\bibfield  {journal} {\bibinfo
   {journal} {EPL}\ }\textbf {\bibinfo {volume} {101}},\ \bibinfo {pages}
  {20010} (\bibinfo {year} {2013})}\BibitemShut {NoStop}%
\bibitem [{\citenamefont {Gaspard}(2004)}]{gaspard2004time}%
  \BibitemOpen
  \bibfield  {author} {\bibinfo {author} {\bibfnamefont {P.}~\bibnamefont
  {Gaspard}},\ }\bibfield  {title} {\bibinfo {title} {Time-reversed dynamical
  entropy and irreversibility in {M}arkovian random processes},\ }\href
  {https://link.springer.com/article/10.1007/s10955-004-3455-1} {\bibfield
  {journal} {\bibinfo  {journal} {J. Stat. Phys.}\ }\textbf {\bibinfo {volume}
  {117}},\ \bibinfo {pages} {599} (\bibinfo {year} {2004})}\BibitemShut
  {NoStop}%
\bibitem [{\citenamefont {Sekimoto}(2012)}]{Sekimoto:2010}%
  \BibitemOpen
  \bibfield  {author} {\bibinfo {author} {\bibfnamefont {K.}~\bibnamefont
  {Sekimoto}},\ }\bibfield  {title} {\bibinfo {title} {Stochastic energetics}\
  }(\bibinfo  {publisher} {Springer-Verlag},\ \bibinfo {address} {Berlin,
  Germany},\ \bibinfo {year} {2012})\ pp.\ \bibinfo {pages} {I--XVIII,
  1--322}\BibitemShut {NoStop}%
\bibitem [{\citenamefont {Cocconi}\ \emph {et~al.}(2020)\citenamefont
  {Cocconi}, \citenamefont {Garcia-Millan}, \citenamefont {Zhen}, \citenamefont
  {Buturca},\ and\ \citenamefont {Pruessner}}]{cocconi2020entropy}%
  \BibitemOpen
  \bibfield  {author} {\bibinfo {author} {\bibfnamefont {L.}~\bibnamefont
  {Cocconi}}, \bibinfo {author} {\bibfnamefont {R.}~\bibnamefont
  {Garcia-Millan}}, \bibinfo {author} {\bibfnamefont {Z.}~\bibnamefont {Zhen}},
  \bibinfo {author} {\bibfnamefont {B.}~\bibnamefont {Buturca}},\ and\ \bibinfo
  {author} {\bibfnamefont {G.}~\bibnamefont {Pruessner}},\ }\bibfield  {title}
  {\bibinfo {title} {Entropy production in exactly solvable systems},\ }\href
  {https://www.mdpi.com/1099-4300/22/11/1252} {\bibfield  {journal} {\bibinfo
  {journal} {Entropy}\ }\textbf {\bibinfo {volume} {22}},\ \bibinfo {pages}
  {1252} (\bibinfo {year} {2020})}\BibitemShut {NoStop}%
\bibitem [{\citenamefont {T{\"a}uber}(2014)}]{Taeuber:2014}%
  \BibitemOpen
  \bibfield  {author} {\bibinfo {author} {\bibfnamefont {U.~C.}\ \bibnamefont
  {T{\"a}uber}},\ }\href@noop {} {\emph {\bibinfo {title} {Critical
  dynamics}}}\ (\bibinfo  {publisher} {Cambridge University Press},\ \bibinfo
  {address} {Cambridge, UK},\ \bibinfo {year} {2014})\ pp.\ \bibinfo {pages}
  {i--xvi,1--511}\BibitemShut {NoStop}%
\bibitem [{Kni()}]{KnightKavehPruessnerSUPPL:2026}%
  \BibitemOpen
  \bibinfo {note} {Supplemental material of \cite{KnightKavehPruessner:2026}
  available at
  \href{http://link.aps.org/supplemental/10.1103/xbk2-ggcf}{http://link.aps.org/supplemental/10.1103/xbk2-ggcf}.}\BibitemShut
  {Stop}%
\bibitem [{PRR()}]{PRRsupplement}%
  \BibitemOpen
  \bibinfo {note} {See Supplemental Material [url] for technical details, which
  includes
  Refs.~\cite{uhlenbeckTheoryBrownianMotion1930a,bauleTwopointCorrelationFunction2007,Mathematica:13.3,EPR_PRR_github}.}\BibitemShut
  {Stop}%
\bibitem [{\citenamefont {Martin}\ \emph {et~al.}(2021)\citenamefont {Martin},
  \citenamefont {O'Byrne}, \citenamefont {Cates}, \citenamefont {Fodor},
  \citenamefont {Nardini}, \citenamefont {Tailleur},\ and\ \citenamefont {van
  Wijland}}]{martin2021statistical}%
  \BibitemOpen
  \bibfield  {author} {\bibinfo {author} {\bibfnamefont {D.}~\bibnamefont
  {Martin}}, \bibinfo {author} {\bibfnamefont {J.}~\bibnamefont {O'Byrne}},
  \bibinfo {author} {\bibfnamefont {M.~E.}\ \bibnamefont {Cates}}, \bibinfo
  {author} {\bibfnamefont {{\'E}.}~\bibnamefont {Fodor}}, \bibinfo {author}
  {\bibfnamefont {C.}~\bibnamefont {Nardini}}, \bibinfo {author} {\bibfnamefont
  {J.}~\bibnamefont {Tailleur}},\ and\ \bibinfo {author} {\bibfnamefont
  {F.}~\bibnamefont {van Wijland}},\ }\bibfield  {title} {\bibinfo {title}
  {Statistical mechanics of active {O}rnstein-{U}hlenbeck particles},\
  }\href@noop {} {\bibfield  {journal} {\bibinfo  {journal} {Phys. Rev. E}\
  }\textbf {\bibinfo {volume} {103}},\ \bibinfo {pages} {032607} (\bibinfo
  {year} {2021})}\BibitemShut {NoStop}%
\bibitem [{\citenamefont {Szamel}(2014)}]{szamel2014}%
  \BibitemOpen
  \bibfield  {author} {\bibinfo {author} {\bibfnamefont {G.}~\bibnamefont
  {Szamel}},\ }\bibfield  {title} {\bibinfo {title} {Self-propelled particle in
  an external potential: Existence of an effective temperature},\ }\href
  {https://doi.org/10.1103/PhysRevE.90.012111} {\bibfield  {journal} {\bibinfo
  {journal} {Phys. Rev. E}\ }\textbf {\bibinfo {volume} {90}},\ \bibinfo
  {pages} {012111} (\bibinfo {year} {2014})}\BibitemShut {NoStop}%
\bibitem [{\citenamefont {Fodor}\ \emph {et~al.}(2016)\citenamefont {Fodor},
  \citenamefont {Nardini}, \citenamefont {Cates}, \citenamefont {Tailleur},
  \citenamefont {Visco},\ and\ \citenamefont {van Wijland}}]{FodorETAL:2016}%
  \BibitemOpen
  \bibfield  {author} {\bibinfo {author} {\bibfnamefont {E.}~\bibnamefont
  {Fodor}}, \bibinfo {author} {\bibfnamefont {C.}~\bibnamefont {Nardini}},
  \bibinfo {author} {\bibfnamefont {M.~E.}\ \bibnamefont {Cates}}, \bibinfo
  {author} {\bibfnamefont {J.}~\bibnamefont {Tailleur}}, \bibinfo {author}
  {\bibfnamefont {P.}~\bibnamefont {Visco}},\ and\ \bibinfo {author}
  {\bibfnamefont {F.}~\bibnamefont {van Wijland}},\ }\bibfield  {title}
  {\bibinfo {title} {How far from equilibrium is active matter?},\ }\href
  {https://doi.org/10.1103/PhysRevLett.117.038103} {\bibfield  {journal}
  {\bibinfo  {journal} {Phys. Rev. Lett.}\ }\textbf {\bibinfo {volume} {117}},\
  \bibinfo {pages} {038103} (\bibinfo {year} {2016})}\BibitemShut {NoStop}%
\bibitem [{\citenamefont {Bothe}\ and\ \citenamefont
  {Pruessner}(2021)}]{BothePruessner:2021}%
  \BibitemOpen
  \bibfield  {author} {\bibinfo {author} {\bibfnamefont {M.}~\bibnamefont
  {Bothe}}\ and\ \bibinfo {author} {\bibfnamefont {G.}~\bibnamefont
  {Pruessner}},\ }\bibfield  {title} {\bibinfo {title} {Doi-peliti field theory
  of free active ornstein-uhlenbeck particles},\ }\href
  {https://doi.org/10.1103/PhysRevE.103.062105} {\bibfield  {journal} {\bibinfo
   {journal} {Phys. Rev. E}\ }\textbf {\bibinfo {volume} {103}},\ \bibinfo
  {pages} {062105} (\bibinfo {year} {2021})}\BibitemShut {NoStop}%
\bibitem [{\citenamefont {{van Kampen}}(1992)}]{vanKampen:1992}%
  \BibitemOpen
  \bibfield  {author} {\bibinfo {author} {\bibfnamefont {N.~G.}\ \bibnamefont
  {{van Kampen}}},\ }\href@noop {} {\emph {\bibinfo {title} {Stochastic
  Processes in Physics and Chemistry}}}\ (\bibinfo  {publisher} {Elsevier
  Science B. V.},\ \bibinfo {address} {Amsterdam, The Netherlands},\ \bibinfo
  {year} {1992})\ \bibinfo {note} {third impression 2001, enlarged and
  revised}\BibitemShut {NoStop}%
\bibitem [{\citenamefont {Dabelow}\ \emph {et~al.}(2021)\citenamefont
  {Dabelow}, \citenamefont {Bo},\ and\ \citenamefont
  {Eichhorn}}]{DabelowBoEichhorn:2021}%
  \BibitemOpen
  \bibfield  {author} {\bibinfo {author} {\bibfnamefont {L.}~\bibnamefont
  {Dabelow}}, \bibinfo {author} {\bibfnamefont {S.}~\bibnamefont {Bo}},\ and\
  \bibinfo {author} {\bibfnamefont {R.}~\bibnamefont {Eichhorn}},\ }\bibfield
  {title} {\bibinfo {title} {How irreversible are steady-state trajectories of
  a trapped active particle?},\ }\href
  {https://doi.org/10.1088/1742-5468/abe6fd} {\bibfield  {journal} {\bibinfo
  {journal} {J. Stat. Mech.}\ }\textbf {\bibinfo {volume} {2021}},\ \bibinfo
  {pages} {033216} (\bibinfo {year} {2021})}\BibitemShut {NoStop}%
\bibitem [{\citenamefont {Dabelow}\ \emph {et~al.}(2019)\citenamefont
  {Dabelow}, \citenamefont {Bo},\ and\ \citenamefont
  {Eichhorn}}]{dabelowIrreversibilityActiveMatter2019}%
  \BibitemOpen
  \bibfield  {author} {\bibinfo {author} {\bibfnamefont {L.}~\bibnamefont
  {Dabelow}}, \bibinfo {author} {\bibfnamefont {S.}~\bibnamefont {Bo}},\ and\
  \bibinfo {author} {\bibfnamefont {R.}~\bibnamefont {Eichhorn}},\ }\bibfield
  {title} {\bibinfo {title} {Irreversibility in {{Active Matter Systems}}:
  {{Fluctuation Theorem}} and {{Mutual Information}}},\ }\href
  {https://doi.org/10.1103/PhysRevX.9.021009} {\bibfield  {journal} {\bibinfo
  {journal} {Phys. Rev. X}\ }\textbf {\bibinfo {volume} {9}},\ \bibinfo {pages}
  {021009} (\bibinfo {year} {2019})}\BibitemShut {NoStop}%
\bibitem [{\citenamefont {Caprini}\ \emph {et~al.}(2019)\citenamefont
  {Caprini}, \citenamefont {Marconi}, \citenamefont {Puglisi},\ and\
  \citenamefont {Vulpiani}}]{capriniEntropyProductionOrnstein2019}%
  \BibitemOpen
  \bibfield  {author} {\bibinfo {author} {\bibfnamefont {L.}~\bibnamefont
  {Caprini}}, \bibinfo {author} {\bibfnamefont {U.~M.~B.}\ \bibnamefont
  {Marconi}}, \bibinfo {author} {\bibfnamefont {A.}~\bibnamefont {Puglisi}},\
  and\ \bibinfo {author} {\bibfnamefont {A.}~\bibnamefont {Vulpiani}},\
  }\bibfield  {title} {\bibinfo {title} {The entropy production of
  {{Ornstein}}--{{Uhlenbeck}} active particles: A path integral method for
  correlations},\ }\href {https://doi.org/10.1088/1742-5468/ab14dd} {\bibfield
  {journal} {\bibinfo  {journal} {J. Stat. Mech.}\ }\textbf {\bibinfo {volume}
  {2019}},\ \bibinfo {pages} {053203} (\bibinfo {year} {2019})}\BibitemShut
  {NoStop}%
\bibitem [{\citenamefont {Garcia-Millan}\ and\ \citenamefont
  {Pruessner}(2021)}]{garcia2021run}%
  \BibitemOpen
  \bibfield  {author} {\bibinfo {author} {\bibfnamefont {R.}~\bibnamefont
  {Garcia-Millan}}\ and\ \bibinfo {author} {\bibfnamefont {G.}~\bibnamefont
  {Pruessner}},\ }\bibfield  {title} {\bibinfo {title} {Run-and-tumble motion
  in a harmonic potential: {F}ield theory and entropy production},\ }\href
  {https://doi.org/10.1088/1742-5468/ac014d} {\bibfield  {journal} {\bibinfo
  {journal} {J. Stat. Mech.}\ }\textbf {\bibinfo {volume} {2021}},\ \bibinfo
  {pages} {063203} (\bibinfo {year} {2021})}\BibitemShut {NoStop}%
\bibitem [{\citenamefont {Pruessner}\ and\ \citenamefont
  {Garcia-Millan}(2025)}]{PruessnerGarcia-Millan:2025}%
  \BibitemOpen
  \bibfield  {author} {\bibinfo {author} {\bibfnamefont {G.}~\bibnamefont
  {Pruessner}}\ and\ \bibinfo {author} {\bibfnamefont {R.}~\bibnamefont
  {Garcia-Millan}},\ }\bibfield  {title} {\bibinfo {title} {Field theories of
  active particle systems and their entropy production},\ }\href
  {https://doi.org/10.1088/1361-6633/adff30} {\bibfield  {journal} {\bibinfo
  {journal} {Rep. Prog. Phys.}\ }\textbf {\bibinfo {volume} {88}},\ \bibinfo
  {pages} {097601} (\bibinfo {year} {2025})}\BibitemShut {NoStop}%
\bibitem [{\citenamefont
  {Marcinkiewicz}(1939)}]{marcinkiewiczProprieteLoiGauss1939}%
  \BibitemOpen
  \bibfield  {author} {\bibinfo {author} {\bibfnamefont {J.}~\bibnamefont
  {Marcinkiewicz}},\ }\bibfield  {title} {\bibinfo {title} {{Sur une
  propri{\'e}t{\'e} de la loi de Gau{\ss}}},\ }\href
  {https://doi.org/10.1007/BF01210677} {\bibfield  {journal} {\bibinfo
  {journal} {Math. Zeit.}\ }\textbf {\bibinfo {volume} {44}},\ \bibinfo {pages}
  {612} (\bibinfo {year} {1939})}\BibitemShut {NoStop}%
\bibitem [{\citenamefont {Doi}(1976)}]{Doi:1976}%
  \BibitemOpen
  \bibfield  {author} {\bibinfo {author} {\bibfnamefont {M.}~\bibnamefont
  {Doi}},\ }\bibfield  {title} {\bibinfo {title} {Second quantization
  representation for classical many-particle system},\ }\href
  {http://stacks.iop.org/0305-4470/9/1465} {\bibfield  {journal} {\bibinfo
  {journal} {J. Phys. A: Math. Gen.}\ }\textbf {\bibinfo {volume} {9}},\
  \bibinfo {pages} {1465} (\bibinfo {year} {1976})}\BibitemShut {NoStop}%
\bibitem [{\citenamefont {Peliti}(1985)}]{Peliti:1985}%
  \BibitemOpen
  \bibfield  {author} {\bibinfo {author} {\bibfnamefont {L.}~\bibnamefont
  {Peliti}},\ }\bibfield  {title} {\bibinfo {title} {Path integral approach to
  birth-death processes on a lattice},\ }\href@noop {} {\bibfield  {journal}
  {\bibinfo  {journal} {J. Phys. (Paris)}\ }\textbf {\bibinfo {volume} {46}},\
  \bibinfo {pages} {1469} (\bibinfo {year} {1985})}\BibitemShut {NoStop}%
\bibitem [{\citenamefont {Cardy}(2008)}]{Cardy:2008}%
  \BibitemOpen
  \bibfield  {author} {\bibinfo {author} {\bibfnamefont {J.}~\bibnamefont
  {Cardy}},\ }\bibfield  {title} {\bibinfo {title} {Reaction-diffusion
  processes},\ }in\ \href@noop {} {\emph {\bibinfo {booktitle} {Non-equilibrium
  Statistical Mechanics and Turbulence}}},\ \bibinfo {editor} {edited by\
  \bibinfo {editor} {\bibfnamefont {S.}~\bibnamefont {Nazarenko}}\ and\
  \bibinfo {editor} {\bibfnamefont {O.~V.}\ \bibnamefont {Zaboronski}}}\
  (\bibinfo  {publisher} {Cambridge University Press},\ \bibinfo {address}
  {Cambridge, UK},\ \bibinfo {year} {2008})\ pp.\ \bibinfo {pages} {108--161},\
  \bibinfo {note} {{London} Mathematical Society Lecture Note Series: 355,
  preprint available from
  \url{http://www-thphys.physics.ox.ac.uk/people/JohnCardy/warwick.pdf}}\BibitemShut
  {NoStop}%
\bibitem [{\citenamefont {Garcia~Millan}(2020)}]{garcia2020interactions}%
  \BibitemOpen
  \bibfield  {author} {\bibinfo {author} {\bibfnamefont {R.}~\bibnamefont
  {Garcia~Millan}},\ }\emph {\bibinfo {title} {Interactions, correlations and
  collective behaviour in non-equilibrium systems}},\ \href@noop {} {Ph.D.
  thesis},\ \bibinfo  {school} {Imperial College London}, \bibinfo {address}
  {London, UK} (\bibinfo {year} {2020})\BibitemShut {NoStop}%
\bibitem [{\citenamefont {Pavliotis}(2014)}]{Pavliotis2014}%
  \BibitemOpen
  \bibfield  {author} {\bibinfo {author} {\bibfnamefont {G.~A.}\ \bibnamefont
  {Pavliotis}},\ }\href {https://doi.org/10.1007/978-1-4939-1323-7} {\emph
  {\bibinfo {title} {Stochastic Processes and Applications: Diffusion
  Processes, the Fokker-Planck and Langevin Equations}}},\ \bibinfo {series}
  {Texts in Applied Mathematics}, Vol.~\bibinfo {volume} {60}\ (\bibinfo
  {publisher} {Springer},\ \bibinfo {address} {New York, NY},\ \bibinfo {year}
  {2014})\BibitemShut {NoStop}%
\bibitem [{\citenamefont {Sch\"uttler}\ \emph {et~al.}(2025)\citenamefont
  {Sch\"uttler}, \citenamefont {Garcia-Millan}, \citenamefont {Cates},\ and\
  \citenamefont {Loos}}]{schuettler}%
  \BibitemOpen
  \bibfield  {author} {\bibinfo {author} {\bibfnamefont {J.}~\bibnamefont
  {Sch\"uttler}}, \bibinfo {author} {\bibfnamefont {R.}~\bibnamefont
  {Garcia-Millan}}, \bibinfo {author} {\bibfnamefont {M.~E.}\ \bibnamefont
  {Cates}},\ and\ \bibinfo {author} {\bibfnamefont {S.~A.~M.}\ \bibnamefont
  {Loos}},\ }\bibfield  {title} {\bibinfo {title} {Active particles in moving
  traps: Minimum work protocols and information efficiency of work
  extraction},\ }\href {https://doi.org/10.1103/4q4f-1dpx} {\bibfield
  {journal} {\bibinfo  {journal} {Phys. Rev. E}\ }\textbf {\bibinfo {volume}
  {112}},\ \bibinfo {pages} {024119} (\bibinfo {year} {2025})}\BibitemShut
  {NoStop}%
\bibitem [{\citenamefont {Garcia-Millan}\ \emph {et~al.}(2025)\citenamefont
  {Garcia-Millan}, \citenamefont {Sch\"uttler}, \citenamefont {Cates},\ and\
  \citenamefont {Loos}}]{garciamillan2025optimalclosedloopcontrolactive}%
  \BibitemOpen
  \bibfield  {author} {\bibinfo {author} {\bibfnamefont {R.}~\bibnamefont
  {Garcia-Millan}}, \bibinfo {author} {\bibfnamefont {J.}~\bibnamefont
  {Sch\"uttler}}, \bibinfo {author} {\bibfnamefont {M.~E.}\ \bibnamefont
  {Cates}},\ and\ \bibinfo {author} {\bibfnamefont {S.~A.~M.}\ \bibnamefont
  {Loos}},\ }\bibfield  {title} {\bibinfo {title} {Optimal closed-loop control
  of active particles and a minimal information engine},\ }\href
  {https://doi.org/10.1103/fbgp-qpvv} {\bibfield  {journal} {\bibinfo
  {journal} {Phys. Rev. Lett.}\ }\textbf {\bibinfo {volume} {135}},\ \bibinfo
  {pages} {088301} (\bibinfo {year} {2025})},\ \Eprint
  {https://arxiv.org/abs/2407.18542} {arXiv:2407.18542} \BibitemShut {NoStop}%
\bibitem [{\citenamefont {Ghosal}\ and\ \citenamefont
  {Bisker}(2022)}]{ghosalInferringEntropyProduction2022}%
  \BibitemOpen
  \bibfield  {author} {\bibinfo {author} {\bibfnamefont {A.}~\bibnamefont
  {Ghosal}}\ and\ \bibinfo {author} {\bibfnamefont {G.}~\bibnamefont
  {Bisker}},\ }\bibfield  {title} {\bibinfo {title} {Inferring entropy
  production rate from partially observed {{Langevin}} dynamics under
  coarse-graining},\ }\href {https://doi.org/10.1039/D2CP03064K} {\bibfield
  {journal} {\bibinfo  {journal} {Phys. Chem. Chem. Phys.}\ }\textbf {\bibinfo
  {volume} {24}},\ \bibinfo {pages} {24021} (\bibinfo {year}
  {2022})}\BibitemShut {NoStop}%
\bibitem [{\citenamefont {Uhlenbeck}\ and\ \citenamefont
  {Ornstein}(1930)}]{uhlenbeckTheoryBrownianMotion1930a}%
  \BibitemOpen
  \bibfield  {author} {\bibinfo {author} {\bibfnamefont {G.~E.}\ \bibnamefont
  {Uhlenbeck}}\ and\ \bibinfo {author} {\bibfnamefont {L.~S.}\ \bibnamefont
  {Ornstein}},\ }\bibfield  {title} {\bibinfo {title} {On the {{Theory}} of the
  {{Brownian Motion}}},\ }\href {https://doi.org/10.1103/PhysRev.36.823}
  {\bibfield  {journal} {\bibinfo  {journal} {Phys. Rev.}\ }\textbf {\bibinfo
  {volume} {36}},\ \bibinfo {pages} {823} (\bibinfo {year} {1930})}\BibitemShut
  {NoStop}%
\bibitem [{\citenamefont {Baule}\ and\ \citenamefont
  {Friedrich}(2007)}]{bauleTwopointCorrelationFunction2007}%
  \BibitemOpen
  \bibfield  {author} {\bibinfo {author} {\bibfnamefont {A.}~\bibnamefont
  {Baule}}\ and\ \bibinfo {author} {\bibfnamefont {R.}~\bibnamefont
  {Friedrich}},\ }\bibfield  {title} {\bibinfo {title} {Two-point correlation
  function of the fractional {{Ornstein-Uhlenbeck}} process},\ }\href
  {https://doi.org/10.1209/0295-5075/79/60004} {\bibfield  {journal} {\bibinfo
  {journal} {EPL}\ }\textbf {\bibinfo {volume} {79}},\ \bibinfo {pages} {60004}
  (\bibinfo {year} {2007})}\BibitemShut {NoStop}%
\bibitem [{\citenamefont {Wolfram~Research}(2023)}]{Mathematica:13.3}%
  \BibitemOpen
  \bibfield  {author} {\bibinfo {author} {\bibfnamefont {I.}~\bibnamefont
  {Wolfram~Research}},\ }\href@noop {} {\bibinfo {title} {Mathematica, version
  13.3}} (\bibinfo {year} {2023}),\ \bibinfo {note} {champaign, IL}\BibitemShut
  {NoStop}%
\bibitem [{\citenamefont {Knight}(2026)}]{EPR_PRR_github}%
  \BibitemOpen
  \bibfield  {author} {\bibinfo {author} {\bibfnamefont {J.}~\bibnamefont
  {Knight}},\ }\href@noop {} {\bibinfo {title} {Github repository
  jacob-w-knight/partial\_epr.git}} (\bibinfo {year} {2026}),\ \bibinfo {note}
  {\href{https://github.com/jacob-w-knight/Partial_EPR.git}{https://github.com/jacob-w-knight/Partial\_EPR.git}}\BibitemShut
  {NoStop}%
\end{thebibliography}%


\providecommand{\noopsort}[1]{}\providecommand{\singleletter}[1]{#1}%
\begin{thebibliography}{14}%
\makeatletter
\providecommand \@ifxundefined [1]{%
 \@ifx{#1\undefined}
}%
\providecommand \@ifnum [1]{%
 \ifnum #1\expandafter \@firstoftwo
 \else \expandafter \@secondoftwo
 \fi
}%
\providecommand \@ifx [1]{%
 \ifx #1\expandafter \@firstoftwo
 \else \expandafter \@secondoftwo
 \fi
}%
\providecommand \natexlab [1]{#1}%
\providecommand \enquote  [1]{``#1''}%
\providecommand \bibnamefont  [1]{#1}%
\providecommand \bibfnamefont [1]{#1}%
\providecommand \citenamefont [1]{#1}%
\providecommand \href@noop [0]{\@secondoftwo}%
\providecommand \href [0]{\begingroup \@sanitize@url \@href}%
\providecommand \@href[1]{\@@startlink{#1}\@@href}%
\providecommand \@@href[1]{\endgroup#1\@@endlink}%
\providecommand \@sanitize@url [0]{\catcode `\\12\catcode `\$12\catcode
  `\&12\catcode `\#12\catcode `\^12\catcode `\_12\catcode `\%12\relax}%
\providecommand \@@startlink[1]{}%
\providecommand \@@endlink[0]{}%
\providecommand \url  [0]{\begingroup\@sanitize@url \@url }%
\providecommand \@url [1]{\endgroup\@href {#1}{\urlprefix }}%
\providecommand \urlprefix  [0]{URL }%
\providecommand \Eprint [0]{\href }%
\providecommand \doibase [0]{https://doi.org/}%
\providecommand \selectlanguage [0]{\@gobble}%
\providecommand \bibinfo  [0]{\@secondoftwo}%
\providecommand \bibfield  [0]{\@secondoftwo}%
\providecommand \translation [1]{[#1]}%
\providecommand \BibitemOpen [0]{}%
\providecommand \bibitemStop [0]{}%
\providecommand \bibitemNoStop [0]{.\EOS\space}%
\providecommand \EOS [0]{\spacefactor3000\relax}%
\providecommand \BibitemShut  [1]{\csname bibitem#1\endcsname}%
\let\auto@bib@innerbib\@empty
\bibitem [{\citenamefont {Martin}\ \emph {et~al.}(2021)\citenamefont {Martin},
  \citenamefont {O'Byrne}, \citenamefont {Cates}, \citenamefont {Fodor},
  \citenamefont {Nardini}, \citenamefont {Tailleur},\ and\ \citenamefont {van
  Wijland}}]{martin2021statistical}%
  \BibitemOpen
  \bibfield  {author} {\bibinfo {author} {\bibfnamefont {D.}~\bibnamefont
  {Martin}}, \bibinfo {author} {\bibfnamefont {J.}~\bibnamefont {O'Byrne}},
  \bibinfo {author} {\bibfnamefont {M.~E.}\ \bibnamefont {Cates}}, \bibinfo
  {author} {\bibfnamefont {{\'E}.}~\bibnamefont {Fodor}}, \bibinfo {author}
  {\bibfnamefont {C.}~\bibnamefont {Nardini}}, \bibinfo {author} {\bibfnamefont
  {J.}~\bibnamefont {Tailleur}},\ and\ \bibinfo {author} {\bibfnamefont
  {F.}~\bibnamefont {van Wijland}},\ }\bibfield  {title} {\bibinfo {title}
  {Statistical mechanics of active {O}rnstein-{U}hlenbeck particles},\
  }\href@noop {} {\bibfield  {journal} {\bibinfo  {journal} {Phys. Rev. E}\
  }\textbf {\bibinfo {volume} {103}},\ \bibinfo {pages} {032607} (\bibinfo
  {year} {2021})}\BibitemShut {NoStop}%
\bibitem [{\citenamefont {Bothe}\ and\ \citenamefont
  {Pruessner}(2021)}]{BothePruessner:2021}%
  \BibitemOpen
  \bibfield  {author} {\bibinfo {author} {\bibfnamefont {M.}~\bibnamefont
  {Bothe}}\ and\ \bibinfo {author} {\bibfnamefont {G.}~\bibnamefont
  {Pruessner}},\ }\bibfield  {title} {\bibinfo {title} {Doi-peliti field theory
  of free active ornstein-uhlenbeck particles},\ }\href
  {https://doi.org/10.1103/PhysRevE.103.062105} {\bibfield  {journal} {\bibinfo
   {journal} {Phys. Rev. E}\ }\textbf {\bibinfo {volume} {103}},\ \bibinfo
  {pages} {062105} (\bibinfo {year} {2021})}\BibitemShut {NoStop}%
\bibitem [{\citenamefont {Uhlenbeck}\ and\ \citenamefont
  {Ornstein}(1930)}]{uhlenbeckTheoryBrownianMotion1930a}%
  \BibitemOpen
  \bibfield  {author} {\bibinfo {author} {\bibfnamefont {G.~E.}\ \bibnamefont
  {Uhlenbeck}}\ and\ \bibinfo {author} {\bibfnamefont {L.~S.}\ \bibnamefont
  {Ornstein}},\ }\bibfield  {title} {\bibinfo {title} {On the {{Theory}} of the
  {{Brownian Motion}}},\ }\href {https://doi.org/10.1103/PhysRev.36.823}
  {\bibfield  {journal} {\bibinfo  {journal} {Phys. Rev.}\ }\textbf {\bibinfo
  {volume} {36}},\ \bibinfo {pages} {823} (\bibinfo {year} {1930})}\BibitemShut
  {NoStop}%
\bibitem [{\citenamefont {Baule}\ and\ \citenamefont
  {Friedrich}(2007)}]{bauleTwopointCorrelationFunction2007}%
  \BibitemOpen
  \bibfield  {author} {\bibinfo {author} {\bibfnamefont {A.}~\bibnamefont
  {Baule}}\ and\ \bibinfo {author} {\bibfnamefont {R.}~\bibnamefont
  {Friedrich}},\ }\bibfield  {title} {\bibinfo {title} {Two-point correlation
  function of the fractional {{Ornstein-Uhlenbeck}} process},\ }\href
  {https://doi.org/10.1209/0295-5075/79/60004} {\bibfield  {journal} {\bibinfo
  {journal} {EPL}\ }\textbf {\bibinfo {volume} {79}},\ \bibinfo {pages} {60004}
  (\bibinfo {year} {2007})}\BibitemShut {NoStop}%
\bibitem [{\citenamefont {Garcia-Millan}\ and\ \citenamefont
  {Pruessner}(2021)}]{garcia2021run}%
  \BibitemOpen
  \bibfield  {author} {\bibinfo {author} {\bibfnamefont {R.}~\bibnamefont
  {Garcia-Millan}}\ and\ \bibinfo {author} {\bibfnamefont {G.}~\bibnamefont
  {Pruessner}},\ }\bibfield  {title} {\bibinfo {title} {Run-and-tumble motion
  in a harmonic potential: {F}ield theory and entropy production},\ }\href
  {https://doi.org/10.1088/1742-5468/ac014d} {\bibfield  {journal} {\bibinfo
  {journal} {J. Stat. Mech.}\ }\textbf {\bibinfo {volume} {2021}},\ \bibinfo
  {pages} {063203} (\bibinfo {year} {2021})}\BibitemShut {NoStop}%
\bibitem [{\citenamefont {Doi}(1976)}]{Doi:1976}%
  \BibitemOpen
  \bibfield  {author} {\bibinfo {author} {\bibfnamefont {M.}~\bibnamefont
  {Doi}},\ }\bibfield  {title} {\bibinfo {title} {Second quantization
  representation for classical many-particle system},\ }\href
  {http://stacks.iop.org/0305-4470/9/1465} {\bibfield  {journal} {\bibinfo
  {journal} {J. Phys. A: Math. Gen.}\ }\textbf {\bibinfo {volume} {9}},\
  \bibinfo {pages} {1465} (\bibinfo {year} {1976})}\BibitemShut {NoStop}%
\bibitem [{\citenamefont {Peliti}(1985)}]{Peliti:1985}%
  \BibitemOpen
  \bibfield  {author} {\bibinfo {author} {\bibfnamefont {L.}~\bibnamefont
  {Peliti}},\ }\bibfield  {title} {\bibinfo {title} {Path integral approach to
  birth-death processes on a lattice},\ }\href@noop {} {\bibfield  {journal}
  {\bibinfo  {journal} {J. Phys. (Paris)}\ }\textbf {\bibinfo {volume} {46}},\
  \bibinfo {pages} {1469} (\bibinfo {year} {1985})}\BibitemShut {NoStop}%
\bibitem [{\citenamefont {Cardy}(2008)}]{Cardy:2008}%
  \BibitemOpen
  \bibfield  {author} {\bibinfo {author} {\bibfnamefont {J.}~\bibnamefont
  {Cardy}},\ }\bibfield  {title} {\bibinfo {title} {Reaction-diffusion
  processes},\ }in\ \href@noop {} {\emph {\bibinfo {booktitle} {Non-equilibrium
  Statistical Mechanics and Turbulence}}},\ \bibinfo {editor} {edited by\
  \bibinfo {editor} {\bibfnamefont {S.}~\bibnamefont {Nazarenko}}\ and\
  \bibinfo {editor} {\bibfnamefont {O.~V.}\ \bibnamefont {Zaboronski}}}\
  (\bibinfo  {publisher} {Cambridge University Press},\ \bibinfo {address}
  {Cambridge, UK},\ \bibinfo {year} {2008})\ pp.\ \bibinfo {pages} {108--161},\
  \bibinfo {note} {{London} Mathematical Society Lecture Note Series: 355,
  preprint available from
  \url{http://www-thphys.physics.ox.ac.uk/people/JohnCardy/warwick.pdf}}\BibitemShut
  {NoStop}%
\bibitem [{\citenamefont {Pruessner}\ and\ \citenamefont
  {Garcia-Millan}(2025)}]{PruessnerGarcia-Millan:2025}%
  \BibitemOpen
  \bibfield  {author} {\bibinfo {author} {\bibfnamefont {G.}~\bibnamefont
  {Pruessner}}\ and\ \bibinfo {author} {\bibfnamefont {R.}~\bibnamefont
  {Garcia-Millan}},\ }\bibfield  {title} {\bibinfo {title} {Field theories of
  active particle systems and their entropy production},\ }\href
  {https://doi.org/10.1088/1361-6633/adff30} {\bibfield  {journal} {\bibinfo
  {journal} {Rep. Prog. Phys.}\ }\textbf {\bibinfo {volume} {88}},\ \bibinfo
  {pages} {097601} (\bibinfo {year} {2025})}\BibitemShut {NoStop}%
\bibitem [{\citenamefont {Garcia~Millan}(2020)}]{garcia2020interactions}%
  \BibitemOpen
  \bibfield  {author} {\bibinfo {author} {\bibfnamefont {R.}~\bibnamefont
  {Garcia~Millan}},\ }\emph {\bibinfo {title} {Interactions, correlations and
  collective behaviour in non-equilibrium systems}},\ \href@noop {} {Ph.D.
  thesis},\ \bibinfo  {school} {Imperial College London}, \bibinfo {address}
  {London, UK} (\bibinfo {year} {2020})\BibitemShut {NoStop}%
\bibitem [{Kni()}]{KnightKavehPruessnerSUPPL:2026}%
  \BibitemOpen
  \bibinfo {note} {Supplemental material of \cite{KnightKavehPruessner:2026}
  available at
  \href{http://link.aps.org/supplemental/10.1103/xbk2-ggcf}{http://link.aps.org/supplemental/10.1103/xbk2-ggcf}.}\BibitemShut
  {Stop}%
\bibitem [{\citenamefont {Knight}\ \emph {et~al.}(2026)\citenamefont {Knight},
  \citenamefont {Kaveh},\ and\ \citenamefont
  {Pruessner}}]{KnightKavehPruessner:2026}%
  \BibitemOpen
  \bibfield  {author} {\bibinfo {author} {\bibfnamefont {J.}~\bibnamefont
  {Knight}}, \bibinfo {author} {\bibfnamefont {F.}~\bibnamefont {Kaveh}},\ and\
  \bibinfo {author} {\bibfnamefont {G.}~\bibnamefont {Pruessner}},\ }\bibfield
  {title} {\bibinfo {title} {Self-propulsion symmetries determine entropy
  production of active particles with hidden states},\ }\href
  {https://doi.org/10.1103/xbk2-ggcf} {\bibfield  {journal} {\bibinfo
  {journal} {Phys. Rev. Lett.}\ }\textbf {\bibinfo {volume} {136}},\ \bibinfo
  {pages} {198302} (\bibinfo {year} {2026})}\BibitemShut {NoStop}%
\bibitem [{\citenamefont {Wolfram~Research}(2023)}]{Mathematica:13.3}%
  \BibitemOpen
  \bibfield  {author} {\bibinfo {author} {\bibfnamefont {I.}~\bibnamefont
  {Wolfram~Research}},\ }\href@noop {} {\bibinfo {title} {Mathematica, version
  13.3}} (\bibinfo {year} {2023}),\ \bibinfo {note} {champaign, IL}\BibitemShut
  {NoStop}%
\bibitem [{\citenamefont {Knight}(2026)}]{EPR_PRR_github}%
  \BibitemOpen
  \bibfield  {author} {\bibinfo {author} {\bibfnamefont {J.}~\bibnamefont
  {Knight}},\ }\href@noop {} {\bibinfo {title} {Github repository
  jacob-w-knight/partial\_epr.git}} (\bibinfo {year} {2026}),\ \bibinfo {note}
  {\href{https://github.com/jacob-w-knight/Partial_EPR.git}{https://github.com/jacob-w-knight/Partial\_EPR.git}}\BibitemShut
  {NoStop}%
\end{thebibliography}%

\end{document}


\preprint{}

\title{\articleTitle: \texorpdfstring{\\}{}Supplementary Material}

\date{26 May 2026}

\maketitle

\renewcommand{\theequation}{S\arabic{equation}}

\renewcommand{\thefigure}{S\arabic{figure}}

\renewcommand{\thesection}{S\Roman{section}}

\tableofcontents

\section{Ornstein-Uhlenbeck correlation function with respect to starred distribution}\seclabel{OU_corr_star}
In this section we outline a derivation of \Eref{starred_corr_aoup}, the two-time correlation function of the Ornstein-Uhlenbeck process with respect to the starred distribution $\PprobWSTAR[\wpath]$ defined implicitly in \Eref{overline_def_main_with_pot}. This is necessary to evaluate the partial EPR of an active Ornstein-Uhlenbeck process (AOUP) \cite{martin2021statistical,BothePruessner:2021} in a potential, \Eref{aoup_EPR}.

The correlation function of the OU process \Eref{AOUPs_langevin_vel} is well-known \cite{uhlenbeckTheoryBrownianMotion1930a, bauleTwopointCorrelationFunction2007}:
\begin{equation}\elabel{ou_correlation}
        \Wave{w(t_1)w(t_2)} = \frac{D_w}{\mu}\exp{-\mu |t_2-t_1|}\;.
\end{equation}
We express this expectation as
\begin{subequations}
    \begin{align}
        \Wave{w(t_1)w(t_2)}
    &=\int\Dint{w}~w(t_1)w(t_2)~
    \PprobW[\wpath]\;,\\
    \text{with}\quad \PprobW[\wpath] &=\frac{1}{\NC} \exP \left\{-\frac{1}{4D_w} \int_0^T \dint{t} \left(\dot{w}(t) + \mu w(t)\right)^2
    \right\}\;,\elabel{path_prob_ou}
    \end{align}
\end{subequations}
where the path probability \Eref{path_prob_ou} is written explicitly as the Onsager-Machlup functional corresponding to the Langevin equation \Eref{AOUPs_langevin_vel}. The required expectation is 
\begin{subequations}
    \begin{align}
        \WSTARave{w(t_1)w(t_2)}
    &=\int\Dint{w}~w(t_1)w(t_2)~
    \PprobWSTAR[\wpath]\;,\\
    \PprobWSTAR[\wpath] &=\frac{1}{\NC\NC^*(\nu)} \exP \left(-\frac{1}{4D_w} \int_0^T \dint{t} \left(\dot{w}(t) + \mu w(t)\right)^2\right)\Exp{-\frac{\nu^2}{4D} \int_0^T\dint{t} w(t)^2 }
    \;,\elabel{path_prob_ou_star}
    \end{align}
\end{subequations}
where the normalisation (ratio) $\NC^*(\nu)$ depends on $\nu$ due to the $\nu$-dependent exponential in \Eref{path_prob_ou_star}. 
After some rearrangement, the exponent of \Eref{path_prob_ou_star} reads
\begin{multline}
    -\frac{1}{4D_w}\int_0^T\dint{t}\left(\dot{w}(t)^2 + w(t)^2 \left(\mu^2 + \frac{\nu^2 D_w}{D}\right) + 2\mu \dot{w}(t) w(t) \right) \\= 
    -\frac{\mu-\mutilde}{4D_w} \big(w(T)^2-w(0)^2\big)
    -\frac{1}{4D_w}\int_0^T\dint{t}\big(\dot{w}(t) + \mutilde w(t)\big)^2  \;,
\end{multline}
where we have introduced $\mutilde^2 = \mu^2 + \nu^2 D_w/D$ and use that $\int \dint{t}2\dot{w}(t)w(t) = w(t)^2+\text{const}$. The distribution $\PprobWSTAR[\wpath]$ is thus given by 
\begin{equation}\elabel{full_PprobWSTAR}
    \PprobWSTAR[\wpath] =\frac{\exp{-\frac{\mu-\mutilde}{4D_w} \big(w(T)^2-w(0)^2\big)}}{\NC\NC^*(\nu)} \exP \left\{-\frac{1}{4D_w} \int_0^T \dint{t} \left(\dot{w}(t) + \mutilde w(t)\right)^2
    \right\}\;.
\end{equation}
Apart from the surface terms in the pre-factor, \Eref{full_PprobWSTAR} is identical to \Eref{path_prob_ou} after the substitution $\mu \to \mutilde$. As such, the starred correlation function in the steady state of $w(t)$ when the surface terms have no bearing is given by \Eref{ou_correlation} with the substitution $\mu \to \mutilde$, \ie
\begin{equation}\elabel{ou_correlation_star}
        \WSTARave{w(t_1)w(t_2)} = \frac{D_w}{\mutilde}\exp{-\mutilde |t_2-t_1|}\;,
\end{equation}
which is \Eref{starred_corr_aoup} in the main text.

\section{Singly-differentiated correlation functions}\seclabel{correlators_one_dot}
In the present section, we consider the $n$-time correlation function $\Xave{ x(t_1) \sdots x(t_n)}$ of a stochastic process $x(t)$ in the steady state and show that a whole class of integrals over $\Xave{ \xdot(t_1) x(t_2) \sdots x(t_n)}$ vanishes. In particular, we want to demonstrate that 
\begin{equation}\elabel{def_I}
    \IC=\int_0^T \dint{t_1}\!\!\sdots\dint{t_n}
    \WSTARaveS{w(t_1)\sdots w(t_n)}{\cbb}
    \Xave{ \xdot(t_1) x(t_2) \sdots x(t_n)} \ ,
\end{equation}
vanishes, where in fact the connected correlation function $\WSTARaveS{w(t_1)\sdots w(t_n)}{\cbb}$ in the integrand of \Eref{def_I} may be replaced by any function invariant under permutations of $t_1,\sdots,t_n$. Using this property, we may write $\IC$ in \Eref{def_I} as
\begin{equation}\elabel{integral_I_rewritten}
    \IC=\int_0^T \dint{t_1}\!\!\sdots\dint{t_n}
    \WSTARaveS{w(t_1)\sdots w(t_n)}{\cbb}
    \frac{1}{n}
    \left(
    \Xave{ \xdot(t_1) x(t_2) \sdots x(t_n)}
    +\Xave{ x(t_1) \xdot(t_2) \sdots x(t_n)}
    + \ldots
    +\Xave{ x(t_1) x(t_2) \sdots \xdot(t_n)}
    \right) \ ,
\end{equation}
as we demand no time-ordering of the dummy variables $t_1,\sdots,t_n$. We will now show that the large bracketed term in the integrand vanishes, by the same reasoning that led to \Eref{harmonic_cancellation}, so that $\IC=0$. To this end, we write it as
\begin{equation}
   \Xave{ \xdot(t_1) x(t_2) \sdots x(t_n)}
    +\Xave{ x(t_1) \xdot(t_2) \sdots x(t_n)}
    + \ldots
    +\Xave{ x(t_1) x(t_2) \sdots \xdot(t_n)}
    =
\left(\partial_{t_1}+\partial_{t_2}+\ldots+\partial_{t_n}\right)\Xave{ x(t_1) \sdots x(t_n)}\;,
\end{equation}
where $\partial_{t_n} = \partial/\partial t_n$. Since the stochastic process $x(t)$ is in steady state, all correlation functions are time-translation invariant, so that we can shift all times by some finite amount $a$
\begin{equation}\elabel{rrelators_2}
    \Xave{ x(t_1) x(t_2) \sdots x(t_n)} = \Xave{ x(t_1-a) x(t_2-a) \sdots x(t_n-a)}\;.
\end{equation}
Setting $a=t_n$ produces the desired result,
\begin{equation}\elabel{sum_deri_zero}
    \begin{split}
    &\left(\partial_{t_1}+\partial_{t_2}+\ldots+\partial_{t_n}\right)\Xave{ x(t_1) \sdots x(t_n)} \\
    &= \left(\partial_{t_1}+\partial_{t_2}+\ldots+\partial_{t_n}\right)\Xave{ x(t_1-t_n)x(t_2-t_n) \sdots x(0)}\;,\\
        &=  \Xave{ \xdot(t_1-t_n)x(t_2-t_n) \sdots x(t_{n-1}-t_n) x(0)} +\ldots + \Xave{ x(t_1-t_n)x(t_2-t_n) \sdots \xdot(t_{n-1}-t_n) x(0)}  \\
        &\quad- \Big( \Xave{ \xdot(t_1-t_n)x(t_2-t_n) \sdots x(t_{n-1}-t_n) x(0)} + \ldots  + \Xave{ x(t_1-t_n)x(t_2-t_n) \sdots \xdot(t_{n-1}-t_n) x(0)} \Big)\;,\\
        &=0\;,
    \end{split}
\end{equation}
as the $n-1$ terms produced by the differentiation with respect to $t_1,\sdots,t_{n-1}$ cancel with the terms due to the differentiation with respect to $t_n$. Using \Eref{sum_deri_zero} in \Eref{integral_I_rewritten} produces $\IC=0$, as defined in \Eref{def_I}.

\section{Four-time correlator of a run-and-tumble particle in a harmonic potential}\seclabel{four_time_correlator}
In this section we calculate the four-time correlation function $\Xave{x(t_1) x(t_2) x(t_3)x(t_4)}$ of an RnT particle in a harmonic potential. We use the field-theoretic framework outlined in \cite{garcia2021run}. To avoid confusion, 
we replace the polarity fields $\nu(x,t)$ and $\tilde{\nu}(x,t)$ used there by $\chi(x,t)$ and $\tilde{\chi}(x,t)$ in our derivations.

Following App.~I of \cite{garcia2021run}, we express the four-time correlation function in the steady state as
\begin{equation}\elabel{four_time_correlation_prop}
    \Xave{x(t_1) x(t_2) x(t_3)x(t_4)} =  \lim_{t_0 \to -\infty}\int \rmd x_1 \rmd x_2 \rmd x_3 \rmd x_4 \; x_1 x_2 x_3 x_4 \; \GC_4(x_4, t_4;x_3, t_3;x_2, t_2;x_1, t_1;x_0, t_0)\;,
\end{equation}
where $\GC_4(x_4, t_4;x_3, t_3;x_2, t_2;x_1, t_1;x_0, t_0)$ is the four-point density in $x_1,x_2,x_3,x_4$ of an RnT particle to be seen at $x_1$ at time $t_1$, at $x_2$ at time $t_2$, at $x_3$ at time $t_3$ and at $x_4$ at time $t_4$. The particle is initialised at $x_0$ at time $t_0$ with equal probability in either of the two self-propulsion states. In the steady state, $t_0\to-\infty$, this density is independent of $x_0$. With time-ordering $t_4>t_3>t_2>t_1$, field theoretically it can be written as
  \begin{equation}\elabel{propagator}
  \begin{split}
     \GC_4(x_4, t_4;x_3, t_3;x_2, t_2;x_1, t_1;x_0, t_0) =  \Big\langle\rho(x_4,t_4) 
     &\big[\tilderho(x_3,t_3)\rho(x_3,t_3)+\tildechi(x_3,t_3)\chi(x_3,t_3) \big] \\
     &\big[\tilderho(x_2,t_2)\rho(x_2,t_2)+\tildechi(x_2,t_2)\chi(x_2,t_2) \big] \\
     &\big[\tilderho(x_1,t_1)\rho(x_1,t_1)+\tildechi(x_1,t_1)\chi(x_1,t_1) \big]\tilderho(x_0,t_0) 
     \Big\rangle\,,
  \end{split}
\end{equation}
where,
in the Doi-Peliti formalism \cite{Doi:1976,Peliti:1985,Cardy:2008,PruessnerGarcia-Millan:2025},
$\rho(x,t)$ and $\chi(x,t)$ represent density and polarity annihilator fields respectively and $\tilderho(x,t)$ and $\tildechi(x,t)$ are the corresponding Doi-shifted creator fields.
Angular brackets denote the path integral
\begin{equation}
\label{eq:def_angular_ave} 
    \ave{\bullet} = 
    \int \Dint{[\rho,\tilderho,\chi,\tildechi]}
    \exp{-\mathcal{A}[\rho,\tilderho,\chi,\tildechi]}\ \bullet
\end{equation}
with action
\begin{equation}
\label{eq:def_action} 
    \mathcal{A} = \int \dint{x} \dint{t}
    \Big\{
    \tilderho \big( \partial_t \rho 
    - k \partial_x (x\rho) 
    - D \partial_x^2 \rho \big)
+    \tildechi \big( \partial_t \chi
    - k \partial_x (x\chi) 
    - D \partial_x^2 \chi \big)
    + \alpha \chitilde \chi \\
    + \nu \tilderho \partial_x \chi
    + \nu \tildechi \partial_x \rho
    \Big\}
\end{equation}
where $k$ denotes the stiffness of the harmonic potential, $D$ is the diffusivity, $\alpha/2$ is the transmutation rate of self-propulsion states and $\nu$ is the self-propulsion speed.
Details of the application of Doi-Peliti field theory to RnT particles in harmonic potentials can be found in Refs.~\cite{garcia2021run,garcia2020interactions}. By design, $\ave{\rho(x_1,t_1)\tilderho(x_0,t_0)}$ is the probability density of finding at time $t_1$ at position $x_1$ any particle, irrespective of whether it is right-moving or left-moving, after a left-moving or a right-moving particle has been deposited with equal probability at $x_0$ at time $t_0$. The term $\tilderho(x,t)\rho(x,t)+\tildechi(x,t)\chi(x,t)$ corresponds to annihilating and thereby probing for a particle at $x$ and then (re-)creating it in the correct state, right-moving or left-moving.

By inspection of the action, \Eref{def_action}, Hermite polynomials $H_n(z)$ (in the probabilists' convention) with $n\in\{0,1,\ldots,\infty\}$ (the ``Hermite-mode'') help diagonalise all terms in the action that do not mix the two fields, using
\begin{subequations}
\elabel{field_transforms} 
\begin{align}
    \rho(x,t)&=\int\dintbar{\omega}\exp{-\imag\omega t}\frac{1}{L} \sum_{n=0}^\infty  u_n(x) \rho_n(\omega)
    &\qquad&
    \tilderho(x,t)&=\int\dintbar{\omega}\exp{-\imag\omega t}\frac{1}{L} \sum_{n=0}^\infty  \tilde{u}_n(x) \tilderho_n(\omega)    \\
    \chi(x,t)&=\int\dintbar{\omega}\exp{-\imag\omega t}\frac{1}{L} \sum_{n=0}^\infty  u_n(x) \chi_n(\omega)
    &\qquad&
    \tildechi(x,t)&=\int\dintbar{\omega}\exp{-\imag\omega t}\frac{1}{L} \sum_{n=0}^\infty  \tilde{u}_n(x) \tildechi_n(\omega)    
\end{align}
\end{subequations}
where
\begin{equation}
\label{eq:def_u} 
    u_n(x) = \exp{-x^2/(2L^2)} He_n\left(\frac{x}{L}\right)
    \qquad\text{and}\qquad
    \tilde{u}_n(x) = \frac{1}{\sqrt{2\pi}n!} He_n\left(\frac{x}{L}\right) \ ,
\end{equation}
$L=\sqrt{D/k}$ and $He_n(z)=(z-\plaind/\plaind z)^n 1$ such that, say, $He_1(z)=z$. 

Diagrams which contribute to the four-time correlation function are given in \Erefs{diagrams_OU} and \Tref{diagrams_fourth_order}. Each diagram, to be read from right to left, is made up of a sequence of propagators, shown as distinct lines or ``legs'' which connect two consecutive times, as indicated by $t_i$. The endpoints of the legs are further characterised by an index (or Hermite-mode) to the left of the time. Examples of expressions corresponding to different legs can be found in App.~C of \cite{garcia2021run}. Each diagram represents a single contribution to the expectation of the fields on the right-hand side of \Eref{propagator}. As an illustration, the diagrams at order $\OC(\nu^0)$, corresponding to pure Ornstein-Uhlenbeck dynamics, are:
\begin{subequations}\elabel{diagrams_OU}
    \begin{align}
     &\straightstraight{1}{t_4}{1}{t_3}
     \straightstraight{0}{t_3}{0}{t_2}
     \straightstraight{1}{t_2}{1}{t_1}
     \straightstraight{0}{t_1}{0}{t_0}\;,\elabel{first_diag}\\
     &\straightstraight{1}{t_4}{1}{t_3}
     \straightstraight{2}{t_3}{2}{t_2}
     \straightstraight{1}{t_2}{1}{t_1}
     \straightstraight{0}{t_1}{0}{t_0}\;.\elabel{second_diag}
    \end{align}
\end{subequations}

\newcounter{savedequation}
\setcounter{savedequation}{\value{equation}}
\begin{table}[ht!]
    \centering
    \begin{tabular}{|c|c|}
        \hline
        $ \nu^2D$ & $ \nu^4$ \\
        \hline
         \begin{subequations}\nonumber
\resizebox{.5\textwidth}{!}{%
  $\begin{aligned}     
     \nonumber
     \straightwiggle{1}{t_4}{0}{t_3}
     \wigglestraight{1}{t_3}{0}{t_2}
     \straightstraight{1}{t_2}{1}{t_1}
     \straightstraight{0}{t_1}{0}{t_0}\\
     \nonumber
     \straightwiggle{1}{t_4}{0}{t_3}
     \wigglewiggle{1}{t_3}{1}{t_2}
     \wigglewiggle{0}{t_2}{0}{t_1}
     \wigglestraight{1}{t_1}{0}{t_0}\\
     \nonumber
     \straightwiggle{1}{t_4}{0}{t_3}
     \wigglewiggle{1}{t_3}{1}{t_2}
     \wigglestraight{2}{t_2}{1}{t_1}
     \straightstraight{0}{t_1}{0}{t_0}\\
     \nonumber
     \straightwiggle{1}{t_4}{0}{t_3}
     \wigglewiggle{1}{t_3}{1}{t_2}
     \wigglewiggle{2}{t_2}{2}{t_1}
     \wigglestraight{1}{t_1}{0}{t_0}\\
     \nonumber
     \straightstraight{1}{t_4}{1}{t_3}
     \straightstraight{0}{t_3}{0}{t_2}
     \straightwiggle{1}{t_2}{0}{t_1}
     \wigglestraight{1}{t_1}{0}{t_0}\\
     \nonumber
     \straightstraight{1}{t_4}{1}{t_3}
     \straightstraight{0}{t_3}{0}{t_2}
     \straightstraight{1}{t_2}{1}{t_1}
     \straightwigglestraight{2}{t_1}{0}{t_0}\\
     \nonumber
     \straightstraight{1}{t_4}{1}{t_3}
     \straightwigglestraight{2}{t_3}{0}{t_2}
     \straightstraight{1}{t_2}{1}{t_1}
     \straightstraight{0}{t_1}{0}{t_0}\\
     \nonumber
     \straightstraight{1}{t_4}{1}{t_3}
     \straightwiggle{2}{t_3}{1}{t_2}
     \wigglewiggle{0}{t_2}{0}{t_1}
     \wigglestraight{1}{t_1}{0}{t_0}\\
     \nonumber
     \straightstraight{1}{t_4}{1}{t_3}
     \straightwiggle{2}{t_3}{1}{t_2}
     \wigglestraight{2}{t_2}{1}{t_1}
     \straightstraight{0}{t_1}{0}{t_0}\\
     \nonumber
     \straightstraight{1}{t_4}{1}{t_3}
     \straightwiggle{2}{t_3}{1}{t_2}
     \wigglewiggle{2}{t_2}{2}{t_1}
     \wigglestraight{1}{t_1}{0}{t_0}\\
     \nonumber
     \straightstraight{1}{t_4}{1}{t_3}
     \straightstraight{2}{t_3}{2}{t_2}
     \straightwiggle{1}{t_2}{0}{t_1}
     \wigglestraight{1}{t_1}{0}{t_0}\\
     \nonumber
     \straightstraight{1}{t_4}{1}{t_3}
     \straightstraight{2}{t_3}{2}{t_2}
     \straightstraight{1}{t_2}{1}{t_1}
     \straightwigglestraight{2}{t_1}{0}{t_0}\\
     \nonumber
     \straightstraight{1}{t_4}{1}{t_3}
     \straightstraight{2}{t_3}{2}{t_2}
     \straightwigglestraight{3}{t_2}{1}{t_1}
     \straightstraight{0}{t_1}{0}{t_0}\\
     \nonumber
     \straightstraight{1}{t_4}{1}{t_3}
     \straightstraight{2}{t_3}{2}{t_2}
     \straightwiggle{3}{t_2}{2}{t_1}
     \wigglestraight{1}{t_1}{0}{t_0}\\
     \nonumber
     \straightstraight{1}{t_4}{1}{t_3}
     \straightstraight{2}{t_3}{2}{t_2}
     \straightstraight{3}{t_2}{3}{t_1}
     \straightwigglestraight{2}{t_1}{0}{t_0}
       \end{aligned}$%
}
\end{subequations}
& 
\begin{subequations}
     \nonumber
\resizebox{.5\textwidth}{!}{%
  $\begin{aligned}     
     \nonumber
     \straightwiggle{1}{t_4}{0}{t_3}
     \wigglestraight{1}{t_3}{0}{t_2}
     \straightwiggle{1}{t_2}{0}{t_1}
     \wigglestraight{1}{t_1}{0}{t_0}\\
     \nonumber
     \straightwiggle{1}{t_4}{0}{t_3}
     \wigglestraight{1}{t_3}{0}{t_2}
     \straightstraight{1}{t_2}{1}{t_1}
     \straightwigglestraight{2}{t_1}{0}{t_0}\\
     \nonumber
     \straightwiggle{1}{t_4}{0}{t_3}
     \wigglewiggle{1}{t_3}{1}{t_2}
     \wigglestraightwiggle{2}{t_2}{0}{t_1}
     \wigglestraight{1}{t_1}{0}{t_0}\\
     \nonumber
     \straightwiggle{1}{t_4}{0}{t_3}
     \wigglewiggle{1}{t_3}{1}{t_2}
     \wigglestraight{2}{t_2}{1}{t_1}
     \straightwigglestraight{2}{t_1}{0}{t_0}\\
     \nonumber
     \straightwiggle{1}{t_4}{0}{t_3}
     \wigglewiggle{1}{t_3}{1}{t_2}
     \wigglewiggle{2}{t_2}{0}{t_1}
     \wigglestraightwigglestraight{1}{t_1}{0}{t_0}\\
     \nonumber
     \straightstraight{1}{t_4}{1}{t_3}
     \straightwigglestraight{2}{t_3}{0}{t_2}
     \straightwiggle{1}{t_2}{0}{t_1}
     \wigglestraight{1}{t_1}{0}{t_0}\\
     \nonumber
     \straightstraight{1}{t_4}{1}{t_3}
     \straightwigglestraight{2}{t_3}{0}{t_2}
     \straightstraight{1}{t_2}{1}{t_1}
     \straightwigglestraight{2}{t_1}{0}{t_0}\\
     \nonumber
     \straightstraight{1}{t_4}{1}{t_3}
     \straightwiggle{2}{t_3}{1}{t_2}
     \wigglestraightwiggle{2}{t_2}{0}{t_1}
     \wigglestraight{1}{t_1}{0}{t_0}\\
     \nonumber
     \straightstraight{1}{t_4}{1}{t_3}
     \straightwiggle{2}{t_3}{1}{t_2}
     \wigglestraight{2}{t_2}{1}{t_1}
     \straightwigglestraight{2}{t_1}{0}{t_0}\\
     \nonumber
     \straightstraight{1}{t_4}{1}{t_3}
     \straightwiggle{2}{t_3}{1}{t_2}
     \wigglewiggle{2}{t_2}{2}{t_1}
     \wigglestraightwigglestraight{3}{t_1}{0}{t_0}\\
     \nonumber
     \straightstraight{1}{t_4}{1}{t_3}
     \straightstraight{2}{t_3}{2}{t_2}
     \straightwigglestraightwiggle{3}{t_2}{0}{t_1}
     \wigglestraight{1}{t_1}{0}{t_0}\\
     \nonumber
     \straightstraight{1}{t_4}{1}{t_3}
     \straightstraight{2}{t_3}{2}{t_2}
     \straightwigglestraight{3}{t_2}{1}{t_1}
     \straightwigglestraight{2}{t_1}{0}{t_0}\\
     \nonumber
     \straightstraight{1}{t_4}{1}{t_3}
     \straightstraight{2}{t_3}{2}{t_2}
     \straightwiggle{3}{t_2}{2}{t_1}
     \wigglestraightwigglestraight{3}{t_1}{0}{t_0}\\
     \nonumber
     \straightstraight{1}{t_4}{1}{t_3}
     \straightstraight{2}{t_3}{2}{t_2}
     \straightstraight{3}{t_2}{3}{t_1}
     \straightwigglestraightwigglestraight{4}{t_1}{0}{t_0}
       \end{aligned}$%
}
\end{subequations}\\
        \hline
    \end{tabular}
    \caption{Doi-Peliti diagrams contributing to the four-time correlation function $\Xave{x(t_1) x(t_2) x(t_3)x(t_4)}$ of an RnT particle in a harmonic potential, \Eref{four_time_correlation_prop}. The left column contains diagrams with a prefactor of $\nu^2D$, corresponding to two transmutations between propagator types, while the right column contains diagrams with a prefactor of $\nu^4$, corresponding to four transmutations. Rules for constructing diagrams and expressions for each propagator are given in \cite{garcia2021run,garcia2020interactions}. It turns out that only the diagrams in the right column contribute to the time-differentiated correlator \Eref{sm_triple_diff_correlator}.}
    \tlabel{diagrams_fourth_order}
\end{table}
\setcounter{equation}{\value{savedequation}}

We construct a set of rules based on \cite{garcia2021run} to determine the coefficient accompanying each diagram. Coefficients are determined by the gaps or ``junctions'' between ``legs'' of a diagram (each leg connects two times). We consider the generic junction
\begin{equation}
    \begin{split}
     \straightstraightdemoleft{i}{t_n}
     \straightstraightdemoright{j}{t_n}\;.
    \end{split}
\end{equation}
The junction contributes a factor $f_{ij}$ given by 
\begin{equation}\elabel{def_junction}
    f_{ij} =
  \begin{cases}
    j & \text{if } j=i+1\;, \\
    1 & \text{if } j=i-1\;, \\
    0 & \text{otherwise}\;,
  \end{cases}
\end{equation}
so that only junctions with $|i-j|=1$ contribute. The form of $f_{ij}$ is due to the recurrence relation of Hermite polynomials, $x\mathrm{He}_j(x)=\mathrm{He}_{j+1}(x)+j\mathrm{He}_{j-1}=\sum_if_{ij}\mathrm{He}_i(x)$ and the integrals in \Eref{four_time_correlation_prop}.
The coefficients corresponding to the diagrams \Erefs{first_diag} and \eref{second_diag} are $1\cdot 1 \cdot 1 = 1$ and $2\cdot 1 \cdot 1 = 2$ respectively.

The bare propagators \cite[][Eq.~(15)]{garcia2021run} expressed in direct time are
\begin{subequations}
\elabel{def_propagator_diagrams}
\begin{align}
    \straightstraight{n}{t_2}{m}{t_1} & =
    \ave{\rho_n(t_2)\tilderho_m(t_1)} =
    \sqrt{\frac{D}{k}} \delta_{n,m} \theta(t_2-t_1) \exp{-k n (t_2-t_1) } \\
        \wigglewiggle{n}{t_2}{m}{t_1} & =
        \ave{\chi_n(t_2)\tildechi_m(t_1)} =
    \sqrt{\frac{D}{k}} \delta_{n,m} \theta(t_2-t_1) \exp{-(k n + \alpha) (t_2-t_1) }
\end{align}
\end{subequations}
so that \Erefs{diagrams_OU} give for $t_4>t_3>t_2>t_1$ and $t_0\to-\infty$
\[
\left(\frac{D}{k}\right)^2 
\Big(
\Exp{-k(t_4-t_3+t_2-t_1)}
+
2 \Exp{-k(t_4+t_3-t_2-t_1)}
\Big)
\]
also visible in the first line of \Eref{full_four_time_correlator} and, as expected in a pure Ornstein-Uhlenbeck process ($\nu=0$), effectively amounting to $3!!$ terms. Field transmutations between the two ``species'', density $\rho$ and polarity $\chi$, occur with rate $\nu\sqrt{k/D}$ and are mediated by a vertex that enforces that the index increases by one towards the left, Eq.~(17) in \cite{garcia2021run}. By inspection, the $n$-point correlator $\Xave{x(t_1)\sdots x(t_n)}$ with $2m$ field transmutations therefore consists of terms proportional to $D^{n/2}(\nu^2/D)^m$. Because field transmutations can only ever increase the index towards the left, and only junctions can decrease it, it follows that $2m\le n$. The diagram in the last row of the right column of \Tref{diagrams_fourth_order} shows how $2m=4$ field transportations are ``squeezed'' into the rightmost leg.

The two propagators involving a single such field transmutation are
\begin{subequations}\elabel{single_transmut}
\begin{align}
    \straightwiggle{n}{t_2}{m}{t_1} & =
    \ave{\rho_n(t_2)\tildechi_m(t_1)} =
    \nu \delta_{n-1,m} \theta(t_2-t_1) \frac{1}{k-\alpha}
    \big( \exp{- (k m + \alpha) (t_2-t_1) } -  \exp{- k (m + 1) (t_2-t_1) } \big) \\
    \wigglestraight{n}{t_2}{m}{t_1} & =
    \ave{\chi_n(t_2)\tilderho_m(t_1)} =
    \nu \delta_{n-1,m} \theta(t_2-t_1) \frac{1}{k+\alpha} \big( \exp{- k m (t_2-t_1) } -  \exp{- (k (m + 1) +\alpha) (t_2-t_1) } \big) \ .
\end{align}
\end{subequations}

A further fifteen diagrams contribute with prefactor $\nu^2D$ along with fourteen diagrams with prefactor $\nu^4$. These are listed in  \Tref{diagrams_fourth_order}. The sum of these diagrams with their appropriate coefficients yields the time-ordered four-time correlation function, 
\begin{align}\elabel{full_four_time_correlator}
    \Xave{x(t_1) x(t_2) x(t_3)x(t_4)} &= \left(\frac{D}{k}\right)^2 \left( 2 e^{-k(t_4+t_3-t_2-t_1)} + e^{-k(t_4-t_3+t_2-t_1)}\right)\notag \\
    &\hspace{1em}+ \frac{D}{k}\frac{\nu^2}{k(k+\alpha)(k-\alpha)} \bigg[ 2 \alpha \left(2 e^{-k(t_4+t_3-t_2-t_1)} + e^{-k(t_4-t_3+t_2-t_1)}\right)\notag \\
    &\hspace{10em}\ \  + k \Big(e^{-k(t_4-t_3)-\alpha(t_2-t_1)} + e^{-k(t_4-t_2)-\alpha(t_3-t_1)} \notag \\
    &\hspace{13em} +e^{-k(t_4-t_1)-\alpha(t_3-t_2)} +e^{-k(t_3-t_2)-\alpha(t_4-t_1)} \notag \\
    &\hspace{13em} +e^{-k(t_3-t_1)-\alpha(t_4-t_2)} +e^{-k(t_2-t_1)-\alpha(t_4-t_3)} \Big)\bigg]\;\notag \\
    &\hspace{1em}+ \frac{\nu^4}{k^2(k+\alpha)^2(k-\alpha)^2(\alpha+3k)(\alpha-3k)}\notag \\
    &\hspace{2em}\bigg[ \alpha^4 \left(2 e^{-k(t_4+t_3-t_2-t_1)} + e^{-k(t_4-t_3+t_2-t_1)}\right)\notag \\
    &\hspace{2em} \;+ k \alpha^3\bigg(-6e^{-k(t_4+t_3+t_2-3t_1)} \notag \\
    &\hspace{6em}- 2\Big(e^{-\alpha(t_4-t_1)-k(t_3+t_2-2t_1)}+e^{-\alpha(t_3-t_1)-k(t_4+t_2-2t_1)}\notag \\
    &\hspace{8em}+e^{-\alpha(t_2-t_1)-k(t_4+t_3-2t_1)}-e^{-\alpha(t_2-t_1)-k(t_4+t_3-2t_2)}\Big)\notag \\
    &\hspace{6em}-\Big(e^{-k(t_4-t_3)-\alpha(t_2-t_1)} + e^{-k(t_4-t_2)-\alpha(t_3-t_1)} \notag \\
    &\hspace{8em} +e^{-k(t_4-t_1)-\alpha(t_3-t_2)} +e^{-k(t_3-t_2)-\alpha(t_4-t_1)} \notag \\
    &\hspace{8em} +e^{-k(t_3-t_1)-\alpha(t_4-t_2)} +e^{-k(t_2-t_1)-\alpha(t_4-t_3)} \Big)\bigg)\notag \\
    &\hspace{2em}\; + k^2 \alpha^2\bigg(e^{-\alpha(t_4-t_3+t_2-t_1)} -18e^{-k(t_4+t_3-t_2-t_1)}-9e^{-k(t_4-t_3+t_2-t_1)} \notag \\
    &\hspace{6em}+ 8\Big(e^{-\alpha(t_4-t_1)-k(t_3+t_2-2t_1)}+e^{-\alpha(t_3-t_1)-k(t_4+t_2-2t_1)}\notag \\
    &\hspace{8em}+e^{-\alpha(t_2-t_1)-k(t_4+t_3-2t_1)}+e^{-\alpha(t_2-t_1)-k(t_4+t_3-2t_2)}\Big)\bigg)\;\notag \\
    &\hspace{2em} \;+ k^3 \alpha \bigg(6e^{-k(t_4+t_3+t_2-3t_1)} \notag \\
    &\hspace{6em}- 6\Big(e^{-\alpha(t_4-t_1)-k(t_3+t_2-2t_1)}+e^{-\alpha(t_3-t_1)-k(t_4+t_2-2t_1)}\notag \\
    &\hspace{8em}+e^{-\alpha(t_2-t_1)-k(t_4+t_3-2t_1)}-e^{-\alpha(t_2-t_1)-k(t_4+t_3-2t_2)}\Big)\notag \\
    &\hspace{6em}+9\Big(e^{-k(t_4-t_3)-\alpha(t_2-t_1)} + e^{-k(t_4-t_2)-\alpha(t_3-t_1)} \notag \\
    &\hspace{8em} +e^{-k(t_4-t_1)-\alpha(t_3-t_2)} +e^{-k(t_3-t_2)-\alpha(t_4-t_1)} \notag \\
    &\hspace{8em} +e^{-k(t_3-t_1)-\alpha(t_4-t_2)} +e^{-k(t_2-t_1)-\alpha(t_4-t_3)} \Big)\bigg)\;\notag \\ 
    &\hspace{2em} \;+k^4 \Big(-9e^{-\alpha(t_4-t_3+t_2-t_1)}\Big)\bigg]\,,
\end{align}
which has no higher order corrections.

We now calculate  $\mathcal{L}_d \Xave{x(t_1) x(t_2) x(t_3)x(t_4)} $ which appears in the integrand of \Eref{rnt_epr_correlators}, where $\mathcal{L}_d = \triplePartial{1}{2}{3}+\triplePartial{1}{2}{4}+\triplePartial{1}{3}{4}+\triplePartial{2}{3}{4}$, \Eref{def_Ld}. The correlator \Eref{full_four_time_correlator} results in seven types of terms:
\begin{align}\elabel{terms_under_ld}
    \mathcal{L}_d\, e^{-k(t_4+t_3-t_2-t_1)} &= 0\;,\notag \\
        \mathcal{L}_d\, e^{-k(t_4-t_3+t_2-t_1)} &= 0\;,\notag \\
        \mathcal{L}_d\, e^{-k(t_i-t_j)-\alpha(t_k-t_m)} &= 0\;,\notag \\
        \mathcal{L}_d\, e^{-\alpha(t_4-t_3+t_2-t_1)} &= 0\;,\notag \\
        \mathcal{L}_d\, e^{-\alpha(t_i-t_1)-k(t_j+t_k-2t_1)} &= 2k(k+\alpha)^2 e^{-\alpha(t_i-t_1)-k(t_j+t_k-2t_1)}\;,\notag \\
        \mathcal{L}_d\, e^{-\alpha(t_2-t_1)-k(t_4+t_3-2t_2)} &= 2k(k-\alpha)^2 e^{-\alpha(t_2-t_1)-k(t_4+t_3-2t_2)}\;,\notag \\
        \mathcal{L}_d\, e^{-k(t_4+t_3+t_2-3t_1)} &= 8k^3 e^{-k(t_4+t_3+t_2-3t_1)}\;,
\end{align}
where $i,j,k,m$ are distinct in terms indexed with letters. Their time ordering does not enter at this stage. Any term factorising into functions of (distinct) differences of time, \eg 
\begin{equation}
    \Exp{-k(t_4+t_3-t_2-t_1)}=\Exp{-k(t_4-t_2)}\Exp{-k(t_3-t_1)}\;,
\end{equation}
but not 
\begin{equation}
\Exp{-k(t_4+t_3+t_2-3t_1)}=\Exp{-k(t_4-t_1)}\Exp{-k(t_3-t_1)}\Exp{-k(t_2-t_1)}\;,
\end{equation}
vanishes under $\LC_d$, since $\LC_d$ can be expressed in terms of sums of derivatives with respect to any such pairing of times, 
\begin{subequations}\elabel{forms_of_Ld}
\begin{align}
    \LC_d &= 
     \partial_{t_1}\partial_{t_2}(\partial_{t_3}+\partial_{t_4})
    +\partial_{t_3}\partial_{t_4}(\partial_{t_1}+\partial_{t_2})\\
    &= 
     \partial_{t_1}\partial_{t_3}(\partial_{t_2}+\partial_{t_4})
    +\partial_{t_2}\partial_{t_4}(\partial_{t_1}+\partial_{t_3})\\
    &= 
     \partial_{t_1}\partial_{t_4}(\partial_{t_2}+\partial_{t_3})
    +\partial_{t_2}\partial_{t_3}(\partial_{t_1}+\partial_{t_4}) \ .
\end{align}
\end{subequations}
The differentiated correlator is thus given by
\begin{multline}
\elabel{sm_triple_diff_correlator}
\mathcal{L}_d \Xave{x(t_1) x(t_2) x(t_3)x(t_4)}
= \frac{4 \alpha \nu^4 }{(k+\alpha)(k-\alpha)(\alpha+3k)(\alpha-3k)}
   \bigg[
      12k^2 e^{-k(t_4+t_3+t_2-3t_1)}
\\
      + (\alpha - 3k)(k+\alpha)
         \Big(
            e^{-\alpha(t_4-t_1)-k(t_3+t_2-2t_1)}
            +e^{-\alpha(t_3-t_1)-k(t_4+t_2-2t_1)}
            +e^{-\alpha(t_2-t_1)-k(t_4+t_3-2t_1)}
         \Big)
\\
      + (\alpha + 3k)(k-\alpha) \, e^{-\alpha(t_2-t_1)-k(t_4+t_3-2t_2)}
   \bigg] 
\,,
\end{multline}
containing only terms with four transmutations, \ie the right column of \Tref{diagrams_fourth_order}. We do not know of a physical reason that prevents terms with only two transmutations to enter.
By construction, the right-hand side of this \Eref{sm_triple_diff_correlator} applies for the particular time ordering $t_4>t_3>t_2>t_1$, but as the left-hand side is invariant under all permutations of the time arguments, they can always be re-arranged to meet the requirements of the right. This is similar to, say, $\Wave{w(t_1)w(t_2)}$ in \Eref{ou_correlation}, whose right-hand side may be written without a modulus as $(D_w/\mu)\Exp{-\mu(t_2-t_1)}$ only for $t_2\ge t_1$. Yet, in this form, the exponential still produces the correct result for $t_1\ge t_2$ once the values of $t_1$ and $t_2$ are exchanged, which is allowed by the symmetry of $\Wave{w(t_1)w(t_2)}$.

\section{Partial EPR of a Run-and-Tumble particle in a harmonic potential}\seclabel{rnt_epr_derivation}
In this section we use the correlation function derived in \Sref{four_time_correlator} to calculate the leading-order partial EPR of an RnT particle with self-propulsion speed $\nu$ and transmutation rate $\alpha/2$ (tumble rate $\alpha$, whereby the next self-propulsion velocity is chosen with equal probability) in a harmonic potential $V(x) = k x^2/2$ with stiffness $k$. We start from \Eref{rnt_epr_correlators} in simplified form:
\begin{equation}\elabel{rnt_epr_correlators_sm}
    \EPRpartX = \lim_{T \to \infty} \frac{1}{T} \frac{k}{12}\left(\frac{\nu}{2D}\right)^4\int_0^T \dint{t_1}\!\!\sdots\dint{t_4} \WSTARaveS{w(t_1)\sdots w(t_4)}{\cbb}  \LC_d\Xave{x(t_1)x(t_2)x(t_3)x(t_4)}+ \OC\left(\frac{\nu^{6}}{D^{6}}\right)\;.
\end{equation}
According to Suppl.~SIII.A \cite{KnightKavehPruessnerSUPPL:2026} of
\cite{KnightKavehPruessner:2026} with 
$\alpha_+=\alpha_-=\alpha/2$ and $w_+=-w_-=1$, $\WSTARaveS{w(t_1)\sdots w(t_4)}{\cbb} = \Wave{w(t_1)\sdots w(t_4)}^{\cbb} 
= - 2 e^{-\alpha(t_4 + t_3 - t_2-t_1)}$
without corrections in $\nu^2/D$, as $w(t)^2$ is constant, \Eref{w_cumulant}. We will now substitute the differentiated $x$-correlator available from \Eref{sm_triple_diff_correlator} into \Eref{rnt_epr_correlators_sm} and carry out the integration. The particular time-ordering, $t_4>t_3>t_2>t_1$, of the right-hand side of \Eref{sm_triple_diff_correlator} is implemented in the following by replacing the original, unconstrained integral in \Eref{rnt_epr_correlators_sm} of a non-time-ordered integrand $\bullet$ over all $t_1,t_2,t_3,t_4\in[0,T)$ by a constrained integral \cite[][Eq.~(S28)]{KnightKavehPruessnerSUPPL:2026},
\begin{equation}
    \int_0^T \dint{t_1}\dint{t_2}\dint{t_3}\dint{t_4} \bullet = 
    4! \int_\TC \dint{t_1}\dint{t_2}\dint{t_3}\dint{t_4} \bullet := 
    4! \int_0^T \dint{t_4} \int_0^{t_4} \dint{t_3} \int_0^{t_3} \dint{t_2} \int_0^{t_2} \dint{t_1} \bullet
\end{equation}
enforcing $t_4>t_3>t_2>t_1$ and assuming that $\bullet$ is invariant under permutations of $t_1,t_2,t_3,t_4$. Evaluating \Eref{rnt_epr_correlators_sm} requires the following elementary integrals:
%
\begin{subequations}
\begin{align}
    \lim_{T \to \infty}\frac{1}{T}\int_{\mathcal{T}} \rmd t_1 \ldots \rmd t_4 e^{-\alpha(t_4 + t_3 - t_2-t_1)} e^{-k(t_4+t_3+t_2-3t_1)} &= \frac{1}{2(\alpha+k)^2(\alpha+3k)}\;,\\
    \lim_{T \to \infty}\frac{1}{T}\int_{\mathcal{T}} \rmd t_1 \ldots \rmd t_4 e^{-\alpha(t_4 + t_3 - t_2-t_1)} e^{-\alpha(t_4-t_1)-k(t_3+t_2-2t_1)} &= \frac{1}{4\alpha(\alpha+k)(3\alpha+k)}\;,\\
    \lim_{T \to \infty}\frac{1}{T}\int_{\mathcal{T}} \rmd t_1 \ldots \rmd t_4 e^{-\alpha(t_4 + t_3 - t_2-t_1)} e^{-\alpha(t_3-t_1)-k(t_4+t_2-2t_1)}&= \frac{1}{2(\alpha+k)^2(3\alpha+k)}\;,\\
    \lim_{T \to \infty}\frac{1}{T}\int_{\mathcal{T}} \rmd t_1 \ldots \rmd t_4 e^{-\alpha(t_4 + t_3 - t_2-t_1)} e^{-\alpha(t_2-t_1)-k(t_4+t_3-2t_1)} &= \frac{1}{4(\alpha+k)^3}\;,\\
    \lim_{T \to \infty}\frac{1}{T}\int_{\mathcal{T}} \rmd t_1 \ldots \rmd t_4 e^{-\alpha(t_4 + t_3 - t_2-t_1)} e^{-\alpha(t_2-t_1)-k(t_4+t_3-2t_2)} &= \frac{1}{4\alpha(\alpha+k)}\;.
\end{align}
\end{subequations}
After simplification, and anticipating the result of the next section, this finally yields the partial EPR
\begin{equation}\elabel{rnt2_epr_SM}
    \EPRpartX = 
    \left(\frac{\nu^2}{D}\right)^4 
    \frac{\alpha k }{4(\alpha+k)^3(\alpha+3k)^2} + \OC\left(\frac{\nu^{10}}{D^5}\right)\;,
\end{equation}
as stated in \Eref{rnt2_epr} in the main text.

\section{Evaluation of contributions at subsequent orders}\seclabel{sixth_order}
The leading-order $n$ of the expansion \Eref{partial_EP_expansion_with_potential} of the partial EPR of an RnT particle in a harmonic potential according to \Eref{rnt2_epr} is $n=4$, giving rise to a term $\propto\nu^8/D^4$. As the correlator $\Xave{x(t_1) x(t_2) x(t_3)x(t_4)}$ contains terms proportional only to powers $m\le2$ of $D^2(\nu^2/D)^m$, it will not give rise to further corrections.

In this section, we verify that the next term in the expansion \Eref{partial_EP_expansion_with_potential}, $n=6$, does not contribute to the same or lower order in $\nu$ than $n=4$, \ie that terms of order $\nu^0$ and $\nu^2$ in $\Xave{x(t_1) x(t_2) x(t_3)x(t_4) x(t_5) x(t_6)}$ do not contribute to the partial EPR, as they would otherwise give rise to terms $\propto(\nu/D)^6D^3$ or $\propto(\nu/D)^6D^2\nu^2$ overall. The leading order contribution from the next order in $n$ is therefore of the same order as the next order contribution at the leading order in $n$, was it not for the constraint $2m\le n$. That the $n=8$ term contributes without transmutation, contributing overall $(\nu/D)^8 D^4$, can be ruled out as this is a term due to an equilibrium process. Mathematically, as this is the Ornstein-Uhlenbeck process, the correlation functions can be written as a product of pair-correlation functions, depending on pairwise time-differences, which vanish under the action of time-differentiation operators like \Eref{forms_of_Ld}.

In particular, we demonstrate that
\begin{equation}\elabel{next_order}
    \begin{split}
    \lim_{T \to \infty} \frac{2}{T} \frac{1}{6!}\left(\frac{\nu}{2D}\right)^6 \int \rmd t_1 \ldots \rmd t_6 \WSTARaveS{w(t_1)\ldots w(t_6)}{\cbb} \Bigg(\binom{6}{3}k^3 &\Xave{\dot{x}(t_1)\dot{x}(t_2)\dot{x}(t_3)x(t_4)x(t_5)x(t_6)}\\
    +\binom{6}{5}k &\Xave{\dot{x}(t_1)\dot{x}(t_2)\dot{x}(t_3)\dot{x}(t_4)\dot{x}(t_5)x(t_6)}\Bigg)\\
    &\hspace{-3em}\sim \mathcal{O}\left(\frac{\nu^{10}}{D^5}\right)\;.\\
    \end{split}
\end{equation}
To this end we devise an algorithm, implemented in Python, to generate all possible Doi-Peliti diagrams contributing to $\Xave{x(t_1)\sdots x(t_6)}$, then use a computer algebra system (Mathematica, \cite{Mathematica:13.3}) to calculate their derivatives. In the subsequent section, we use the term ``leg'' to refer to a single segment of a Doi-Peliti diagram connecting two times, and the term ``diagram'' to refer to the whole of all six successive legs from time $t_0$ to time $t_6$.  Using this terminology, the two diagrams in \Erefs{diagrams_OU} comprise four legs each.

\subsection{Generation of diagrams}

The algorithm is used to generate all Doi-Peliti diagrams which contribute to the correlator $\Xave{x(t_1)\sdots x(t_6)}$. Diagrams are generated order by order in $\nu$. The logic for generating all diagrams with n legs of order $\nu^z$ is laid out in the following.

We label the legs that make up a diagram contributing to the $n$-time correlation function $\Xave{x(t_1)\sdots x(t_n)}$ by $i=1,2,\ldots,n$ from right to left. Each such leg contains $s_i\ge0$ field transmutation vertices $\propto\nu$, \eg $s_1=1$ in the two diagrams shown in \Erefs{single_transmut}, producing the constraint $\sum_{i=1}^ns_i=z$. Each field transmutation increases a leg's index (or Hermite-mode) by one \cite{garcia2021run}. Each junction between legs increases or decreases the index by one, \Eref{def_junction}, giving rise to a factor that needs to be tracked. Consistent with that factor, the index cannot become negative. We will further denote by $p_i\in\{-1,1\}$ for $i=1,2,\ldots,n-1$ the difference between the right index of the leg $i+1$ to the left of a junction and the left index of leg $i$ to the right. Every diagram begins on the right with index $0$ and terminates on the left with index $1$, so that $\sum_{i=1}^n s_i+ \sum_{i=1}^{n-1} p_i=1$. As $|\sum_{i=1}^{n-1}p_i|\le n-1$ it follows that $z\le n$, \ie the $n$-time correlation function contains no terms beyond $\nu^n$.

In a ``brute-force'' algorithm, we generate all $\{s_1,\sdots,s_n\}\in\{0,1,\sdots,z\}^n$ and all $\{p_1,\sdots,p_{n-1}\}\in\{-1,1\}^{n-1}$, in total $2^{n-1}(z+1)^n$ configurations and retain those if
\begin{itemize}
\item the diagram has the desired order, $\sum_{i=1}^ns_i=z$.
\item the index increases overall by unity, $\sum_{i=1}^ns_i+\sum_{i=1}^{n-1} p_i=1$.
\item the index never becomes negative,
$\sum_{i=1}^j(s_i+p_i)\ge0$ for $j=1,2,\sdots,n-1$.
\end{itemize}

\begin{table}
\begin{tabular}{|b{0.48\textwidth}|b{0.48\textwidth}|}
\hline
\resizebox{0.4777\textwidth}{!}{\straightwiggle{1}{t_6}{0}{t_5}\wigglestraight{1}{t_5}{0}{t_4}\straightstraight{1}{t_4}{1}{t_3}\straightstraight{0}{t_3}{0}{t_2}\straightstraight{1}{t_2}{1}{t_1}\straightstraight{0}{t_1}{0}{t_0}} & \resizebox{0.4777\textwidth}{!}{\straightwiggle{1}{t_6}{0}{t_5}\wigglestraight{1}{t_5}{0}{t_4}\straightstraight{1}{t_4}{1}{t_3}\straightstraight{2}{t_3}{2}{t_2}\straightstraight{1}{t_2}{1}{t_1}\straightstraight{0}{t_1}{0}{t_0}} \\
\resizebox{0.4777\textwidth}{!}{\straightwiggle{1}{t_6}{0}{t_5}\wigglewiggle{1}{t_5}{1}{t_4}\wigglestraight{2}{t_4}{1}{t_3}\straightstraight{0}{t_3}{0}{t_2}\straightstraight{1}{t_2}{1}{t_1}\straightstraight{0}{t_1}{0}{t_0}} & \resizebox{0.4777\textwidth}{!}{\straightwiggle{1}{t_6}{0}{t_5}\wigglewiggle{1}{t_5}{1}{t_4}\wigglestraight{2}{t_4}{1}{t_3}\straightstraight{2}{t_3}{2}{t_2}\straightstraight{1}{t_2}{1}{t_1}\straightstraight{0}{t_1}{0}{t_0}} \\
\resizebox{0.4777\textwidth}{!}{\straightwiggle{1}{t_6}{0}{t_5}\wigglewiggle{1}{t_5}{1}{t_4}\wigglewiggle{0}{t_4}{0}{t_3}\wigglestraight{1}{t_3}{0}{t_2}\straightstraight{1}{t_2}{1}{t_1}\straightstraight{0}{t_1}{0}{t_0}} & \resizebox{0.4777\textwidth}{!}{\straightwiggle{1}{t_6}{0}{t_5}\wigglewiggle{1}{t_5}{1}{t_4}\wigglewiggle{2}{t_4}{2}{t_3}\wigglestraight{1}{t_3}{0}{t_2}\straightstraight{1}{t_2}{1}{t_1}\straightstraight{0}{t_1}{0}{t_0}} \\
\resizebox{0.4777\textwidth}{!}{\straightwiggle{1}{t_6}{0}{t_5}\wigglewiggle{1}{t_5}{1}{t_4}\wigglewiggle{2}{t_4}{2}{t_3}\wigglestraight{3}{t_3}{2}{t_2}\straightstraight{1}{t_2}{1}{t_1}\straightstraight{0}{t_1}{0}{t_0}} & \resizebox{0.4777\textwidth}{!}{\straightwiggle{1}{t_6}{0}{t_5}\wigglewiggle{1}{t_5}{1}{t_4}\wigglewiggle{0}{t_4}{0}{t_3}\wigglewiggle{1}{t_3}{1}{t_2}\wigglestraight{2}{t_2}{1}{t_1}\straightstraight{0}{t_1}{0}{t_0}} \\
\resizebox{0.4777\textwidth}{!}{\straightwiggle{1}{t_6}{0}{t_5}\wigglewiggle{1}{t_5}{1}{t_4}\wigglewiggle{2}{t_4}{2}{t_3}\wigglewiggle{1}{t_3}{1}{t_2}\wigglestraight{2}{t_2}{1}{t_1}\straightstraight{0}{t_1}{0}{t_0}} & \resizebox{0.4777\textwidth}{!}{\straightwiggle{1}{t_6}{0}{t_5}\wigglewiggle{1}{t_5}{1}{t_4}\wigglewiggle{2}{t_4}{2}{t_3}\wigglewiggle{3}{t_3}{3}{t_2}\wigglestraight{2}{t_2}{1}{t_1}\straightstraight{0}{t_1}{0}{t_0}} \\
\resizebox{0.4777\textwidth}{!}{\straightwiggle{1}{t_6}{0}{t_5}\wigglewiggle{1}{t_5}{1}{t_4}\wigglewiggle{0}{t_4}{0}{t_3}\wigglewiggle{1}{t_3}{1}{t_2}\wigglewiggle{0}{t_2}{0}{t_1}\wigglestraight{1}{t_1}{0}{t_0}} & \resizebox{0.4777\textwidth}{!}{\straightwiggle{1}{t_6}{0}{t_5}\wigglewiggle{1}{t_5}{1}{t_4}\wigglewiggle{0}{t_4}{0}{t_3}\wigglewiggle{1}{t_3}{1}{t_2}\wigglewiggle{2}{t_2}{2}{t_1}\wigglestraight{1}{t_1}{0}{t_0}} \\
\resizebox{0.4777\textwidth}{!}{\straightwiggle{1}{t_6}{0}{t_5}\wigglewiggle{1}{t_5}{1}{t_4}\wigglewiggle{2}{t_4}{2}{t_3}\wigglewiggle{1}{t_3}{1}{t_2}\wigglewiggle{0}{t_2}{0}{t_1}\wigglestraight{1}{t_1}{0}{t_0}} & \resizebox{0.4777\textwidth}{!}{\straightwiggle{1}{t_6}{0}{t_5}\wigglewiggle{1}{t_5}{1}{t_4}\wigglewiggle{2}{t_4}{2}{t_3}\wigglewiggle{1}{t_3}{1}{t_2}\wigglewiggle{2}{t_2}{2}{t_1}\wigglestraight{1}{t_1}{0}{t_0}} \\
\resizebox{0.4777\textwidth}{!}{\straightwiggle{1}{t_6}{0}{t_5}\wigglewiggle{1}{t_5}{1}{t_4}\wigglewiggle{2}{t_4}{2}{t_3}\wigglewiggle{3}{t_3}{3}{t_2}\wigglewiggle{2}{t_2}{2}{t_1}\wigglestraight{1}{t_1}{0}{t_0}} & \resizebox{0.4777\textwidth}{!}{\straightstraight{1}{t_6}{1}{t_5}\straightwigglestraight{2}{t_5}{0}{t_4}\straightstraight{1}{t_4}{1}{t_3}\straightstraight{0}{t_3}{0}{t_2}\straightstraight{1}{t_2}{1}{t_1}\straightstraight{0}{t_1}{0}{t_0}} \\
\resizebox{0.4777\textwidth}{!}{\straightstraight{1}{t_6}{1}{t_5}\straightwigglestraight{2}{t_5}{0}{t_4}\straightstraight{1}{t_4}{1}{t_3}\straightstraight{2}{t_3}{2}{t_2}\straightstraight{1}{t_2}{1}{t_1}\straightstraight{0}{t_1}{0}{t_0}} & \resizebox{0.4777\textwidth}{!}{\straightstraight{1}{t_6}{1}{t_5}\straightwiggle{2}{t_5}{1}{t_4}\wigglestraight{2}{t_4}{1}{t_3}\straightstraight{0}{t_3}{0}{t_2}\straightstraight{1}{t_2}{1}{t_1}\straightstraight{0}{t_1}{0}{t_0}} \\
\resizebox{0.4777\textwidth}{!}{\straightstraight{1}{t_6}{1}{t_5}\straightwiggle{2}{t_5}{1}{t_4}\wigglestraight{2}{t_4}{1}{t_3}\straightstraight{2}{t_3}{2}{t_2}\straightstraight{1}{t_2}{1}{t_1}\straightstraight{0}{t_1}{0}{t_0}} & \resizebox{0.4777\textwidth}{!}{\straightstraight{1}{t_6}{1}{t_5}\straightwiggle{2}{t_5}{1}{t_4}\wigglewiggle{0}{t_4}{0}{t_3}\wigglestraight{1}{t_3}{0}{t_2}\straightstraight{1}{t_2}{1}{t_1}\straightstraight{0}{t_1}{0}{t_0}} \\
\resizebox{0.4777\textwidth}{!}{\straightstraight{1}{t_6}{1}{t_5}\straightwiggle{2}{t_5}{1}{t_4}\wigglewiggle{2}{t_4}{2}{t_3}\wigglestraight{1}{t_3}{0}{t_2}\straightstraight{1}{t_2}{1}{t_1}\straightstraight{0}{t_1}{0}{t_0}} & \resizebox{0.4777\textwidth}{!}{\straightstraight{1}{t_6}{1}{t_5}\straightwiggle{2}{t_5}{1}{t_4}\wigglewiggle{2}{t_4}{2}{t_3}\wigglestraight{3}{t_3}{2}{t_2}\straightstraight{1}{t_2}{1}{t_1}\straightstraight{0}{t_1}{0}{t_0}} \\
\resizebox{0.4777\textwidth}{!}{\straightstraight{1}{t_6}{1}{t_5}\straightwiggle{2}{t_5}{1}{t_4}\wigglewiggle{0}{t_4}{0}{t_3}\wigglewiggle{1}{t_3}{1}{t_2}\wigglestraight{2}{t_2}{1}{t_1}\straightstraight{0}{t_1}{0}{t_0}} & \resizebox{0.4777\textwidth}{!}{\straightstraight{1}{t_6}{1}{t_5}\straightwiggle{2}{t_5}{1}{t_4}\wigglewiggle{2}{t_4}{2}{t_3}\wigglewiggle{1}{t_3}{1}{t_2}\wigglestraight{2}{t_2}{1}{t_1}\straightstraight{0}{t_1}{0}{t_0}} \\
\resizebox{0.4777\textwidth}{!}{\straightstraight{1}{t_6}{1}{t_5}\straightwiggle{2}{t_5}{1}{t_4}\wigglewiggle{2}{t_4}{2}{t_3}\wigglewiggle{3}{t_3}{3}{t_2}\wigglestraight{2}{t_2}{1}{t_1}\straightstraight{0}{t_1}{0}{t_0}\;,} & \resizebox{0.4777\textwidth}{!}{\straightstraight{1}{t_6}{1}{t_5}\straightstraight{2}{t_5}{2}{t_4}\straightwiggle{1}{t_4}{0}{t_3}\wigglewiggle{1}{t_3}{1}{t_2}\wigglewiggle{0}{t_2}{0}{t_1}\wigglestraight{1}{t_1}{0}{t_0}} \\
\resizebox{0.4777\textwidth}{!}{\straightstraight{1}{t_6}{1}{t_5}\straightstraight{2}{t_5}{2}{t_4}\straightwiggle{1}{t_4}{0}{t_3}\wigglewiggle{1}{t_3}{1}{t_2}\wigglewiggle{2}{t_2}{2}{t_1}\wigglestraight{1}{t_1}{0}{t_0}} & \resizebox{0.4777\textwidth}{!}{\straightstraight{1}{t_6}{1}{t_5}\straightstraight{2}{t_5}{2}{t_4}\straightwiggle{3}{t_4}{2}{t_3}\wigglewiggle{1}{t_3}{1}{t_2}\wigglewiggle{0}{t_2}{0}{t_1}\wigglestraight{1}{t_1}{0}{t_0}} \\
\resizebox{0.4777\textwidth}{!}{\straightstraight{1}{t_6}{1}{t_5}\straightstraight{2}{t_5}{2}{t_4}\straightwiggle{3}{t_4}{2}{t_3}\wigglewiggle{1}{t_3}{1}{t_2}\wigglewiggle{2}{t_2}{2}{t_1}\wigglestraight{1}{t_1}{0}{t_0}} & \resizebox{0.4777\textwidth}{!}{\straightstraight{1}{t_6}{1}{t_5}\straightstraight{2}{t_5}{2}{t_4}\straightwiggle{3}{t_4}{2}{t_3}\wigglewiggle{3}{t_3}{3}{t_2}\wigglewiggle{2}{t_2}{2}{t_1}\wigglestraight{1}{t_1}{0}{t_0}} \\
\resizebox{0.4777\textwidth}{!}{\straightstraight{1}{t_6}{1}{t_5}\straightstraight{0}{t_5}{0}{t_4}\straightstraight{1}{t_4}{1}{t_3}\straightwigglestraight{2}{t_3}{0}{t_2}\straightstraight{1}{t_2}{1}{t_1}\straightstraight{0}{t_1}{0}{t_0}} & \resizebox{0.4777\textwidth}{!}{\straightstraight{1}{t_6}{1}{t_5}\straightstraight{2}{t_5}{2}{t_4}\straightstraight{1}{t_4}{1}{t_3}\straightwigglestraight{2}{t_3}{0}{t_2}\straightstraight{1}{t_2}{1}{t_1}\straightstraight{0}{t_1}{0}{t_0}} \\
\resizebox{0.4777\textwidth}{!}{\straightstraight{1}{t_6}{1}{t_5}\straightstraight{2}{t_5}{2}{t_4}\straightstraight{3}{t_4}{3}{t_3}\straightwigglestraight{2}{t_3}{0}{t_2}\straightstraight{1}{t_2}{1}{t_1}\straightstraight{0}{t_1}{0}{t_0}} & \resizebox{0.4777\textwidth}{!}{\straightstraight{1}{t_6}{1}{t_5}\straightstraight{2}{t_5}{2}{t_4}\straightstraight{3}{t_4}{3}{t_3}\straightwigglestraight{4}{t_3}{2}{t_2}\straightstraight{1}{t_2}{1}{t_1}\straightstraight{0}{t_1}{0}{t_0}} \\
\resizebox{0.4777\textwidth}{!}{\straightstraight{1}{t_6}{1}{t_5}\straightstraight{0}{t_5}{0}{t_4}\straightstraight{1}{t_4}{1}{t_3}\straightwiggle{2}{t_3}{1}{t_2}\wigglestraight{2}{t_2}{1}{t_1}\straightstraight{0}{t_1}{0}{t_0}} & \resizebox{0.4777\textwidth}{!}{\straightstraight{1}{t_6}{1}{t_5}\straightstraight{2}{t_5}{2}{t_4}\straightstraight{1}{t_4}{1}{t_3}\straightwiggle{2}{t_3}{1}{t_2}\wigglestraight{2}{t_2}{1}{t_1}\straightstraight{0}{t_1}{0}{t_0}} \\
\resizebox{0.4777\textwidth}{!}{\straightstraight{1}{t_6}{1}{t_5}\straightstraight{2}{t_5}{2}{t_4}\straightstraight{3}{t_4}{3}{t_3}\straightwiggle{2}{t_3}{1}{t_2}\wigglestraight{2}{t_2}{1}{t_1}\straightstraight{0}{t_1}{0}{t_0}} & \resizebox{0.4777\textwidth}{!}{\straightstraight{1}{t_6}{1}{t_5}\straightstraight{2}{t_5}{2}{t_4}\straightstraight{3}{t_4}{3}{t_3}\straightwiggle{4}{t_3}{3}{t_2}\wigglestraight{2}{t_2}{1}{t_1}\straightstraight{0}{t_1}{0}{t_0}} \\
\resizebox{0.4777\textwidth}{!}{\straightstraight{1}{t_6}{1}{t_5}\straightstraight{0}{t_5}{0}{t_4}\straightstraight{1}{t_4}{1}{t_3}\straightwiggle{2}{t_3}{1}{t_2}\wigglewiggle{0}{t_2}{0}{t_1}\wigglestraight{1}{t_1}{0}{t_0}} & \resizebox{0.4777\textwidth}{!}{\straightstraight{1}{t_6}{1}{t_5}\straightstraight{0}{t_5}{0}{t_4}\straightstraight{1}{t_4}{1}{t_3}\straightwiggle{2}{t_3}{1}{t_2}\wigglewiggle{2}{t_2}{2}{t_1}\wigglestraight{1}{t_1}{0}{t_0}} \\
\resizebox{0.4777\textwidth}{!}{\straightstraight{1}{t_6}{1}{t_5}\straightstraight{2}{t_5}{2}{t_4}\straightstraight{1}{t_4}{1}{t_3}\straightwiggle{2}{t_3}{1}{t_2}\wigglewiggle{0}{t_2}{0}{t_1}\wigglestraight{1}{t_1}{0}{t_0}} & \resizebox{0.4777\textwidth}{!}{\straightstraight{1}{t_6}{1}{t_5}\straightstraight{2}{t_5}{2}{t_4}\straightstraight{1}{t_4}{1}{t_3}\straightwiggle{2}{t_3}{1}{t_2}\wigglewiggle{2}{t_2}{2}{t_1}\wigglestraight{1}{t_1}{0}{t_0}} \\
\resizebox{0.4777\textwidth}{!}{\straightstraight{1}{t_6}{1}{t_5}\straightstraight{2}{t_5}{2}{t_4}\straightstraight{3}{t_4}{3}{t_3}\straightwiggle{2}{t_3}{1}{t_2}\wigglewiggle{0}{t_2}{0}{t_1}\wigglestraight{1}{t_1}{0}{t_0}} & \resizebox{0.4777\textwidth}{!}{\straightstraight{1}{t_6}{1}{t_5}\straightstraight{2}{t_5}{2}{t_4}\straightstraight{3}{t_4}{3}{t_3}\straightwiggle{2}{t_3}{1}{t_2}\wigglewiggle{2}{t_2}{2}{t_1}\wigglestraight{1}{t_1}{0}{t_0}} \\
\resizebox{0.4777\textwidth}{!}{\straightstraight{1}{t_6}{1}{t_5}\straightstraight{2}{t_5}{2}{t_4}\straightstraight{3}{t_4}{3}{t_3}\straightwiggle{4}{t_3}{3}{t_2}\wigglewiggle{2}{t_2}{2}{t_1}\wigglestraight{1}{t_1}{0}{t_0}} & \resizebox{0.4777\textwidth}{!}{\straightstraight{1}{t_6}{1}{t_5}\straightstraight{0}{t_5}{0}{t_4}\straightstraight{1}{t_4}{1}{t_3}\straightstraight{2}{t_3}{2}{t_2}\straightwigglestraight{3}{t_2}{1}{t_1}\straightstraight{0}{t_1}{0}{t_0}} \\
\resizebox{0.4777\textwidth}{!}{\straightstraight{1}{t_6}{1}{t_5}\straightstraight{2}{t_5}{2}{t_4}\straightstraight{1}{t_4}{1}{t_3}\straightstraight{2}{t_3}{2}{t_2}\straightwigglestraight{3}{t_2}{1}{t_1}\straightstraight{0}{t_1}{0}{t_0}} & \resizebox{0.4777\textwidth}{!}{\straightstraight{1}{t_6}{1}{t_5}\straightstraight{2}{t_5}{2}{t_4}\straightstraight{3}{t_4}{3}{t_3}\straightstraight{2}{t_3}{2}{t_2}\straightwigglestraight{3}{t_2}{1}{t_1}\straightstraight{0}{t_1}{0}{t_0}} \\
\resizebox{0.4777\textwidth}{!}{\straightstraight{1}{t_6}{1}{t_5}\straightstraight{2}{t_5}{2}{t_4}\straightstraight{3}{t_4}{3}{t_3}\straightstraight{4}{t_3}{4}{t_2}\straightwigglestraight{3}{t_2}{1}{t_1}\straightstraight{0}{t_1}{0}{t_0}} & \resizebox{0.4777\textwidth}{!}{\straightstraight{1}{t_6}{1}{t_5}\straightstraight{0}{t_5}{0}{t_4}\straightstraight{1}{t_4}{1}{t_3}\straightstraight{0}{t_3}{0}{t_2}\straightwiggle{1}{t_2}{0}{t_1}\wigglestraight{1}{t_1}{0}{t_0}} \\
\resizebox{0.4777\textwidth}{!}{\straightstraight{1}{t_6}{1}{t_5}\straightstraight{0}{t_5}{0}{t_4}\straightstraight{1}{t_4}{1}{t_3}\straightstraight{2}{t_3}{2}{t_2}\straightwiggle{1}{t_2}{0}{t_1}\wigglestraight{1}{t_1}{0}{t_0}} & \resizebox{0.4777\textwidth}{!}{\straightstraight{1}{t_6}{1}{t_5}\straightstraight{0}{t_5}{0}{t_4}\straightstraight{1}{t_4}{1}{t_3}\straightstraight{2}{t_3}{2}{t_2}\straightwiggle{3}{t_2}{2}{t_1}\wigglestraight{1}{t_1}{0}{t_0}} \\
\resizebox{0.4777\textwidth}{!}{\straightstraight{1}{t_6}{1}{t_5}\straightstraight{2}{t_5}{2}{t_4}\straightstraight{1}{t_4}{1}{t_3}\straightstraight{0}{t_3}{0}{t_2}\straightwiggle{1}{t_2}{0}{t_1}\wigglestraight{1}{t_1}{0}{t_0}} & \resizebox{0.4777\textwidth}{!}{\straightstraight{1}{t_6}{1}{t_5}\straightstraight{2}{t_5}{2}{t_4}\straightstraight{1}{t_4}{1}{t_3}\straightstraight{2}{t_3}{2}{t_2}\straightwiggle{1}{t_2}{0}{t_1}\wigglestraight{1}{t_1}{0}{t_0}} \\
\resizebox{0.4777\textwidth}{!}{\straightstraight{1}{t_6}{1}{t_5}\straightstraight{2}{t_5}{2}{t_4}\straightstraight{1}{t_4}{1}{t_3}\straightstraight{2}{t_3}{2}{t_2}\straightwiggle{3}{t_2}{2}{t_1}\wigglestraight{1}{t_1}{0}{t_0}} & \resizebox{0.4777\textwidth}{!}{\straightstraight{1}{t_6}{1}{t_5}\straightstraight{2}{t_5}{2}{t_4}\straightstraight{3}{t_4}{3}{t_3}\straightstraight{2}{t_3}{2}{t_2}\straightwiggle{1}{t_2}{0}{t_1}\wigglestraight{1}{t_1}{0}{t_0}} \\
\resizebox{0.4777\textwidth}{!}{\straightstraight{1}{t_6}{1}{t_5}\straightstraight{2}{t_5}{2}{t_4}\straightstraight{3}{t_4}{3}{t_3}\straightstraight{2}{t_3}{2}{t_2}\straightwiggle{3}{t_2}{2}{t_1}\wigglestraight{1}{t_1}{0}{t_0}} & \resizebox{0.4777\textwidth}{!}{\straightstraight{1}{t_6}{1}{t_5}\straightstraight{2}{t_5}{2}{t_4}\straightstraight{3}{t_4}{3}{t_3}\straightstraight{4}{t_3}{4}{t_2}\straightwiggle{3}{t_2}{2}{t_1}\wigglestraight{1}{t_1}{0}{t_0}} \\
\resizebox{0.4777\textwidth}{!}{\straightstraight{1}{t_6}{1}{t_5}\straightstraight{0}{t_5}{0}{t_4}\straightstraight{1}{t_4}{1}{t_3}\straightstraight{0}{t_3}{0}{t_2}\straightstraight{1}{t_2}{1}{t_1}\straightwigglestraight{2}{t_1}{0}{t_0}} & \resizebox{0.4777\textwidth}{!}{\straightstraight{1}{t_6}{1}{t_5}\straightstraight{0}{t_5}{0}{t_4}\straightstraight{1}{t_4}{1}{t_3}\straightstraight{2}{t_3}{2}{t_2}\straightstraight{1}{t_2}{1}{t_1}\straightwigglestraight{2}{t_1}{0}{t_0}} \\
\resizebox{0.4777\textwidth}{!}{\straightstraight{1}{t_6}{1}{t_5}\straightstraight{0}{t_5}{0}{t_4}\straightstraight{1}{t_4}{1}{t_3}\straightstraight{2}{t_3}{2}{t_2}\straightstraight{3}{t_2}{3}{t_1}\straightwigglestraight{2}{t_1}{0}{t_0}} & \resizebox{0.4777\textwidth}{!}{\straightstraight{1}{t_6}{1}{t_5}\straightstraight{2}{t_5}{2}{t_4}\straightstraight{1}{t_4}{1}{t_3}\straightstraight{0}{t_3}{0}{t_2}\straightstraight{1}{t_2}{1}{t_1}\straightwigglestraight{2}{t_1}{0}{t_0}} \\
\resizebox{0.4777\textwidth}{!}{\straightstraight{1}{t_6}{1}{t_5}\straightstraight{2}{t_5}{2}{t_4}\straightstraight{1}{t_4}{1}{t_3}\straightstraight{2}{t_3}{2}{t_2}\straightstraight{1}{t_2}{1}{t_1}\straightwigglestraight{2}{t_1}{0}{t_0}} & \resizebox{0.4777\textwidth}{!}{\straightstraight{1}{t_6}{1}{t_5}\straightstraight{2}{t_5}{2}{t_4}\straightstraight{1}{t_4}{1}{t_3}\straightstraight{2}{t_3}{2}{t_2}\straightstraight{3}{t_2}{3}{t_1}\straightwigglestraight{2}{t_1}{0}{t_0}} \\
\resizebox{0.4777\textwidth}{!}{\straightstraight{1}{t_6}{1}{t_5}\straightstraight{2}{t_5}{2}{t_4}\straightstraight{3}{t_4}{3}{t_3}\straightstraight{2}{t_3}{2}{t_2}\straightstraight{1}{t_2}{1}{t_1}\straightwigglestraight{2}{t_1}{0}{t_0}} & \resizebox{0.4777\textwidth}{!}{\straightstraight{1}{t_6}{1}{t_5}\straightstraight{2}{t_5}{2}{t_4}\straightstraight{3}{t_4}{3}{t_3}\straightstraight{2}{t_3}{2}{t_2}\straightstraight{3}{t_2}{3}{t_1}\straightwigglestraight{2}{t_1}{0}{t_0}} \\
\resizebox{0.4777\textwidth}{!}{\straightstraight{1}{t_6}{1}{t_5}\straightstraight{2}{t_5}{2}{t_4}\straightstraight{3}{t_4}{3}{t_3}\straightstraight{4}{t_3}{4}{t_2}\straightstraight{3}{t_2}{3}{t_1}\straightwigglestraight{2}{t_1}{0}{t_0}} & \resizebox{0.4777\textwidth}{!}{\straightstraight{1}{t_6}{1}{t_5}\straightwiggle{2}{t_5}{1}{t_4}\wigglewiggle{0}{t_4}{0}{t_3}\wigglewiggle{1}{t_3}{1}{t_2}\wigglewiggle{0}{t_2}{0}{t_1}\wigglestraight{1}{t_1}{0}{t_0}} \\
\resizebox{0.4777\textwidth}{!}{\straightstraight{1}{t_6}{1}{t_5}\straightwiggle{2}{t_5}{1}{t_4}\wigglewiggle{0}{t_4}{0}{t_3}\wigglewiggle{1}{t_3}{1}{t_2}\wigglewiggle{2}{t_2}{2}{t_1}\wigglestraight{1}{t_1}{0}{t_0}} & \resizebox{0.4777\textwidth}{!}{\straightstraight{1}{t_6}{1}{t_5}\straightwiggle{2}{t_5}{1}{t_4}\wigglewiggle{2}{t_4}{2}{t_3}\wigglewiggle{1}{t_3}{1}{t_2}\wigglewiggle{0}{t_2}{0}{t_1}\wigglestraight{1}{t_1}{0}{t_0}} \\
\resizebox{0.4777\textwidth}{!}{\straightstraight{1}{t_6}{1}{t_5}\straightwiggle{2}{t_5}{1}{t_4}\wigglewiggle{2}{t_4}{2}{t_3}\wigglewiggle{1}{t_3}{1}{t_2}\wigglewiggle{2}{t_2}{2}{t_1}\wigglestraight{1}{t_1}{0}{t_0}} & \resizebox{0.4777\textwidth}{!}{\straightstraight{1}{t_6}{1}{t_5}\straightwiggle{2}{t_5}{1}{t_4}\wigglewiggle{2}{t_4}{2}{t_3}\wigglewiggle{3}{t_3}{3}{t_2}\wigglewiggle{2}{t_2}{2}{t_1}\wigglestraight{1}{t_1}{0}{t_0}} \\
\resizebox{0.4777\textwidth}{!}{\straightstraight{1}{t_6}{1}{t_5}\straightstraight{2}{t_5}{2}{t_4}\straightwigglestraight{3}{t_4}{1}{t_3}\straightstraight{0}{t_3}{0}{t_2}\straightstraight{1}{t_2}{1}{t_1}\straightstraight{0}{t_1}{0}{t_0}} & \resizebox{0.4777\textwidth}{!}{\straightstraight{1}{t_6}{1}{t_5}\straightstraight{2}{t_5}{2}{t_4}\straightwigglestraight{3}{t_4}{1}{t_3}\straightstraight{2}{t_3}{2}{t_2}\straightstraight{1}{t_2}{1}{t_1}\straightstraight{0}{t_1}{0}{t_0}} \\
\resizebox{0.4777\textwidth}{!}{\straightstraight{1}{t_6}{1}{t_5}\straightstraight{0}{t_5}{0}{t_4}\straightwiggle{1}{t_4}{0}{t_3}\wigglestraight{1}{t_3}{0}{t_2}\straightstraight{1}{t_2}{1}{t_1}\straightstraight{0}{t_1}{0}{t_0}} & \resizebox{0.4777\textwidth}{!}{\straightstraight{1}{t_6}{1}{t_5}\straightstraight{2}{t_5}{2}{t_4}\straightwiggle{1}{t_4}{0}{t_3}\wigglestraight{1}{t_3}{0}{t_2}\straightstraight{1}{t_2}{1}{t_1}\straightstraight{0}{t_1}{0}{t_0}} \\
\resizebox{0.4777\textwidth}{!}{\straightstraight{1}{t_6}{1}{t_5}\straightstraight{2}{t_5}{2}{t_4}\straightwiggle{3}{t_4}{2}{t_3}\wigglestraight{1}{t_3}{0}{t_2}\straightstraight{1}{t_2}{1}{t_1}\straightstraight{0}{t_1}{0}{t_0}} & \resizebox{0.4777\textwidth}{!}{\straightstraight{1}{t_6}{1}{t_5}\straightstraight{2}{t_5}{2}{t_4}\straightwiggle{3}{t_4}{2}{t_3}\wigglestraight{3}{t_3}{2}{t_2}\straightstraight{1}{t_2}{1}{t_1}\straightstraight{0}{t_1}{0}{t_0}} \\
\resizebox{0.4777\textwidth}{!}{\straightstraight{1}{t_6}{1}{t_5}\straightstraight{0}{t_5}{0}{t_4}\straightwiggle{1}{t_4}{0}{t_3}\wigglewiggle{1}{t_3}{1}{t_2}\wigglestraight{2}{t_2}{1}{t_1}\straightstraight{0}{t_1}{0}{t_0}} & \resizebox{0.4777\textwidth}{!}{\straightstraight{1}{t_6}{1}{t_5}\straightstraight{2}{t_5}{2}{t_4}\straightwiggle{1}{t_4}{0}{t_3}\wigglewiggle{1}{t_3}{1}{t_2}\wigglestraight{2}{t_2}{1}{t_1}\straightstraight{0}{t_1}{0}{t_0}} \\
\resizebox{0.4777\textwidth}{!}{\straightstraight{1}{t_6}{1}{t_5}\straightstraight{2}{t_5}{2}{t_4}\straightwiggle{3}{t_4}{2}{t_3}\wigglewiggle{1}{t_3}{1}{t_2}\wigglestraight{2}{t_2}{1}{t_1}\straightstraight{0}{t_1}{0}{t_0}} & \resizebox{0.4777\textwidth}{!}{\straightstraight{1}{t_6}{1}{t_5}\straightstraight{2}{t_5}{2}{t_4}\straightwiggle{3}{t_4}{2}{t_3}\wigglewiggle{3}{t_3}{3}{t_2}\wigglestraight{2}{t_2}{1}{t_1}\straightstraight{0}{t_1}{0}{t_0}} \\
\resizebox{0.4777\textwidth}{!}{\straightstraight{1}{t_6}{1}{t_5}\straightstraight{0}{t_5}{0}{t_4}\straightwiggle{1}{t_4}{0}{t_3}\wigglewiggle{1}{t_3}{1}{t_2}\wigglewiggle{0}{t_2}{0}{t_1}\wigglestraight{1}{t_1}{0}{t_0}} & \resizebox{0.4777\textwidth}{!}{\straightstraight{1}{t_6}{1}{t_5}\straightstraight{0}{t_5}{0}{t_4}\straightwiggle{1}{t_4}{0}{t_3}\wigglewiggle{1}{t_3}{1}{t_2}\wigglewiggle{2}{t_2}{2}{t_1}\wigglestraight{1}{t_1}{0}{t_0}} \\
\hline
\end{tabular}
\caption{\tlabel{diagrams_sixth_order} All 84 Doi-Peliti diagrams contributing with pre-factor $\nu^2D^2$ ($z=2$) to the six-time correlation
function $\Xave{x(t_1) x(t_2) x(t_3)x(t_4) x(t_5) x(t_6)}$ of an RnT
particle in a harmonic potential. Rules for constructing diagrams and
expressions for each propagator are given in
\cite{garcia2021run,garcia2020interactions}.}
\end{table}

For $n=4$, this algorithm reproduces the diagrams shown in \Eref{diagrams_OU} for $z=0$ and in \Tref{w_symmetries} for $z=2$ and $z=4$. To verify that  \Erefs{rnt2_epr_SM} correctly states all leading order terms or equivalently \eref{next_order}, diagrams must be calculated for $n=6$ and $z=0,2$. The algorithm produces five diagrams to order $\nu^0D^3$, corresponding to $z=0$:
\begin{align*}
\straightstraightX{1}{t_6}{1}{t_5}{s_6=0,p_5=1}
\straightstraightX{0}{t_5}{0}{t_4}{s_5=0,p_4=-1}
\straightstraightX{1}{t_4}{1}{t_3}{s_4=0,p_3=1}
\straightstraightX{0}{t_3}{0}{t_2}{s_3=0,p_2=-1}
\straightstraightX{1}{t_2}{1}{t_1}{s_2=0,p_1=1}
\straightstraightX{0}{t_1}{0}{t_0}{s_1=0}\;,\\ 
\straightstraightX{1}{t_6}{1}{t_5}{s_6=0,p_5=1}
\straightstraightX{0}{t_5}{0}{t_4}{s_5=0,p_4=-1}
\straightstraightX{1}{t_4}{1}{t_3}{s_4=0,p_3=-1}
\straightstraightX{2}{t_3}{2}{t_2}{s_3=0,p_2=1}
\straightstraightX{1}{t_2}{1}{t_1}{s_2=0,p_1=1}
\straightstraightX{0}{t_1}{0}{t_0}{s_1=0}\;,\\ 
\straightstraightX{1}{t_6}{1}{t_5}{s_6=0,p_5=-1}
\straightstraightX{2}{t_5}{2}{t_4}{s_5=0,p_4=1}
\straightstraightX{1}{t_4}{1}{t_3}{s_4=0,p_3=1}
\straightstraightX{0}{t_3}{0}{t_2}{s_3=0,p_2=-1}
\straightstraightX{1}{t_2}{1}{t_1}{s_2=0,p_1=1}
\straightstraightX{0}{t_1}{0}{t_0}{s_1=0}\;,\\ 
\straightstraightX{1}{t_6}{1}{t_5}{s_6=0,p_5=-1}
\straightstraightX{2}{t_5}{2}{t_4}{s_5=0,p_4=1}
\straightstraightX{1}{t_4}{1}{t_3}{s_4=0,p_3=-1}
\straightstraightX{2}{t_3}{2}{t_2}{s_3=0,p_2=1}
\straightstraightX{1}{t_2}{1}{t_1}{s_2=0,p_1=1}
\straightstraightX{0}{t_1}{0}{t_0}{s_1=0}\;,\\ 
\straightstraightX{1}{t_6}{1}{t_5}{s_6=0,p_5=-1}
\straightstraightX{2}{t_5}{2}{t_4}{s_5=0,p_4=-1}
\straightstraightX{3}{t_4}{3}{t_3}{s_4=0,p_3=1}
\straightstraightX{2}{t_3}{2}{t_2}{s_3=0,p_2=1}
\straightstraightX{1}{t_2}{1}{t_1}{s_2=0,p_1=1}
\straightstraightX{0}{t_1}{0}{t_0}{s_1=0}\;. 
\end{align*}
We do not expect the diagrams above to contribute, as they characterise the equilibrium process, the six-time correlation function of a Brownian particles in a harmonic trap, $\nu=0$. As a sanity check, we identify effectively $1+2+2+4+6=5!!$ terms from the pre-factors due to the junctions. That these terms do not contribute is confirmed below, \Erefs{deris_of_six}. If they were to contribute, they would produce a partial EPR of order $\nu^6D^3$.

There are a further 84 diagrams corresponding to $z=2$, as shown in \Tref{diagrams_sixth_order}. Generating those using the algorithm outlined above was the aim of the present section.

\subsection{Differentiation using a computer algebra system}

Each diagram is evaluated algebraically using Mathematica \cite{Mathematica:13.3}. Diagrams are then differentiated according to \Eref{next_order} and then summed. After symmetrising both terms in the large bracket of \Eref{next_order} vanish to order $\nu^2$, \ie at $z=0$ and $z=2$,
\begin{subequations}\elabel{deris_of_six}
\begin{align}
    \left(\partial_{t_1}\partial_{t_2}\partial_{t_3} + \ldots + \partial_{t_4}\partial_{t_5}\partial_{t_6}\right) \sum \big[\text{diagrams at $z\in\{0,2\}$}\big]=0\;,\\
    \left(\partial_{t_1}\partial_{t_2}\partial_{t_3}\partial_{t_4}\partial_{t_5} + \ldots + \partial_{t_2}\partial_{t_3}\partial_{t_4}\partial_{t_5}\partial_{t_6}\right) \sum \big[\text{diagrams at $z\in\{0,2\}$}\big]=0\;.
\end{align}
\end{subequations}
If order $z=4$ contributes, then the resulting contribution to the partial EPR is $\propto(\nu/D)^6 D^3(\nu^2/D)^2=\nu^{10}/D^5$ overall, confirming \Eref{next_order}.

Mathematica and Python files can be found in our public Github repository \cite{EPR_PRR_github}. 

\bibliography{references}